\documentclass[12pt]{article}
\usepackage{amsmath,amsfonts,amssymb,graphics,psfrag}
\usepackage{mathrsfs, dsfont}
\usepackage{multirow,array,epsfig,multirow,stmaryrd,graphicx}
\usepackage{comment}
\usepackage{sectsty}
\usepackage[letterpaper,includefoot,top=1in,bottom=1in,left=1in,right=1in]{geometry}
\usepackage[format=hang,margin=1.5cm,font=small,labelfont=bf]{caption}
\usepackage{subfigure}

\usepackage[dvips,bookmarks,breaklinks]{hyperref}		
\usepackage[all,cmtip]{xy}

\def\hybrid{\topmargin -20pt    \oddsidemargin 0pt
        \headheight 0pt \headsep 0pt
        \textwidth 6.25in       
        \textheight 9 in       
        \marginparwidth .875in
        \parskip 5pt plus 1pt 
          \jot = 1.5ex
   }
\hybrid
\numberwithin{equation}{section}
\numberwithin{table}{section}

\sectionfont{\large\bfseries}
\subsectionfont{\normalsize\bfseries}
\subsubsectionfont{\normalsize\it}

\newcommand{\beq}{\begin{equation}}
\newcommand{\eeq}{\end{equation}}
\newcommand{\be}{\begin{equation}} 
\newcommand{\ee}{\end{equation}}
\newcommand{\bea}{\begin{eqnarray}}
\newcommand{\eea}{\end{eqnarray}}   
\newcommand{\ben}{\begin{eqnarray*}}
\newcommand{\een}{\end{eqnarray*}}                  
\newcommand{\ba}{\begin{aligned}}
\newcommand{\ea}{\end{aligned}}
\newcommand{\bt}{\begin{tabular}}
\newcommand{\et}{\end{tabular}}
\newcommand{\bc}{\begin{center}}
\newcommand{\ec}{\end{center}}

\newcommand{\cO}{\mathcal{O}}

\newcommand{\cE}{\mathcal{E}}

\newcommand{\cC}{\mathcal{C}}

\newcommand{\cL}{\mathcal{L}}

\newcommand{\cK}{\mathcal{K}}
\newcommand{\cN}{\mathcal{N}}

\newcommand{\cR}{\mathcal{R}}

\newcommand{\I}{\text{Im}}

\newcommand{\bbZ}{\mathbb{Z}}

\newcommand{\bbC}{\mathbb{C}}
\newcommand{\bbP}{\mathds{P}}

\newcommand{\bbL}{\mathbb{L}}

\newcommand{\nn}{\nonumber}

\newcommand{\cref}{{\bf [check ref]}}

\newcommand{\zI}{z^{\text{\tiny I}}}        
\newcommand{\zII}{z^{\text{\tiny II}}}

\newcommand{\Li}{{\rm Li}}
\psfrag{n1}{$\nu_1$}
\psfrag{n2}{$\nu_2$}
\psfrag{n1'}{$\nu_1'$}
\psfrag{n2'}{$\nu_2'$}
\psfrag{n9}{$\nu_9$}
\psfrag{n10'}{$\nu_{10}'$}
\psfrag{t1}{$\tilde{\nu}_1$}
\psfrag{t2}{$\tilde{\nu}_2$}
\psfrag{t9}{$\tilde{\nu}_9$}
\psfrag{t1'}{$\tilde{\nu}_1'$}
\psfrag{t2'}{$\tilde{\nu}_2'$}
\psfrag{t10'}{$\tilde{\nu}_{10}'$}

\entrymodifiers={+!!<0pt,\fontdimen22\textfont2>}		

\def\P{\mathds{P}}

\def\mcal{\mathcal}

\newcommand{\ds}{\displaystyle} 
\newcommand{\Div}{D}
\newcommand{\Jiv}{J}

\numberwithin{equation}{section}			

\makeatletter
\renewcommand*\env@matrix[1][*\c@MaxMatrixCols c]{%
  \hskip -\arraycolsep
  \let\@ifnextchar\new@ifnextchar
  \array{#1}}
\makeatother

\begin{document}

\baselineskip=17pt

\begin{titlepage}
\begin{flushright}
\parbox[t]{1.8in}{
BONN-TH-2009-06\\
0909.2025\ [hep-th]
}
\end{flushright}

\begin{center}

\vspace*{ 1.2cm}

{\large \bf Computing Brane and Flux Superpotentials in F-theory Compactifications
}

\vskip 1.2cm

\begin{center}
 \bf{Thomas W.~Grimm, Tae-Won Ha, Albrecht Klemm and Denis Klevers} \footnote{\texttt{grimm, tha, aklemm, klevers@th.physik.uni-bonn.de}}
\end{center}
\vskip .2cm

{\em Bethe Center for Theoretical Physics, Universit\"at Bonn, \\[.1cm]
Nussallee 12, 53115 Bonn, Germany}
 \vspace*{1cm}

\end{center}

\vskip 0.2cm

\begin{center} {\bf ABSTRACT } \end{center}

In four-dimensional F-theory compactifications with N=1 supersymmetry
the fields describing the dynamics of space-time filling 7-branes
are part of the complex structure moduli space of
the internal Calabi-Yau fourfold. We explicitly compute
the flux superpotential in F-theory depending on all complex structure
moduli, including the 7-brane deformations and the field corresponding to the axio-dilaton.
Since fluxes on the 7-branes induce 5-brane charge, a local limit
allows to effectively match the F-theory results to a D5-brane in a non-compact Calabi-Yau threefold
with threeform fluxes. We analyze the classical and instanton contributions to the F-theory
superpotential using mirror symmetry for Calabi-Yau fourfolds.
The F-theory compactifications under consideration also admit
heterotic dual descriptions and
we discuss the identification of the moduli in this non-perturbative duality.

\hfill September 2009
\end{titlepage}

\tableofcontents
\newpage

\section{Introduction} 

The study of four-dimensional string compactifications leading to effective
supergravity theories with $\cN=1$ supersymmetry is  crucial to connect string
theory with phenomenology. A prominent scenario yielding such minimally
supersymmetric effective theories are Type II string compactifications with
space-time filling D-branes \cite{Lust:2004ks,fluxrev,blumenhagen-flux,Denef:2008wq}. 
At the present stage it remains challenging to
compute the characteristic functions encoding the four-dimensional physics
explicitly without restricting to specific limits in space of compactification
manifolds. In part this is due to the fact that the low amount of supersymmetry
does not significantly restrict the form of most of the couplings in the
effective theory. The situation improves, however, if one focuses on holomorphic
couplings such as the $\cN=1$ superpotential and gauge-coupling function.

The explicit computation of the $\mathcal{N}=1$ superpotential $W$ in a string
compactification allows to infer some crucial information about the vacuum
structure in the effective theory. In particular, in exploring the possible
vacua of Type II string theory, the superpotential induced by non-trivial
background fluxes has been studied intensively for various examples \cite{fluxrev,blumenhagen-flux}.
This task has been tractable since in this case $W$ can be explicitly computed
by solving the Picard-Fuchs system of differential equations 
which determine the moduli dependence of the holomorphic three-form on 
the Calabi-Yau manifold. The superpotential is then expressed in
terms of period integrals of the internal Calabi-Yau threefold which encode the
dependence on the closed string moduli, the complex structure moduli and the
complex axio-dilaton. Much less explored is the dependence of
the superpotential on the open and closed moduli in the presence of D-branes.
The D-brane superpotential $W$ is generically induced by D5-brane charge \cite{Witten:1997ep}, and
hence can arise on D5-branes or on higher dimensional branes with gauge flux on
their world-volume which induces D5-brane charge.  Formally, the brane superpotential can be calculated by
considering reductions of Witten's holomorphic Chern-Simons action
\cite{Witten:1992fb}. However, its explicit computation is more involved and requires 
the study of the full open-closed moduli space.

 A natural generalization to evaluate the open-closed superpotential would be to
find an extended Picard-Fuchs system for the closed and open periods. 
Its solutions then encode the full superpotential in the vicinity of a D-brane. 
So far, this has not been achieved in generality, but only in specific
D-brane settings \cite{AV,Aganagic:2001nx,LMW,Walcher,KSch,Jockers}. 
In \cite{Lerche:2001cw, Alim:2009rf} it was proposed to use (non-compact) Calabi-Yau fourfolds 
to compute the D5-brane superpotential and a connection with F-theory was indicated.
On the other hand, to compute D5-brane superpotentials, we have proposed in \cite{Grimm:2008dq} 
a constructive method involving the blow-up of D5-brane curve in the 
Calabi-Yau threefold. A further alternative to the above methods is 
studied in \cite{Baumgartl:2007an}.

In this work we will study the computation of the open-closed superpotential for  
space-time filling seven-branes with gauge flux on the internal part of their 
worldvolume.   In order to do that we use the
fact that seven-branes admit a natural geometrization in F-theory on an
elliptically fibered Calabi-Yau fourfold. Here the seven-branes wrap four-cycles 
in the base of the elliptic fibration over which the fiber degenerates. 
Moreover, in F-theory the seven-brane
deformations, the complex structure deformations corresponding 
to closed Type IIB moduli, and the axio-dilaton are on the same footing. 
They all arise as complex structure deformations of the Calabi-Yau fourfold.
The superpotential for these fields is the famous Gukov-Vafa-Witten superpotential 
induced by four-form fluxes \cite{Gukov:1999ya}. 
Hence, their appearance in the superpotential is parameterized by the 
holomorphic four-form on the Calabi-Yau fourfold and the open-closed 
superpotential is obtained from the periods of this higher-dimensional 
Calabi-Yau geometry. In \cite{Lust:2005bd} this computation of the Type II open-closed superpotential has been carried out for the case of Type IIB orientifold compactifications on $K3\times T^2/\mathds{Z}_2$ with D3/D7-branes via an embedding into F-theory on $K3\times K3$.

In order to study the F-theory superpotential $W$ we use several powerful 
tools. We start with the construction of the mirror fourfolds which admit a small number of 
K\"ahler moduli and allow us to compute the classical terms in the superpotential 
by evaluating elementary topological data of the fourfold. The mirror elliptic fourfolds 
thus have a small number of complex structure moduli and we are able to 
compute the moduli dependence of the holomorphic four-form explicitly. 
To identify the properly normalized moduli directions, we also 
make use of the duality of F-theory and heterotic 
$E_8 \times E_8$ string theory. It allows us to identify the dependence of
$W$ on the closed string moduli as well as heterotic bundle moduli which map 
to deformations of 7-branes. 
Moreover, in the local limit the
superpotential of a 7-brane with gauge flux should coincide with the
superpotential of a D5-brane in local geometries.  In the
refs.~\cite{AV,Aganagic:2001nx}, the authors have computed the superpotential
for D5-branes in non-compact geometries by evaluating the periods and the chain
integrals directly.  Mirror symmetry allows one to map the result back to the
obstruction problem in the moduli space of special Lagrangians $L$ in Calabi-Yau threefolds 
where the superpotential in the large complex structure limit becomes a generating
function of Gromov-Witten invariants for holomorphic disks ending on $L$.  We use these invariants to identify the correct periods and we thereby determine the
four-form flux element which corresponds to the 7-brane gauge flux.  
By doing so we effectively have 
embedded the deformation problem of a D5-brane into the complex structure deformation 
of a compact Calabi-Yau fourfold.

This work is organized as follows. In section \ref{SuperpotsMirrors} we outline the general
structure of the flux and brane superpotential in the context of mirror symmetry
which we will exploit in our later computations. Then we provide the required technical tools
of toric geometry to realize Calabi-Yau mirror pairs as well as D-branes. In section
\ref{fluxinF} we turn to the discussion of the F-theory fourfold compactifications  
and introduce the geometric realization of seven-branes as well as the form of the flux superpotential. 
Then we describe the construction of fourfold geometries
that admit a particularly rich fibration structure. 
Furthermore, we present a very brief account of the non-perturbative heterotic/F-theory duality 
exploiting the spectral cover construction. Our focus there lies on the identification of F-theory 
moduli in terms of the moduli of the heterotic spectral cover and complex structure. 

In section \ref{ffConstruction}
we explicitly construct mirror pairs of elliptic Calabi-Yau fourfolds. 
Since we aim to compare local Calabi-Yau threefold setups with branes with the 
F-theory compactifications, we first discuss the local geometry and 
its brane content for the local Calabi-Yau threefold $\mathcal O(-3)\rightarrow \mathds{P}^2$. 
Then we construct a compact threefold $\tilde{Y}_3$ containing the local threefold and 
finally obtain the fourfold by fibering $\tilde Y_3$ over a $\mathds{P}^1$-base. 
Our whole analysis there will be guided by 
the F-theory interpretation of the constructed geometries, 
its seven-brane content and the comparison to the D-brane picture.
We also discuss the heterotic/F-theory duality using the spectral cover 
construction in the context of the considered example.
By doing so, we can explicitly identify the threefold $Y_3$, the mirror of $\tilde Y_3$, 
as the heterotic dual Calabi-Yau threefold and the open modulus as the bundle modulus.

We proceed with a general analysis of mirror symmetry on fourfolds and readily apply the introduced 
concepts to our main example in section \ref{FFMirrors}. We show that the classical terms of the 
F-theory superpotential can be fixed by computing topological data of the Calabi-Yau fourfold. 
Crucially, to fix the normalizations of the classical terms, we perform a monodromy analysis 
for the fourfold conifold point by analytic continuation of the fourfold periods. 
As a byproduct, we discover a new conifold monodromy and 
show that the conifold periods admit an interesting leading behavior which is different from 
Calabi-Yau threefolds.
In section \ref{FFMirrors} we also state our main results for the computation of the F-theory 
flux superpotential and its interpretation in terms of the flux and brane superpotential. 
We are able to deduce the form of the classical terms 
in the maximally logarithmic period at the large complex structure point. 
Then the computation of the flux superpotential relies on the identification of the 
threefold periods among the fourfold periods what is supported by the physical 
relevance of $Y_3$ for heterotic/F-theory duality. 
Furthermore, we single out the fourfold 
period that matches the open superpotential. Finally, we are 
able to deduce appropriate $G_4$-flux which induces a F-theory superpotential 
matching both the brane and flux superpotential. 

We conclude with section \ref{EnumGeo} where we provide the basic
background necessary to understand enumerative interpretation of the brane, flux and F-theory superpotentials.
Furthermore, in appendix \ref{FurtherP2} we summarize more detailed geometrical data of the example fourfold
studied in the main text. Finally, in appendix \ref{appa} we study two more fourfold geometries with more moduli.

\section{D7-brane superpotentials and mirror symmetry} \label{SuperpotsMirrors}

In recent years compactifications of Type IIB string 
theory yielding a four-dimensional effective theory with $\cN=1$ supersymmetry
have been studied intensively. One prominent setup are Calabi-Yau orientifold 
compactifications with space-time filling D7-branes and O7-planes \cite{Denef:2008wq,fluxrev,blumenhagen-flux}.
Here the extended D-branes wrap four-cycles, i.e.~divisors, in the internal 
compactification space. The orientifold geometry is obtained by dividing out 
a $\bbZ_2$ symmetry of a compact Calabi-Yau threefold $Z_3$.

In order to study the vacua of such $\cN=1$ compactifications, it is crucial 
to determine the effective four-dimensional superpotential. We will focus on the 
part of the superpotential which is generated by three-from fluxes 
$F_3 = \langle dC_2 \rangle $ and $H_3 = \langle dB_2 \rangle $, where $C_2,B_2$ are the R-R and NS-NS two-forms, 
as well as two-form fluxes $F_2 = \langle dA\rangle$ for the field strength of the $U(1)$ gauge-potential $A$
on the internal part of a D7-brane. The $U(1)$ flux $F_2$ is thus an element of $H^2(D,\bbZ)$, where 
$D$ is the divisor in $Z_3$ wrapped by the D7-brane.
The respective superpotentials are given by \cite{Gukov:1999ya,Giddings:2001yu,Thomas:2001ve,Denef:2008wq}
\beq \label{oc_super}
  W_{\rm flux}(z) = \int_{Z_3} (F_3 - \tau H_3)\wedge \Omega_3  \ , \qquad \quad W_{\rm D7}(z,\zeta) = \int_{\cC_5} F_2 \wedge \Omega_3\ ,
\eeq
where $\tau = C_0+i e^{-\phi}$ is the axio-dilaton, and $\Omega_3$ is the holomorphic three-form on the Calabi-Yau manifold $Z_3$. Note that 
$W_{\rm flux}$ only depends on the complex structure deformations of $Z_3$ due to the 
appearance of $\Omega_3$, while $W_{\rm D7}$ will also depend on the deformation moduli $\zeta$
of the D7-brane. To see the latter, one notes that $\cC_5$ is a five-chain which ends on the 
divisor $D$, i.e.~one has $ D\subset \partial\cC_5$, and carries the information about the 
embedding of the D7-brane into $Z_3$. In the following we will discuss the 
two superpotentials \eqref{oc_super} in more detail.

\subsection{The flux superpotential for Calabi-Yau orientifolds \label{flux_onCY3}}

Let us first discuss the flux superpotential $W_{\rm flux}$. It can be 
evaluated in terms of the periods $(X^A,F_A)$ of the holomorphic three-form $\Omega_3$
\beq \label{flux_threefold}
  W_{\rm flux} = \hat N_A X^A(z) - \hat M^A F_A(z)\ ,\qquad  X^A = \int_{A^A} \Omega_3 \ , \qquad F_A = \int_{B_A} \Omega_3\ .
\eeq
where $(\hat M_A,\hat N^A) = (M_A -\tau \tilde M_A,N^A - \tau \tilde N^A)$ are complex numbers 
with flux quantum numbers $(M_A,N^A)$ of $F_3$ and $(\tilde M_A,\tilde N^A)$ of $H_3$. 
We have also introduced a symplectic basis $(A^A,B_A)$ of 
three-cycles in $H_3(Z_3,\bbZ)$. The dependence of the $2h^{2,1}(Z_3)+2$ periods $(X^A,F_A)$ on the
$h^{2,1}(Z_3)$ complex structure moduli $z^i$ can be evaluated by solving a system of partial differential 
equations, the Picard-Fuchs equations $\cL_a\, (X^A,F_B) =0$. Here $\cL_a$ are linear differential 
operators in the complex structure moduli $z^i$, which can be determined as
reviewed, for example, in~\cite{Hosono:1993qy}.

It is important to point out that by special geometry the non-trivial
information about the $F_A$ periods can be encoded by a single holomorphic
function, the  prepotential $F^0(X^A)$, which is homogeneous of degree two in the 
periods $X^A$, such that the $F_A$ can be written as $F_A = \partial F^0 /\partial X^A$.

There are only a few general observations which can be made about the flux superpotential $W_{\rm flux}$,
since the form of $W_{\rm flux}$ will highly depend on the point at which it 
is evaluated on the complex structure moduli space. One particularly interesting 
point is the large complex structure point which by mirror symmetry corresponds to 
a large volume compactification of Type IIA string theory.
By the known monodromy of the NS-NS $B$-field shift in the Type IIA theory 
one knows that the point must be of maximal unipotent monodromy, which implies on the B-model side a maximal 
logarithmic degeneration of periods near this point $z_i=0$ \cite{Candelas:1990rm,Hosono:1993qy}. 
More specifically, one can identify $X^0 \propto \cO(z)$ as the fundamental period having 
no logarithmic dependence on $z^i$, while $X^i \propto \log z + \cO(z)$, 
$F_i \propto (\log z)^2 + \cO(z)$ and $F_0 \propto (\log z)^3 + \cO(z)$
are always logarithmic in the $z^i$. Mirror symmetry maps the log-terms to classical 
large-volume contributions while the regular terms in the $F_i$ encode the closed 
string world-sheet instantons corrections. 
To see this one notes that the mirror map takes the form $z_i=e^{2\pi it_i} + \ldots$, 
where $t^i = X^i/X^0$ is the world-sheet volume complexified with the NS-NS B-field on the 
Type IIA side. The prepotential $F^0$ encodes the classical couplings as well
as the genus zero world-sheet instantons and takes the general form 
\beq \label{pre_largeV}
  F^0 = -\tfrac{1}{3!} \cK_{ijk}\, t^i t^j t^k - \tfrac{1}{2!}\cK_{ij}\, t^i t^j + \cK_{i} t^i + \tfrac12 \cK_0  + \sum_{\beta} n_\beta^0\,\text{Li}_3(q^\beta)
\eeq
where $q^\beta = e^{2\pi i\beta_j t^j}$ for a vector  $\beta$ with entries  $\bbZ_{\ge 0}$.
Here 
\begin{align} \label{classic_terms_threefold}
  &\cK_{ijk} = \int_{\tilde Z_3} J_i \wedge J_j \wedge J_k\ ,\qquad 
  &&\cK_{ij}=\frac{1}{2}\int_{\tilde Z_3} \imath_*(c_1(J_j)) \wedge J_i \ , \\ 
  &\cK_j = \frac{1}{2^2 3!}\int_{\tilde Z_3} c_2(T_{\tilde Z_3})\wedge J_j\ , \qquad 
  &&\cK_0= \frac{\zeta(3)}{(2 \pi i)^3}\int_{\tilde Z_3} c_3(T_{\tilde Z_3}) \ . \nn
\end{align}
are determined by the classical intersections of the mirror Calabi-Yau threefold $\tilde Z_3$ of $Z_3$.
Note that by $c_1(J_j)$ we mean the first Chern class of the divisor associated to $J_j$ and $\imath_*=P_{\tilde Z_3}i_*P^{-1}_{J_j}$ is the 
Gysin homomorphism where $P_{\tilde Z_3}$ ($P_{\tilde J_j}$) is the Poincar\'e-duality map on $\tilde Y$ ($J_j$) and $i_*$ is the push-forward on the homology. 
Thus, $\imath_*(c_1(J_j))$ is a four-form. The constants $n_m^0$ are the integral Gopakuma-Vafa invariants (BPS numbers) which 
can be computed explicitly for a given example by solving the Picard-Fuchs differential 
equation. Inserting the form of the pre-potential \eqref{pre_largeV} into the flux 
superpotential \eqref{flux_threefold} with $\hat M^0=0$ one finds
\beq \label{flux_ori}
   W_{\rm flux} = \hat N_0 + \hat N_i t^i -  \hat M^i \big[\tfrac12 \cK_{ijk}\, t^j t^k + \cK_{ij}\, t^j + \cK_{i}+ \sum_{\beta} \beta_i n_\beta^0\,\text{Li}_2(q^\beta)\big]\ .
\eeq
 The equation \eqref{flux_ori} means that in addition to a cubic classical polynomial, also  instanton 
correction terms proportional to $\text{Li}_2(q) = \sum_{k=1}^\infty \frac{z^k}{k^2}$ are 
induced by non-vanishing flux $M^i$.

\subsection{The D7-brane superpotentials}

Let us now turn to the superpotential for the D7-brane. Ideally one
would like to compute the functional dependence of $W_{\rm D7}$ on the
D7-deformations $\zeta$ and complex structure moduli $z$ by also evaluating 
a set of open-closed Picard-Fuchs equations. As we will see in section \ref{fluxinF}
this can be indeed achieved if one lifts the setup to an F-theory compactification
on a Calabi-Yau fourfold. One can, however, already infer some crucial property 
of $W_{\rm D7}$ by applying mirror symmetry at the large complex structure/large volume
point. Recall that under mirror symmetry, a Type IIB compactification with D7-branes is mapped 
to a Type IIA compactification with D6-branes. In a supersymmetric configuration these 
D6-branes wrap special Lagrangian cycles $L$ in the mirror Calabi-Yau space $\tilde Z_3$.
The dimension of $H^1(L,\bbZ)$ gives precisely the number of classical deformations $\hat t$ of $L$. 
The superpotential on the mirror side is then induced by string world-sheet 
discs which end on $L$. 

We summarize the enumerative geometry involved in the  
counting problem for discs ending on $L$ in more detail in 
section \ref{EnumGeo}. Here, let us note that the superpotential 
induced by the open string world-sheets takes the form 
\begin{equation} 
   W_{\rm D7}= C_{i} t^i \hat t + C_{ij} t^i t^j +C \hat t^2+  \sum_{\beta,\, n} n^0_{\beta,n}\, {\rm Li}_2(q^\beta Q^n)\ ,\qquad Q = e^{2\pi i\hat t}\ .
\label{wgromovwitten}
\end{equation}     
with constants $C,C_i,C_{ij}$ and $n^0_{\beta,n}$ determined by the D-brane and $Z_3$ geometry, as
well as the flux $F_2$. One finds that the full D7 superpotential contains both classical
terms as well as instanton corrections which again has the Li$_2$ structure as in \eqref{flux_ori}. 

While the D7-superpotential \eqref{oc_super}, \eqref{wgromovwitten} has not been studied extensively 
its D5-brane analogue is studied in many works. 
Clearly, since both are induced by D5-brane charge, it is important to summarize 
some of the literature on the D5-brane superpotential.
The most thoroughly studied cases are mirror geometries of non-compact toric Calabi-Yau manifolds 
with Harvey-Lawson type branes \cite{AV,Aganagic:2001nx}.
In these examples the brane superpotential, i.e.\ the chain integral, is calculated directly using a 
meromorphic differential naturally given in the mirror geometry.
For compact geometries, only recently there has been much progress in computing the superpotential.
For D5-branes which are mirror to D6-branes wrapping a rigid involution special Lagrangian cycle, the 
superpotential can be calculated by deriving an inhomogeneous Picard-Fuchs system and solving it \cite{Walcher,KSch}.
Extending the works \cite{LMW} to compact examples, a method to derive a Picard-Fuchs system for open/closed 
moduli space is proposed using an auxiliary hypersurface \cite{Jockers}. 
In \cite{Alim:2009rf} it is proposed to compute the superpotential for toric branes by extending the polyhedron 
describing the ambient toric variety to one dimension higher polyhedron.
The authors of \cite{Alim:2009rf,Jockers} propose that the two last methods are equivalent. 
Unfortunately, at least up to now, neither of the two methods provides a constructive algorithm 
to tackle a given D5-brane configuration, since in both formalisms one effectively 
has to work on an auxiliary divisor and pick the correct linear combination for 
the D5-superpotential by hand. 
In \cite{Grimm:2008dq} we have proposed another more constructive method to compute the superpotential for 
D5-brane. Concretely, we argued that the deformation problem for a D5-brane in a Calabi-Yau threefold with variable complex 
structure is equivalent to considering the complex structure of a 
non-Calabi-Yau threefold which is canonically obtained by blowing up the original Calabi-Yau manifold 
along the curve which the D5-brane wraps. Work in computing explicit examples is in progress.

\subsection{Mirror symmetry with branes in toric geometry \label{mirror_toric_branes}}
After giving a general overview of the superpotential and its relation 
to mirror symmetry, we now discuss the toric realization of mirror symmetry and branes \cite{Hori,AV,Aganagic:2001nx}. 
At this point we will also introduce notions and techniques of toric geometry that are inevitable for the rest of this work.

\subsubsection{Calabi-Yau manifolds as hypersurfaces \label{CY_hyper}}

A powerful tool to construct Calabi-Yau manifolds $Y_n$ and their mirrors $\tilde Y_n$ for an arbitrary complex dimension $n$ is by realizing them as hypersurfaces in toric ambient spaces. These hypersurfaces are 
specified by reflexive polyhedra \cite{Batyrev}. 

For threefolds we start with IIA on a Calabi-Yau $\tilde Y_3$ which is mirror dual to Type IIB on $Y_3$. 
We realize the compact Calabi-Yau $\tilde Z_3$ as the hypersurface in a toric ambient variety $\tilde V_4$ 
constructed from a pair of reflexive polyhedra $\Delta_4^{\tilde Y}$ and $\Delta_4^{Y}$ in lattices $N$, 
$M$, that are dual, i.e.~$\Delta_4^{\tilde Y}=(\Delta_4^Y)^\ast$.\footnote{For convenience we denote quantities 
in Type IIA always with `$\sim$' in order to omit them for their mirror quantities on 
the Type IIB side where we will mainly work.}
In general, the dual polyhedron $\Delta^\ast$ of a given polyhedron $\Delta$ in a lattice $M$ is defined 
as the set of points $p$ in the real span $N_\mathbb{R}=N\otimes \mathbb{R}$ of the dual 
lattice $N$ of $M$ such that
\begin{equation}
 	\Delta_4^\ast=\{p\in N_\mathbb{R}|\langle q,p\rangle\geq -1\text{ for all } q\in\Delta_4 \}.
\end{equation}
Let us assume that the combinatorics of $\tilde V_4$ associated to the polyhedron $\Delta_4^{\tilde Y}$ is 
encoded in $k$ charge vectors $\ell^{(j)}$ describing the relations among the $m=k+4$ vertices $\tilde v_i$.
The Calabi-Yau $\tilde Y_3$ is then given as the hypersurface $\{\tilde f=0\}$ in 
$\tilde V_4$ where $\tilde f$ is given as the following polynomial \cite{Batyrev} 
\begin{equation} \label{Y3typeIIA}
 	\tilde f=\sum_{q\in \Delta^Y_4\cap M}\tilde a_q\prod_{i}\tilde x_i^{\langle \tilde v_i,q\rangle+1}
\end{equation}
in the $m$ projective coordinates $\tilde x_j$ of $\tilde V$ associated to each vertex $\tilde v_j$. 
This formula provides a direct way to count the number of complex structure parameters $\tilde a_q$ 
(up to automorphisms of $\tilde V_4$) by counting the integral points $q\in\Delta_4^Y$.
Furthermore, $\tilde Y_3$ is Calabi-Yau since \eqref{Y3typeIIA} contains the monomial 
$\tilde x_1\ldots\tilde x_m$ corresponding to the origin in $\Delta_4^Y$ so that $f$ is a 
section of the anti-canonical bundle $K_{\tilde V_4}^\ast=\mathcal{O}(\sum_i D_i)$, 
where $D_i=\{\tilde x_i=0\}$ is a toric divisor.

For the case of hypersurfaces in toric varieties, (closed string) mirror symmetry \cite{Batyrev} 
is realized in a very elegant way. The mirror threefold $Y_3$ on the Type IIB side is obtained 
by simply interchanging the roles of $\Delta_4^{\tilde Y}$ and $\Delta_4^Y$ so that \eqref{Y3typeIIA} 
describes $Y_3$ as the hypersurface in the toric variety $V_4$ associated to the polyhedron $\Delta_4^Y$, 
\begin{equation} \label{Y3typeIIB}
 	 f=\sum_{p\in \Delta^{\tilde Y}_4\cap N}a_p\prod_{i}x_i^{\langle v_i,p\rangle+1}.
\end{equation}
Here, we again associated the projective coordinates $x_i$ to each vertex $v_i$ of $\Delta_4^Y$.  
Indeed, the necessary requirements for mirror symmetry, $h^{1,1}(Y_3)=h^{2,1}(\tilde Y_3)$ and 
$h^{2,1}(Y_3)=h^{1,1}(\tilde Y_3)$, are fulfilled for this construction. This is obvious from 
Batyrev's formula for the Hodge numbers \cite{Batyrev} of a given $n$-fold $Y_n$ in a toric 
ambient space specified by $\Delta_{n+1}^{Y}$
\begin{eqnarray} 
  h^{n-1,1}(Y_n) &=& 
h^{1,1}(\tilde Y_n)\label{Hodgenumbers1}\\ &=&
\ l(\Delta_{n+1}^{\tilde{Y}}) - (n+2) -\sum_{\dim \tilde\theta =n} l'(\tilde\theta)+
              \sum_{\text{codim} \tilde\theta_i =2} l'(\tilde \theta_i) l'(\theta_i)\ ,\nn\\
  h^{1,1}(Y_n) &=& h^{n-1,1}(\tilde Y_n)  \label{Hodgenumbers2}\\ &=&  
\ l(\Delta_{n+1}^Y) - (n+2) -\sum_{\dim \theta =n} l'(\theta)+
              \sum_{\text{codim}  \theta_i =2} l'(\theta_i) l'(\tilde \theta_i)\ .\nn
\end{eqnarray}
In this expression $\theta$ ($\tilde \theta$)
denote faces of $\Delta^Y_{4}$ ($\Delta^{\tilde Y}_{4}$), while the sum is over 
pairs $(\theta_i,\tilde \theta_i)$
of dual faces. The $l(\theta)$ and $l'(\theta)$ count the total number of integral points of a 
face $\theta$ and the number inside the face $\theta$, respectively. Finally, 
$l(\Delta)$ is the total number of integral points in the polyhedron $\Delta$. 
Using these formulas one notes that polyhedra with a small 
number of points will correspond to Calabi-Yau fourfolds with few K\"ahler moduli, i.e.~small 
$h^{1,1}$, and many complex structure moduli $h^{3,1}$. Since $h^{1,1}$ and
$h^{3,1}$ are exchanged by mirror symmetry Calabi-Yau fourfolds with small $h^{3,1}$ are 
obtained as mirror manifolds of hypersurfaces specified by a small number of lattice points in the polyhedron. 

For the case of Calabi-Yau fourfolds $(\tilde{X}_4,X_4)$ the complete list of model dependent Hodge numbers is
$h^{1,1}(X_4)$, $h^{3,1}(X_4)$, $h^{2,1}(X_4)$ and $h^{2,2}(X_4)$, However only three of these are independent due to the Hirzebruch-Riemann-Roch index theorem implying \cite{Klemm:1996ts} 
\begin{equation} \label{HodgeRel}
 	h^{2,2}(X_4)=2(22 + 2h^{1,1}(X_4) + 2h^{3,1}(X_4) - h^{2,1}(X_4))\,. 
\end{equation}
Therefore, only $h^{2,1}(X_4)$ has to be calculated in addition to fix the basic topological data of $(\tilde{X}_4,X_4)$. Analogously to \eqref{Hodgenumbers1} it is readily given by the symmetric expression
\begin{equation} \label{Hodgenumbers3}
	h^{2,1}(X_4)= h^{2,1}(\tilde{X}_4)=\sum_{\text{codim} \tilde\theta_i =3}l'(\tilde \theta_i) l'(\theta_i)\,.
\end{equation}
This finally enables us to calculate the Euler number of fourfolds by
\begin{equation}\label{FFEulerNumb}
 	\chi(X_4)=\chi(\tilde{X}_4)=6(8+h^{3,1}+h^{1,1}-h^{2,1})\,.
\end{equation}

\subsubsection{Toric branes and their mirrors}

To a setup of mirror pairs $(Z_3,\tilde Z_3)$ of toric Calabi-Yau hypersurfaces we 
want to add open string degrees of freedom which introduce the so-called Harvey-Lawson 
type branes. This type of branes have been studied intensively in local toric Calabi-Yau threefolds 
in refs.~\cite{AV, Aganagic:2001nx}. Note that in these examples the toric variety itself is a non-compact 
Calabi-Yau threefold. Since we want to cover also the non-compact case we will work in the 
following with a toric space of dimension $m-k$ denoted by $\tilde V_{m-k}$, where $m-k=3$ in the non-compact case, 
and $m-k=4$ in the compact case as in section \ref{CY_hyper}. 

The toric variety 
$\tilde V_{m-k}$ is represented as a symplectic quotient $\mathbb{C}^{m}//G$ defined first by 
imposing vanishing moment maps 
\begin{equation} \label{DTermConstraint}
 	\sum_{j=1}^m \ell^{(i)}_j |\tilde x_j|^2= r^i
\end{equation}
and second by dividing by the isometry or gauge group $G=U(1)^k$ as 
$\tilde{x}_j\mapsto e^{i\ell^{(i)}_j\epsilon_i}x_j$  \cite{Witten:1993yc}. Then solving \eqref{DTermConstraint} 
and using coordinates $p_j=|\tilde x_j|^2$, $\theta_j$ the toric variety $\tilde V_{m-k}$ 
can be visualized as a $T^{m-k}$-fibration over a real $(m-k)$-dimensional base $\mathbb{B}_{m-k}$ \cite{Leung:1997tw, AV}. 
The degeneration loci of the $T^{m-k}$-fibration where one or more $S^1$ shrink are on the boundary of 
$\mathbb{B}_{m-k}$ which is determined by $p_j=0$ or intersections thereof since $p_j\geq 0$.

In Type IIA the Harvey-Lawson 
type branes wrap special Langrangian cycles $L$ which can be specified 
by $r$ additional brane charge vectors $\hat \ell^{(a)}$ restricting the $|x_j|^2$ and 
their angles $\theta_i$ in the toric ambient variety $\tilde V_{m-k}$ so that \cite{AV}
\begin{equation} \label{ABranes}
 	\sum_{j=1}^{m} \hat\ell^{(a)}_j |\tilde x_j|^2=c^a\ ,\qquad \theta_i=\sum_{a=1}^r\hat\ell^{(a)}_i\phi_a\ ,
\end{equation}
for angular parameters $\phi_a$. To fulfill the `special' condition of $L$, which is 
equivalent to $\sum_i\theta_i=0$, one demands $\sum_j\hat\ell_j^{(a)}=0$.

This construction was used in non-compact Calabi-Yau threefolds 
$\tilde Z_3=\mathbb{C}^{k+3}//G$ for which the Calabi-Yau condition 
$\sum_j\ell^{(i)}_j=0$ has to hold \cite{AV, Aganagic:2001nx}. 
Then A-branes introduced by \eqref{ABranes} are graphically represented 
as real codimension $r$ subspaces of the toric base $\mathbb{B}_{3}$. 
The case which was considered for the non-compact examples in 
 \cite{Aganagic:2001nx} is $r=2$ where the non-compact three-cycle $L$ 
is represented by a straight line ending on a point when projected onto the base $\mathbb{B}_3$.
The generic fiber is a $T^2$ so that the topology of $L$ is just $\mathbb{R}\times S^1\times S^1$. 
However, upon tuning the moduli $c^a$ it is most convenient to move the $L_a$ to the boundary 
of $\mathbb{B}_3$ where two $\{p_j=0\}$-planes intersect. Then one of the two moduli is frozen, and 
one $S^1$ pinches such that the topology becomes $\mathbb{C}\times S^1$. 
These D6-branes are mirror to non-compact D5-branes which intersect a Riemann surface at a point. 
Later on, we will use the D5-brane results of \cite{AV, Aganagic:2001nx} in order to study 
the superpotential \eqref{oc_super}
of D7-branes with gauge flux $F_2$ on compact Calabi-Yau manifolds. 
The gauge flux induces an effective D5-charge on the D7-brane 
and we will be able to compare the D5-brane superpotential of \cite{AV, Aganagic:2001nx} to 
the D7-brane superpotential with appropriate $F_2$ in the local limit.

The mirror Type IIB description \cite{Hori,AV} can be obtained as follows.
First the mirror Calabi-Yau $Y_{3}$ is determined by a polynomial $W$ and $n$ constraints 
given by 
\begin{equation} \label{eqn:HVmirror}
 	W=\sum_{j=0}^m y_j\ ,\qquad \prod_{j=0}^m y_j^{\ell^{(i)}_j}=z^i\ ,\qquad i=1,\ldots, n\ ,
\end{equation}
where the $z^i$ denote the complex structure moduli of $Z_3$ that are related to the (complexified) K\"ahler moduli $t^i$ of $\tilde Y_3$ by $z^i=e^{2\pi it^i}$.
We note that we introduced a further 
coordinate $y_0$ for which we also have to include a zeroth component $\ell^{(i)}_0=-\sum^m_{j=1} \ell^{(i)}_j$. 
For compact threefolds $Y_3$ is then obtained as the orbifolded hypersurface
\begin{equation}
 	\{W(x_i)=0\}/\Gamma
\end{equation}
in the mirror toric variety $V_4$ with homogeneous coordinates $x_i$. These are introduced by a change 
of coordinates such that the constraints in \eqref{eqn:HVmirror} are automatically solved. The map from $y_i$ 
to $x_i$ as well as the orbifold group $\Gamma$ is determined by the etal\'{e} map which is worked 
out for weighted projective spaces in \cite{Hosono:1993qy,Batyrev}.
In the non-compact case, $Z_3$ is similarly
\begin{equation}\label{localMirror}
 	xz=W(x_i)
\end{equation}
in affine coordinates $x,z$ of $\mathbb{C}$ and $x_i\in\mathbb{C}^*$, respectively.

Analogously the B-branes on holomorphic submanifolds $\mathcal{C}$ in $Z_3$ are specified by
\begin{equation} \label{BBrane}
 	\prod_{j=0}^m y_j^{\hat\ell^{(a)}_j}=\epsilon^a e^{-c^a},\quad a=1,\ldots, r\ , 
\end{equation}
that can also be re-expressed in terms of the coordinates $x_i$.
The phases $\epsilon^a$ are dual to the Wilson line background of the flat U(1)-connection on the 
special Lagrangian $L$ and complexify the moduli $c^a$ to the open moduli 
$\zeta^a$ \cite{Strominger:1996it}.
As is clear from \eqref{BBrane} the B-brane is supported over a holomorphic cycle 
$\mathcal{C}$ of complex codimension $r$.  Thus for the configuration $r=2$ the mirror 
is a D5-Brane. Other cases can be considered as well leading to mirrors given by D7-branes 
on divisors ($r=1$) or D3-branes on points ($r=3$).

\section{Flux superpotentials in F-theory \label{fluxinF}}

F-theory provides a geometrization of $\mathcal{N}=1$ Type IIB backgrounds with holomorphically 
varying complexified string coupling constant $\tau$ \cite{Vafa:1996xn}. The parameter $\tau$ is interpreted
as the complex structure modulus of a two-torus which can be fibered over the 
spatial dimensions of the Type IIB target space that is a K\"ahler manifold with positive curvature.  
The $(3$+$1)$-complex-dimensional geometry obtained this way captures non-trivial monodromies of 
$\tau$ around degeneration loci of the two-torus. Precisely this provides a geometrization 
of (non-perturbative) seven-branes, which include D7-branes and O7-branes as 
special cases. In this section we discuss compactifications of F-theory to four space-time 
dimensions with a focus on the induced flux superpotential \cite{Gukov:1999ya} inherited from 
the M-theory description of F-theory \cite{Denef:2008wq}.

\subsection{Elliptic fourfolds and seven-branes in F-theory}

Let us study the four-dimensional effective $\cN=1$ theory which arises 
by compactification of F-theory on an elliptically fibered fourfold 
$X_4 \rightarrow B_3^X$ over a complex three-dimensional base $B_3^X$. 
This corresponds to Type IIB string theory compactified on $B_3^X$ with an axio-dilaton 
$\tau = C_0 + i e^{-\phi}$ varying holomorphically over the K\"ahler base $B^X_3$, i.e.~one has 
\beq \label{FonX4}
   \text{F-theory on}\ X_4 \ =\ \text{Type IIB on}\ B_3^X\ .
\eeq
The Weierstrass form of the elliptic fibration of $X_4$ is 
given by 
\beq \label{Weierstrass}
  y^2 = x ^3 + f(u) x z^4 + g(u) z^6\ ,
\eeq
where $f(u)$ and $g(u)$ vary over the base $B_3^X$ with coordinates $u$. To ensure 
that \eqref{Weierstrass} is well-defined, $(x,y,z)$ are sections of $(\mathcal O_B,\mathcal O_B,K_B)$  
and $(f,g)$ are sections of $(K_{B}^{-4},K^{-6}_B)$, where $K_B$ is the 
canonical bundle of the base $B_3^{X}$. 
Equation \eqref{Weierstrass}
with the defined scalings for the coordinates $(x,y,z)$ implies that the generic elliptic 
fiber is $\bbP^2(1,2,3)[6]$, 
i.e.~a degree $6$ hypersurface in weighted projective space $\bbP^2(1,2,3)$. This will be 
the case for all examples considered in this work even if $X_4$ becomes singular.

As the axio-dilaton of Type IIB string theory $\tau$ corresponds  to the 
complex structure of the elliptic fiber, it can be specified by the value of the classical 
$SL(2,\bbZ)$ modular invariant $j$-function which is expressed through the 
functions $f$ and $g$ in \eqref{Weierstrass} as
\beq \label{def-j}
  j(\tau) = \frac{4(24\,f)^3}{\Delta}\ , \qquad \Delta = 27\, g^2 + 4 f^3\ .
\eeq
The function $j(\tau)$ admits a large $\I\, \tau$ expansion 
$j(\tau) = e^{-2\pi i \tau}+744+\cO(e^{2\pi i \tau})$ from which we can read off the 
monodromy of $\tau$ around a $7$-brane. 

In general, the elliptic fibration will be singular 
over the discriminant $\Delta$.
It can factorize into 
several components which individually correspond to divisors $D_i$ in $B^X_3$ which are
wrapped by seven-branes including the well-known D7-branes and O7-planes. 
 The singularities of the elliptic fibration over the $D_i$ 
determine the gauge group on the seven-branes. These can be determined explicitly using 
generalizations of the Tate formalism \cite{Bershadsky:1996nh}. 
The weak string coupling limit of F-theory is given by $\I\, \tau \rightarrow \infty$ and yields 
a consistent orientifold setup with D7-branes on a Calabi-Yau manifold \cite{Sen:1997gv}. 

It is important to note that the degeneration of the elliptic fibration 
can be so severe that the Calabi-Yau fourfold $X_4$ as given in \eqref{Weierstrass} 
becomes singular. In this case it is not possible to work with the singular space 
directly since the topological quantities such as the Euler characteristic and 
intersection numbers are not well-defined. To remedy this problem the singularities 
can be systematically blown up to obtain a smooth geometry \cite{Bershadsky:1996nh}. 
In the cases considered in this 
paper this is done using the methods of toric geometry \cite{Bershadsky:1996nh,Candelas:1996su,Candelas:1997eh}.  
The resulting smooth geometry still contains the information about the gauge-groups on the seven-branes and 
allows to analyze the compactification in detail. 

In this work we will entirely focus on the complex structure sector of the Calabi-Yau fourfold $X_4$.
We will consider smooth spaces $X_4$ which only admit a small number $h^{3,1}(X_4)$ of complex structure deformations, 
but are obtained from singular elliptically fibered Calabi-Yau fourfolds by multiple blow-ups. 
This affects only the number of K\"ahler moduli, which we will not discuss in the following. 
In order to compare to the Type IIB weak coupling picture the complex structure moduli can be split into three classes \cite{Denef:2008wq}:
\begin{enumerate}
	\item[(1)]One complex modulus corresponding to the complex axio-dilaton $\tau$
parametrizing the complex structure of the elliptic fiber.
\item[(2)] The moduli corresponding to the deformations of the seven-branes wrapped on 
divisors on $B^X_3$.
\item[(3)] The complex structure moduli corresponding to the deformations of the 
basis and its double covering Calabi-Yau threefold obtained in the orientifold limit.
\end{enumerate}

\subsection{The flux superpotential \label{F-flux_sup}}

It is well-known that F-theory admits a superpotential upon switching on four-form flux $G_4$.
For even second Chern class of $X_4$ this flux is integer 
quantized\footnote{To be precise \cite{Witten:1996md} we note that $G_4$ is quantized such that $[G_4-c_2(X_4)/2]\in H^4(X_4,\mathbb{Z})$.}  
$G_4 \in H^4(X_4,\bbZ)$ \cite{Witten:1996md}. To determine 
the F-theory superpotential one uses the duality between 
M-theory and F-theory \cite{Denef:2008wq,Haack:2001jz}. In an M-theory compactification on $X_4$ 
one encounters the famous Gukov-Vafa-Witten superpotential 
\beq \label{GVW-super}
  W(z) = \int_{X_4}  G_4 \wedge \Omega \ ,
\eeq
where $\Omega$ is the holomorphic $(4,0)$ form on $X_4$.
The superpotential $W(z)$ depends on the complex structure 
deformations $z$ of the fourfold $X_4$. As we will 
discuss momentarily, upon imposing restrictions 
on the allowed fluxes $G_4$, the superpotential \eqref{GVW-super}
also provides the correct expression for an F-theory compactification. 
The goal of this work is to explicitly compute \eqref{GVW-super}
for specific elliptically fibered Calabi-Yau fourfolds. The 
result is then matched with the superpotentials \eqref{oc_super} at weak string 
coupling such that 
\beq \label{SuperpotLimit}
  \int_{X_4}  G_4 \wedge \Omega \quad \rightarrow \quad \int_{Z_3} (F_3-\tau H_3) \wedge \Omega_3 
                  + \sum_{m} \int_{\cC^{m}_5} F^m_2 \wedge \Omega_3\ ,
\eeq
where $m$ labels all D7-branes on divisors $D_m$ carrying two-form fluxes $F_2^m$.

Note that already by a pure counting of the flux quanta encoded by $G_4 \in H^{4}(Y_4,\bbZ)$, as well 
as $F_3,H_3 \in H^3(Z_3,\bbZ)$ and $F_2^m \in H^{2}(D_m,\bbZ)$ one will generically encounter 
a mismatch. This can be traced back to the fact that not all fluxes $G_4$ are actually 
allowed in an F-theory compactification, since in the duality between M-theory on $X_4$ and F-theory on $X_4$ 
one of the dimensions of the elliptic fiber will become to a space-time dimension \cite{Denef:2008wq}.
The simplest case with only D7-branes with abelian gauge groups 
is discussed, for example, in ref.~\cite{Denef:2008wq}. In this situation the flux $G_4$ is 
allowed to have only 
components satisfying 
\beq
   \int G_4 \wedge \omega_1 \wedge \omega_2 = 0 \ ,\qquad  \quad \forall\, \omega_1,\omega_2 \in H^{1,1}(X_4)\ .
\eeq
Our examples are, however, significantly more complicated and admit D7-branes with 
rather large gauge groups. This is due to the fact that we need to analyze fourfolds $X_4$ 
with few complex structure moduli, which typically have  order a thousand elements 
of $H^{1,1}(X_4)$. Many of the elements in $H^{1,1}(X_4)$ will correspond to 
blow-ups of a singular elliptic fibration and signal the presence of enhanced gauge groups. 
Therefore, it will be more practical to discuss the matching of the moduli dependence 
encoded by the periods of $\Omega$ on $X_4$.

In the next step we want to extract the moduli dependence of 
the F-theory superpotential. Here we are aiming to state some general 
features of the flux superpotential. More details on the moduli dependence 
of $W$ will be presented in section \ref{FFMirrors}. 
As in the threefold case \eqref{flux_threefold} the 
fourfold superpotential can be expressed through the 
periods of $\Omega$. However, in the fourfold case the variations of the 
$(4,0)$ form $\Omega$ do not span the full cohomology $H^4(X_4)$, but 
rather only a subspace $H_H^4(X_4)$, known as the primary horizontal subspace of $H^4(X_4)$ \cite{Greene:1993vm}.
It takes the form
\beq \label{horiz}
  H^4_H(X_4,\bbC) = H^{4,0} \oplus H^{3,1} \oplus H^{2,2}_H  \oplus H^{1,3} \oplus H^{0,4}\ ,
\eeq
where $H^{2,2}_H$ consists of the elements in $H^{2,2}$ which 
can be obtained as second variation of $\Omega$ with respect to the 
complex structure on $X_4$. The numbers $h^{4-i,i}_H(X_4)$ denote the dimension of the respective cohomologies in \eqref{horiz}. Note that 
the fact that not all $H^{4}(X_4)$ can be reached as variations 
of $\Omega$ is in contrast to the Calabi-Yau threefold case. In 
the threefold case one can simply define the periods of $\Omega_3$
as in \eqref{flux_threefold} by introducing an integral homology basis of 
$H_3(Y_3,\bbZ)$. In the fourfold case, however, one has 
to introduce a basis $\gamma^{(i)}_{a}$ of $H_4^H(X_4,\bbZ)$,
in order to define the periods  
\beq \label{def-periods}
    \Pi^{(i)\, a}   =   \int_{\gamma^{(i)}_{a}} \Omega \ , \qquad \quad i = 0,\ldots,4\ , 
\eeq
where ${\gamma}^{(i)}_a$ for $a=1,\ldots,h^{4}_H(X_4)$ denote a graded basis of $H_{4}^H(X_4)$ with grade $i=0,\ldots,4$.
Here we also introduced the dual basis $\hat{\gamma}^{(i)}_{a}$ of $H_H^{4}(X_4,\mathbb{Z})$ with pairing
\begin{equation} \label{dualbasis}
 	\int_{\gamma^{(i)}_{a}}\hat{\gamma}^{(j)}_{b} 
                                              = \delta^{ij} \delta_{a b}\ .
\end{equation}
This cohomology basis satisfies 
\bea \label{def-eta}
  \int_{X_4} \hat{\gamma}^{(i)}_{a} \wedge  \hat{\gamma}^{(4-i)}_{b} &=& \eta^{(i)}_{a b}\ ,\\
   \int_{X_4} \hat{\gamma}^{(i)}_{a} \wedge \hat{\gamma}^{(j)}_{b}\ \  &=& 0  \quad\text{for}\quad i+j> 4\ .
     \nn
\eea
In this basis we expand the holomorphic four-form $\Omega=\sum_i {\Pi}^{(i)\, a}\hat{\gamma}_{a}^{(i)}$.
Analogously, the flux quantization condition of $G_4$ reads in this basis of integral cycles 
\begin{equation}
    G_4= \sum_i N^{(i)\, a}\ \hat{\gamma}^{(i)}_{a}\ ,\qquad \quad  N^{(i)\, a} = \int_{\gamma^{(i)}_{a}} G_4  \ ,  
\end{equation} 
where $N^{(i)\,a}$ are integral flux numbers.  
Then in terms of these definitions the flux superpotential \eqref{GVW-super} can be expanded as 
\begin{equation}
 	W= \sum_{i} N^{(i)\, a}\,   \Pi^{(4-i)\, b}\, \eta^{(i)}_{a b} \equiv \sum_{i} N^{(i)\, a}\,   \Pi^{(4-i)}_a \ ,
\end{equation}
with the moduli 
independent intersection matrix $\eta^{(i)}_{a b}$ defined in \eqref{def-eta}.
Note that as in the threefold case a direct definition of the integral 
basis $\gamma^{(i)}_{a}$ is impossible. However, the existence of such a 
basis can be inferred by using mirror symmetry at the large complex structure 
point. In fact, as in section \ref{flux_onCY3} the periods $\Pi^{(i)\, a}$ can be selected 
according to their leading logarithmic behavior at this point as we will discuss in more detail in section
\ref{FFMirrors}.

\subsection{Constructing elliptic fourfolds \label{Kreuzermethods}}

In the following we will discuss the construction of elliptically fibered 
Calabi-Yau fourfolds for which we want to compute the 
F-theory superpotential \eqref{GVW-super}. Our strategy is to 
find fourfold examples $X_4$ which admit a small number of complex structure 
moduli such that we can evaluate the Picard-Fuchs equations 
determining the holomorphic four-form $\Omega$. Candidate examples have already been 
considered in refs.~\cite{Mayr:1996sh,Klemm:1996ts}. Moreover, we construct the 
fourfolds in such a way that they contain a local Calabi-Yau patch in which the 
effective D5-brane superpotential has been computed explicitly \cite{AV,Aganagic:2001nx}. 

The Calabi-Yau fourfolds studied in this paper will be obtained as mirror dual 
to a Calabi-Yau threefold fibration over $\bbP^1$. Denoting by $\tilde Y_3$
the Calabi-Yau threefold fiber we can write this as
\beq \label{tildeYoverP1}
  \begin{array}{cccc} \text{fiber}  & \rightarrow & \text{total space} & \\
         && \downarrow &\\ 
         && \text{base} & 
   \end{array}\qquad \qquad
   \begin{array}{cccc} \tilde Y_3  & \rightarrow & \tilde X_4 & \\
         && \downarrow &\\ 
         && \bbP^1 & 
   \end{array} \quad .
\eeq
We will later pick Calabi-Yau threefolds $\tilde Y_3$ which are obtained
by compactifications of local Calabi-Yau geometries which can 
support Harvey-Lawson type D6-branes as introduced in section \ref{mirror_toric_branes}.
The compact Calabi-Yau threefolds $\tilde Y_3$ have small numbers $h^{1,1}$ of 
K\"ahler moduli, which is a feature inherited by $\tilde X_4$. Moreover, since 
we want to study F-theory on the mirror $X_4$ of $\tilde X_4$,
the Calabi-Yau threefold $\tilde Y_3$ as well as the fibration 
structure of $\tilde X_4$ will be chosen carefully, such that $X_4$ has an elliptic fibration. This is achieved, for example, by choosing $Y_3$ elliptically fibered \cite{Berglund:1998ej},
\beq
   \begin{array}{cccc} \cE  & \rightarrow & Y_3 & \\
         && \downarrow &\\ 
         && B_2^Y & 
   \end{array}   
  \qquad \qquad  \begin{array}{cccc} \cE  & \rightarrow & X_4 & \\
         && \downarrow &\\ 
         && B_3^X & 
   \end{array}   \quad ,
\eeq 
for which $X_4$ also admits a $K3$-fibration with a heterotic dual on $Y_3$. Here $\cE$ is the generic elliptic fiber and $B_2^Y$ and $B_3^X$ are the two-dimensional 
and three-dimensional base spaces of the 
fibrations, respectively. As we will show for the explicit examples, $\cE = \bbP^2(1,2,3)[6]$ is 
the generic elliptic fiber shared by $Y_3$ and $X_4$.

To detect these fibration structures of a given mirror pair of 
Calabi-Yau fourfolds $(\tilde X_{4},X_{4})$ and in order to understand our construction more thoroughly it turns out to be sufficient 
to study the toric data in the corresponding reflexive polyhedra $(\Delta_{\tilde X_{4}},\Delta_{X_{4}})$ 
without computing the intersection numbers \cite{Kreuzer}. 
In fact, in our examples of spaces $X_4$ used in \eqref{FonX4} the intersection numbers will be hard to compute 
because of their huge number of K\"ahler moduli.
In the following we will recall the general theorem of ref.~\cite{Kreuzer} and later, in section \ref{ffConstruction}, apply it to our main example.

Suppose $(\tilde X_{4},X_{4})$ are given as hypersurfaces in the toric varieties 
constructed from the reflexive pair 
$(\Delta_5^X,\Delta^{\tilde X}_5)$ in the pair of dual lattices $(M,N)$.
The statement of \cite{Kreuzer} gives two equivalent conditions for the existence of a Calabi-Yau fibration 
structure of the given fourfold $X_{4}$ once in terms of $\Delta^X_5$ and another time in terms of its dual $\Delta^{\tilde X}_5$.  
Assume there exists a $(n-k)$-dimensional lattice hyperplane in $N$ through the origin such that $\Delta_k^{F}:=H\cap\Delta^{X}_5$ is a $k$-dimensional reflexive polyhedron.
Then this is equivalent with the existence of a projection $P$ to a $k$-dimensional sublattice of $M$ such that 
$P\Delta^{\tilde X}_5$ is a $k$-dimensional reflexive polyhedron $\Delta^{\tilde F}_k$ which is the dual of $\Delta^{F}_k$.
If these conditions are satisfied, then the Calabi-Yau manifold $X_{4}$ which is obtained as a 
hypersurface of $\Delta^X_5$ has a Calabi-Yau fibration whose $(k-1)$-dimensional fiber $F_{k-1}$ is given by $\Delta^F_k$. 
The crucial point of these two equivalent criteria is that we can turn things around 
and analyze $X_{4}$ by not looking at hyperplanes $H$ in the complicated polyhedron $\Delta^X_5$, 
but at projections $P$ in $\Delta^{\tilde X}_5$ which is simple by construction. 
In both cases the base of the fibration can be found by considering the 
quotient polyhedron $\Delta^X_5/\Delta^F_k$  
\cite{Candelas:1997eh}. 
Here this quotient polyhedron is obtained by first determining the quotient lattice in $M\supset\Delta_5^X$ by dividing out the lattice generated by the integral points of $\Delta^F_k$. Then the integral points of $\Delta^X_5/\Delta^F_k$ are the equivalence classes of integral points in $\Delta^X_5$ in this quotient lattice.
Schematically the analysis of the fibration structure can be summarized as
\beq \label{fibrations}
\rule[-2.0cm]{0cm}{4.2cm}\begin{array}{|c|l|c|l|c|}
   \cline{1-1}\cline{3-3} \cline{5-5}  \rule[-3mm]{0mm}{5mm}
   	 \rule[-3mm]{0mm}{8mm} \text{Fibration structure} & &(\Delta_5^{\tilde X},\tilde X_4)  & \leftrightarrow&   (\Delta_5^{ X}, X_4) \\ \cline{1-1}\cline{3-3} \cline{5-5}
	\tilde{X}_4\text{ admits} & &\text{Injection} & \leftrightarrow&\text{Projection} \rule[-3mm]{0mm}{8mm}  \\
   	\text{CY}_{m-1}-\text{fiber } \tilde{f}_{m-1}& & \Delta^{\tilde{f}}_{m}=\tilde{H}\cap \Delta_5^{\tilde{X}}&  & \Delta^f_{m}=P\Delta_5^X\\
	\cline{1-1}\cline{3-3} \cline{5-5}
	X_4\text{ admits} & &\text{Projection} &\leftrightarrow &\text{Injection} \rule[-3mm]{0mm}{8mm}\\
   	\text{CY}_{k-1}-\text{fiber } F_{k-1}&  & \Delta^{\tilde F}_{k}=P\Delta_5^{\tilde{X}}&  &\Delta^{F}_{k}=H\cap \Delta_5^{X}\\
\cline{1-1}\cline{3-3} \cline{5-5}
\end{array}
\eeq
where the arrow `$\leftrightarrow$' indicates the action of mirror symmetry interchanging projection and injection.
Clearly, this analysis can be also used to determine Calabi-Yau fibers $\tilde f_{m-1}$ 
of the mirror $\tilde X_4$. In general, it is not the case that mirror symmetry preserves 
fibration structures. However, in the constructions which we will analyze in the section \ref{ffConstruction}, we 
will find that both $X_4$ and $\tilde X_4$ admit an intriguingly rich fibration structure 

\subsection{Heterotic/F-theory duality \label{het-Fdual}}

In section \ref{Kreuzermethods} we have presented a construction of $X_4$ as the mirror of a 
fibration of the Calabi-Yau manifold $\tilde Y_3$ over $\bbP^1$. Specifically, we 
are interested in considering F-theory on the elliptic fourfold $X_4$. If $X_4$ also admits 
a K3 fibration, one can fiberwise apply the duality of F-theory on K3 with heterotic 
$E_8\times E_8$ strings on $T^2$ \cite{Vafa:1996xn,Sen:1996vd,MV,Andreas:1998zf}. For the four-dimensional compactification
this implies that F-theory on $X_4$ is dual to the heterotic string on $Y_3$~\cite{MV, Bershadsky:1997zs}. 
The gauge theory on the seven-branes arises either perturbatively 
from the ten-dimensional $E_8 \times E_8$ or through non-perturbative effects in the 
heterotic string. In particular, the perturbative gauge group is encoded by 
two bundles $V_1$ and $V_2$ which encode the breaking of $E_8 \times E_8$ 
to smaller gauge groups.

We are interested in the map of the complex structure moduli of $X_4$ to heterotic 
moduli. One first notes that the complex structure moduli of $Y_3$ are identified 
with complex structure moduli of $X_4$ under duality. In fact, one expects that 
the heterotic flux superpotential 
\beq
  W_{\rm het} = \int_{Y_3}  H_3 \wedge \Omega_3\ ,
\eeq
with $H_3$ being the threeform flux background of the heterotic B-field, is mapped to a part of the 
F-theory flux superpotential. For the concrete examples considered in the rest of this 
paper we will show, that there indeed exist periods $\Pi_a^{(i)}$ of the holomorphic fourform $\Omega$
of $X_4$ which are identified with the threefold periods of $\Omega_3$ on $Y_3$. This provides an explicit map 
from the F-theory to the heterotic setup. 

A second set of heterotic moduli which are mapped to the complex structure moduli of $X_4$ 
is a subset of the bundle moduli of $V_1,V_2$. To make this more precise one notes that 
one can specify certain non-trivial bundles on an elliptically fibered $Y_3 \rightarrow B_2^Y$ by 
a spectral cover construction \cite{Friedman:1997yq}. One first defines a flat bundle $V|_{T^2}$ on the elliptic fiber $T^2$ 
and fibers these data over $B_2^Y$. Let us focus on $SU(N)$-bundles in the following. Such 
bundles can be specified by $N$ line bundles on $T^2$, or, in the dual picture, $N$ points on the 
dual torus \cite{Friedman:1997yq}. Fibering these $N$ points over $B_2^Y$ one obtains a divisor in $Y_3$ that is an $N$-fold
cover of $B_2^Y$, called the spectral cover. Concretely, for a Calabi-Yau threefold $Y_3$ of the Weierstrass form 
\beq \label{def-p0}
  p_0 =  y^2 + x^3 + f x z^4 + g z^6
\eeq
the data of the $N$ points on each elliptic fiber are specified by solutions to \cite{Friedman:1997yq,Berglund:1998ej}
\beq \label{def-p+}
  p_+ = b_0 z^N + b_2 x z^{N-2} + b_3 y z^{N-3} + \ldots + \left\{ \begin{array}{l} b_N x^{N/2}\\ b_N y x^{(N-3)/2}\end{array} \right. \ ,
\eeq
where one distinguishes the cases $N$ even and $N$ odd. The $b_i$ are sections
of a line bundle on $B_2^Y$. In general, they depend on moduli fields $\hat z_i$
which encode the deformations of the spectral cover and, hence, the bundle $V$.\footnote{In general, there 
can be also Wilson line moduli. These are, however, absent for our examples.}

In the duality between the heterotic string on $(Y_3,V_1,V_2)$ and F-theory on $X_4$
the moduli $\hat z_i$ of the spectral cover are also mapped to complex structure moduli 
of $X_4$. This identification was made precise in ref.~\cite{Friedman:1997yq,Bershadsky:1997zs,Curio:1998bva}. Roughly speaking, 
for an $SU(N)\times SU(M)$ bundle on the heterotic side, the dual F-theory fourfold is given 
locally by the constraint of the form  \cite{Berglund:1998ej}
\beq \label{spectral_cov}
  \tilde \mu = p_0 + v p_+ + v^{-1} p_- =0 \ ,
\eeq 
where $p_0 = 0$ specifies the threefold $Y_3$ as in \eqref{def-p0}, $p_+ = 0$ specifies an  
$SU(N)$ bundle $V_1$ as in \eqref{def-p+}, and $p_-$ specifies an $SU(M)$ bundle $V_2$. The coordinate $v$ 
is the coordinate on the $\bbP^1$-basis in the $K3$-fibers. For our concrete examples we will show that the 
open string moduli of the seven-branes in F-theory are precisely mapped to the coefficients of the 
spectral cover $p_+$. Switching on four-form fluxes $G_4$ in F-theory 
generates a superpotential for these fields which we will determine explicitly.


\section{Elliptically fibered Calabi-Yau fourfold mirror pairs} \label{ffConstruction}

The computation of the F-theory superpotential \eqref{GVW-super} will be done for a class of Calabi-Yau fourfolds $X_4$ that we will introduce here. Our strategy in constructing a fourfold $X_4$ with a low number of complex structure moduli is to construct its mirror $\tilde X_4$ as a Calabi-Yau threefold fibration $\tilde{Y}_3$ over a $\mathds{P}^1$-base. 
The threefolds $\tilde{Y}_3$ we are interested in are themselves elliptically fibered and admit a local limit yielding the non-compact geometries $\mathcal{O}(K)\rightarrow B_2^{\tilde Y}$ studied in \cite{Aganagic:2001nx} where mirror symmetry with branes was analyzed in detail. This fact will be exploited when we analyze the seven-brane content of the F-theory setup $X_4$ and later on determine the F-theory flux superpotential which we split into flux and brane superpotential as in \eqref{SuperpotLimit} .

Due to this chain of geometries it is natural to introduce the three geometries separately, where we will throughout our whole presentation focus on one concrete example for simplicity. We first summarize in section \ref{localThreefold} the local geometry $\mathcal{O}(K)\rightarrow B_2^{\tilde Y}$ for the example of $B_2^{\tilde Y}=\mathds{P}^2$ with D5-branes on its mirror geometry given by the Riemann surface $\Sigma$ of \cite{Aganagic:2001nx}. This non-compact geometry has both one complex structure modulus as well as one brane modulus of the D5-brane.
Then we consider the threefold $\tilde Y_3$ obtained by compactifying this geometry in section \ref{threefold}. We will put emphasis on the singularities of the elliptic fibration of $Y_3$ and its interpretation when going back from $\tilde Y_3$ to the local geometry. One of the two complex structure moduli of $Y_3$ is fixed in the local limit whereas the second one appears as a parameter of one component of the discriminant of the elliptic fibration of $Y_3$. Finally in section \ref{fourfolds} we will construct the fourfold $\tilde{X}_4$ and its mirror $X_4$ used for our four-dimensional F-theory compactification. There we will analyze the seven-brane content in detail. We will find that among the four complex structure moduli of $X_4$ one matches the complex structure of $\Sigma$ of the non-compact geometry and a second one introduces a brane modulus. Here the crucial point will be the geometrical interpretation of the appearance of this new modulus in F-theory opposed to the naive expectation from perturbative IIB with branes on $Y_3$. 

\subsection{The non-compact Calabi-Yau geometry with D-branes \label{localThreefold}}

In the following we will discuss the local Calabi-Yau geometries in which the explicit computations of
open and closed BPS numbers can be performed for the example of local $\mathds{P}^2$, i.e.~${\cal O}(-3)\rightarrow \mathds{P}^2$. Then we will consider the elliptically fibered Calabi-Yau threefold in the weighted 
projective space $\mathds{P}^4(1,1,1,6,9)$ that contains the non-compact geometry 
in the limit of large elliptic fiber.  

\begin{figure}[!ht]
\begin{center} 
\includegraphics[height=6cm]{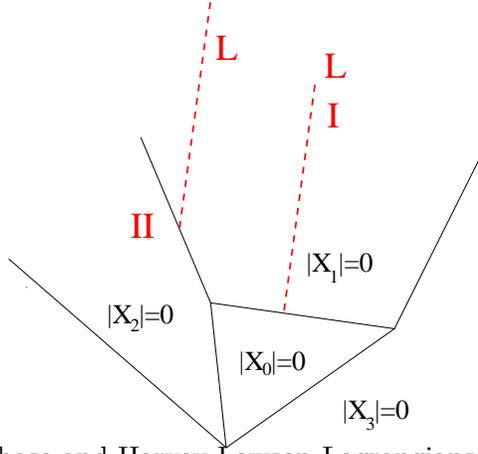}
\vspace*{-.5cm} 
\caption{\small  \label{two_branes_phases} Toric base and Harvey-Lawson Lagrangians for
  non-compact $\mathds{P}^2$  }
\end{center} 
\end{figure}

In~\cite{Aganagic:2001nx} the non-compact 
${\cal O}(-3)\rightarrow \mathds{P}^2$ geometry with non-compact 
Harvey-Lawson branes was considered. 
The local Calabi-Yau is defined as the toric variety $\tilde V_3$ characterized by the polyhedron
\begin{equation}\label{localp2}
	\begin{pmatrix}[c|ccc|c|c]
	    	&   &  \Delta_3    &   	&  \ell^{(1)}& \\ \hline
		v_1 & 0 & 0 &  1 	   &-3 &  X_0\\
		v^b_1 & 1 & 1 &  1 	   & 1 & X_1\\
		v^b_2 &-1 & 0 &  1 	   & 1 & X_2\\
		v^b_3 & 0 &-1 &  1 	   & 1 & X_3\\
	\end{pmatrix},
\end{equation}
where the superscript $^b$ denotes the two-dimensional basis $\mathds{P}^2$ and the $X_i$ denote homogeneous coordinates.
The D-term constraint for this geometry reads
\begin{equation}
-3|X_0|^2+|X_1|^2+|X_2|^2+|X_3|^2=0\, 	
\end{equation}
and $\tilde V_3$ can be viewed as a $(S^1)^3$-fibration over a three-dimensional base $\mathbb{B}_3$. The degeneration loci of the fiber, $|X_i|=0$, are shown in figure \ref{two_branes_phases}. 
The brane is defined torically by the brane charge vectors  
\beq \label{def-ellhat}
    \hat\ell^{(1)}=(1,0,-1,0) \ , \qquad  \hat\ell^{(2)}=(1,0,0,-1)\ . 
\eeq    
This leads to the two constraints
\begin{equation} \label{ABrane}
 	|X_0|^2-|X_2|^2=c^1\,,\qquad |X_0|^2-|X_3|^2=c^2\,,
\end{equation}
where the $c^a$ denote the open string moduli.
The brane geometry is $\mathbb{C}\times
S^1$ and can be described by a one dimensional half line in the three real 
dimensional toric base geometry $\mathbb{B}_3$ ending on a line where two of the three
$\mathbb{C}^*$-fibers degenerate. The A-brane has two inequivalent brane phases I and II as indicated in Figure \ref{two_branes_phases}.\footnote{Note that our phase II is precisely phase III of \cite{Aganagic:2001nx}. The phase 
II of \cite{Aganagic:2001nx} has been omitted since it is equivalent to phase I by symmetry of $\mathds{P}^2$.} 

Mirror Symmetry for this geometry was analyzed in \cite{Aganagic:2001nx} where the disk instantons of the A-model were calculated exploiting the fact that the mirror geometry of $\mathcal{O}(-3)\rightarrow \mathds{P}^2$ effectively reduces to the Riemann surface $\Sigma$ defined by $W(x_1,x_2)=0$ of \eqref{localMirror}. The D6-brane is mapped under mirror symmetry 
to a D5-brane which intersects $\Sigma$ in a point. It will be this D5-brane picture 
which can be reformulated as a seven-brane with flux and embedded into an F-theory compactification.

\subsection{The compact elliptic Calabi-Yau threefold\label{threefold}}

This local Calabi-Yau can easily be embedded into a compact Calabi-Yau threefold. 
The compactification can be understood as a replacement of the non-compact $\mathbb{C}$-fiber
corresponding to $v_1$ of ${\cal O}(-3)\rightarrow \mathds{P}^2$ by an elliptic fiber. 
Here we choose the generic fiber to be the elliptic curve in $\mathds{P}^2(1,2,3)$ which we fiber over the $\mathds{P}^2$-basis the same way as the non-compact $\mathbb{C}$-fiber before. Thus, the polyhedron of this compact threefold $\tilde Y_{3}$, its charge vectors, the homogeneous coordinates $\tilde{x}_i$ as well as the corresponding monomials for the mirror geometry $Y_3$, cf.~\eqref{mirror3foldellp2}, are given by
\begin{equation}\label{3foldellp2}
	\begin{pmatrix}[c|cccc|cc|c|c]
	    	&   &  \Delta_4^{\tilde Y} &   &    &  \ell^{(1)} & \ell^{(2)} &  & \\ \hline
		v_0 & 0 & 0 & 0 & 0 	  &  0  &  -6 & \tilde x_0 & zxyu_1u_2u_3\\
		v_1 & 0 & 0 & 2 & 3 	  &  -3   &1  & \tilde x_1 & z^6 u_1^6 u_2^6 u_3^6\\
		v^b_1 & 1 & 1 & 2 & 3 	&  1   & 0  & \tilde x_2 & z^6 u_3^{18}\\
		v^b_2 &-1 & 0 & 2 & 3 	&  1   & 0  & \tilde x_3 & z^6 u_1^{18}\\
		v^b_3 & 0 &-1 & 2 & 3 	&  1   & 0  & \tilde x_4 & z^6 u_2^{18}\\
	        v_2 & 0 & 0 &-1 & 0 	&  0   & 2  & \tilde x_5 & x^3\\
		v_3 & 0 & 0 & 0 &-1 	&  0   & 3  & \tilde x_6 & y^2 
	\end{pmatrix}.
\end{equation}
Here the points $v_1,v_2,v_3$ carry the information of the elliptic fiber where we added the inner point $v_1$ in order to recover the $\mathds{P}^2(1,2,3)$, in particular its homogeneous coordinate $\tilde{x}_1$ with weight one 
under the new $\mathbb{C}^*$-action $\ell^{(2)}$. 
Furthermore, applying the insights of \eqref{fibrations}, the elliptic 
fibration structure of $\tilde{Y}_3$ is obvious from the fact, that the polyhedron of $\mathds{P}^2(1,2,3)$ 
occurs in the hyperplane $H=\{(0,0,a,b)\}$, but also as a projection $P$ on the (3-4)-plane indicating an 
elliptic fibration of the mirror $Y_3$, too.\footnote{Besides the above chosen $(2,3)$, which leads to an 
elliptic fibration with one section, the values $(1,2)$ and $(1,1)$ are also 
admissible in the sense that these choices lead to reflexive polyhedra. 
The corresponding elliptic fibration has two and three sections, respectively. }

The polyhedron (\ref{3foldellp2}) corresponds to the degree $18$ hypersurface in the
weighted projective space $\mathds{P}^4(1,1,1,6,9)$ blown up along the singular 
curve $\tilde{x}_2=\tilde{x}_3=\tilde{x}_4=0$ with exceptional divisor $v_1$. Its Euler number is $\chi=-540$ whereas $h^{1,1}=2, \ h^{2,1}=272$. Denoting the toric divisors $\tilde x_i =0$ by $D_i$, the
two K\"ahler classes $J_1 = D_2$ and $J_2 = 3 D_2 + D_1$ correspond to the
Mori vectors $\ell^{(1)}$ and $\ell^{(2)}$ in \eqref{3foldellp2}. They represent a curve 
in the hyperplane class of the $\mathds{P}^2$ base and a curve in the elliptic
fiber, respectively. The intersections of the dual divisors and the second Chern class 
are respectively computed to be\footnote{In performing these 
toric computations we have used the Maple package {\tt Schubert}.}
\bea \label{intersectionsY}
  {\mcal C}_0 &=& 9 J_2^3+3 J_2^2 J_1 + J_2 J_1^2\ , \\
  {\mcal C}_2 &=& 102\, J_2+36\, J_1\ .\nn
\eea
In this notation the coefficients 
of the top intersection ring $\mcal C_0$ are the cubic
intersection numbers $J_i\cap J_j\cap J_k$, while the coefficients
of $\mcal C_2$ are $[c_2(T_{\tilde X})]\cap J_i$. 

Mirror symmetry for this example has been studied
in~\cite{Hosono:1993qy,Candelas:1994hw}. 
In order to construct the mirror pair $(Y_3,\tilde Y_3)$ as well as their constraints \eqref{Y3typeIIA}, \eqref{Y3typeIIB} we need the dual polyhedron 
\begin{equation}\label{mirror3foldellp2}
	\begin{pmatrix}[c|cccc|c]
	      &    &  \Delta_4^{Y} &   & &   \\ \hline
	v_1   &  0 &   0 &  1 &  1 & z\\
	v_1^b &-12 &   6 &  1 &  1 & u_1\\
	v_2^b &  6 & -12 &  1 &  1 & u_2\\
 	v_3^b &  6 &   6 &  1 &  1 & u_3\\
	v_2   &  0 &   0 & -2 &  1 & x\\
 	v_3   &  0 &   0 &  1 & -1 & y
	\end{pmatrix},
\end{equation}
where again the basis was indicated by a superscript $^b$. 
Again we added the inner point $v_1$ to recover the polyhedron of $\mathds{P}^2(1,2,3)$ as the injection with $H=\{0,0,a,b\}$, thus confirming the elliptic fibration of the mirror $Y_3$.
Here we distinguish between the two-dimensional basis $B_2^Y=\mathds{P}^2$ and the elliptic fiber by denoting the homogeneous coordinates of $\mathds{P}^2(1,2,3)$ by $(z,x,y)$ and of $B_2^Y$ by $(u_1,u_2,u_3)$. 
The elliptic fibration structure reflects in particular in the constraint of $Y_3$ which takes a Weierstrass form
\begin{equation} \label{Weierst3fold}
 	p_0=a_6 y^2 +a_5 x^3+ a_0 z x y u_1 u_2 u_3+z^6 (a_3 u_1^{18}+a_4 u_2^{18}+a_1 u_1^6 u_2^6 u_3^6+ a_2 u_3^{18})=0\,.
\end{equation}
The generic fiber can be seen by setting the coordinates $u_i$ of the basis $B_2^Y$ to some reference point, such that $p_0$ takes the form of a degree six hypersurface in $\mathds{P}^2(1,2,3)$. The basis itself is obtained as the section $z=0$ of the elliptic fibration over $B_2^Y$.

The complex structure dependence of $Y_3$ is evident from the dependence of $p_0$ on the 
parameters $a_i$ which are coordinates on $\mathds{P}^6$. However, they redundantly 
parameterize the complex structure of $Y$ due to the symmetries of $\mathds{P}^4(1,1,1,6,9)$. 
Indeed there is a $(\mathbb{C}^\ast)^6/(\mathbb{C}^\ast)^2$ rescaling symmetry of the coordinates 
that enables us to eliminate four of the $a_i$ recovering the two complex structure parameters 
that match $h^{1,1}(\tilde Y_3)=h^{2,1}(Y_3)=2$. The appropriate coordinates $z_i$ obeying $z_i=0$ 
at the large complex structure/large volume point are completely determined by the phase 
of the A-model, i.e.~the choice of charge vectors $\ell^{(i)}$ of $\Delta_4^{\tilde{Y}}$. They are 
given in general by 
\begin{equation} \label{LargeComplexStr}
 	z_{i}=(-1)^{\ell^{(a)}_{0}}\prod_{j=0}^m a_j^{\ell_j^{(i)}}\,,
\end{equation}
which we readily apply for the situation at hand to obtain
\begin{equation} \label{zLargeRadius}
 	z_1=\frac{a_2a_3a_4}{a_1^3}\ ,\qquad z_2=\frac{a_1a_5^2a_6^3}{a_0}\ .
\end{equation}
Thus, we can use the $(\mathbb{C}^\ast)^4$ action and the overall scaling 
to set $a_i=1,\, i=2,\ldots,6$ for five parameters to obtain
\beq \label{MirrorEtale}
 	p_0  =  y^2+x^3+z x y m_1 +z^6 m_6 \ ,
\eeq
where we have abbreviated 
\beq \label{defm_i}
 	m_1 = z_2^{-1/6} z_1^{-1/18} u_1 u_2 u_3\ ,\qquad \quad m_6 =u_1^{18}+u_2^{18}+u_3^{18}+z_1^{-1/3}u_1^6 u_2^6 u_3^6\ .
\eeq
Alternatively, this result can be obtained more directly by the mirror 
construction  \eqref{eqn:HVmirror}. In this case one needs the 
following assignment of coordinates $y_i$ to points of $\Delta_4^{\tilde Y}$ and monomials
\begin{equation}  \label{etalep2}
 	\begin{pmatrix}[c|c|c] 	 	
	y_0 &v_0  &  a_0\, z x y u_1 u_2 u_3 \\
	y_1 &v_1  &  a_1\, z^6 u_1^6 u_2^6 u_3^6 \\
	y_2 &v^b_1&  a_2\, z^6 u_3^{18} \\ 
	y_3 &v^b_2&  a_3\, z^6 u_1^{18} \\ 
	y_4 &v^b_3&  a_4\, z^6 u_2^{18} \\
	y_5 &v_2  &  a_5\, x^3 \\
	y_6 &v_3  &  a_6\,  y^2 
 	\end{pmatrix}\ . 
\end{equation}
This defines the etal\'{e}-map that solves the constraints of \eqref{eqn:HVmirror} automatically when \eqref{zLargeRadius} holds. By setting $a_0=z_2^{-1/6}z_1^{-1/18}$, $a_1=z_1^{-1/3}$ and $a_i=1,\, i=2,\ldots,6$ we solve \eqref{zLargeRadius} and $W=\sum_jy_j$ immediately reproduces $p_0$ in \eqref{MirrorEtale}.

Next we show that \eqref{MirrorEtale} indeed gives back the local geometry which is a 
conic over a genus one Riemann surface.
The local limit in the A-model geometry is given by making the 
elliptic fiber infinitely large.
This corresponds to $z_2\rightarrow 0$ in the B-model geometry. 
We parameterize $z_2$ by $\varepsilon\equiv z_2$ such that the local limit 
is given by $\varepsilon\rightarrow 0$.
At the end we should obtain an affine equation, thus, using the two $\mathbb C^*$-action 
we set the coordinates $z$ and $u_3$ to one.
By redefining the coordinates $x$ and $y$ as follows
\begin{equation} \label{epslimit}
	y\rightarrow \varepsilon^{-1/2}y+k_1^{1/2}\quad ,\quad x\rightarrow \varepsilon^{-1/3}x+k_2^{2/3},
\end{equation}
the hypersurface equation $p_0=0$ becomes
\begin{equation}
	p_0=\frac{1}{\varepsilon} \tilde p_0+k_1^2+k_2^2+m_6=0
\end{equation}
where we set $z=1$ and $u_3=1$.
Now we split the above equation
\begin{equation}
	\tilde p_0=\varepsilon\quad  ,\quad k_1^2+k_2^2+m_6 =-1.
\end{equation}
If we now take the $\varepsilon\rightarrow 0$ limit we obtain after appropriately redefining the $k_i$ the equation for the local geometry of the form
\begin{equation}
	uv=H(x,y)=x+1-\phi \frac{x^3}{y}+y.
\end{equation}
The Riemann surface defined by $H(x,y)=0$ is isomorphic to the surface $m_6=0$ up to isogeny.

As discussed in section \ref{het-Fdual} considering heterotic 
string theory on the elliptically fibered Calabi-Yau threefold $Y_3$ 
is expected to be dual to F-theory on $X_4$ if the fourfold admits a K3 fibration 
and its mirror is constructed as in \eqref{tildeYoverP1} \cite{Berglund:1998ej}. We have shown that $Y_3$ is indeed 
an elliptic fibration, and will confirm in the next section 
that $X_4$ is a K3 fibration. However, it is crucial to point 
out that there will be a large non-perturbative gauge group from 
the blown-up singularities of the elliptic fibration of $Y_3$.
Upon introducing the full set of coordinates, i.e.~introducing the inner points in $\Delta_4^Y$, 
one notes that the elliptic fibration not only degenerates over the curves $m_6=0$ and 
$432 m_6 + m_1^6=0$ in the base of $Y_3$, but also over many curves described by the 
additional coordinates. Even though the determination of the non-perturbative 
gauge group will be not of importance in our analysis, let us point out 
that we will similarly find a large gauge group in the F-theory compactification on 
$X_4$. However, the identification of the moduli of the gauge bundles with the complex 
structure moduli of $X_4$ can still be performed by extracting the spectral cover constraint \eqref{spectral_cov}.

Before continuing with the construction of the Calabi-Yau fourfold, let us close 
with another comment on the use of the vectors $\hat \ell^{(1)}$ and $\hat \ell^{(2)}$ 
given in \eqref{def-ellhat}. On the compact threefold they translate to 
\beq
   \hat \ell^{(1)} = (0,1,0,-1,0,0,0)\ ,\qquad \hat \ell^{(2)} = (0,1,0,0,-1,0,0)\ , 
\eeq
due to the new origin in the polyhedron \eqref{3foldellp2}.
In fact, applying \eqref{BBrane} and using \eqref{etalep2}, they define the divisors 
\begin{equation}\label{divisors}
 	   z_1^{-1/3}  u_1^6 u_2^6 u_3^6 =\hat{z}_1  u_1^{18}\,,\qquad 
 	   z_1^{-1/3}  u_1^6 u_2^6 u_3^6 =\hat{z}_2  u_2^{18}\,,
\end{equation}
in the compact $Y_3$. Here we introduced the moduli $\hat{z}_a$ corresponding to the 
charge vector $\hat{\ell}^{(a)}$. Note that in our F-theory compactification of the next section 
we will not consider seven-branes naively wrapped on these divisors as one would in a compactification 
of Type IIB on Calabi-Yau orientifolds. 
Rather we will construct a Calabi-Yau fourfold with 7-branes on its discriminant which 
possess additional moduli. These additional fields correspond to either $\hat z_1$ or $\hat z_2$
and allow deformations of the seven-brane constraint by the additional terms \eqref{divisors}.
Hence, $\hat z_i$ can be interpreted as deformations of the seven-brane divisors in $X_4$, or as
spectral cover moduli in the heterotic dual.

\subsection{Construction of the elliptically fibered Calabi-Yau fourfold} \label{fourfolds}

Having discussed the threefold geometry, we are now in the position to construct and analyze the 
elliptically fibered Calabi-Yau fourfold $X_4$ which is used in the F-theory compactification.
Again, we start by constructing the mirror $\tilde X_4$ first. It is obtained 
by fibering the Calabi-Yau threefold $\tilde Y_3$ over a $\bbP^1$ such that one 
of the D-brane vectors $\hat \ell^{(i)}$ of the local model \eqref{localp2} 
appears as a new charge vector. As we will see later on, this new charge 
vector dictates the location of the moving seven-brane, while the second vector not used in the construction of the fourfold controls the volume of the $\mathds{P}^1$-basis of the dual fourfold $\tilde{X}_4$ in \eqref{tildeYoverP1}.
We will discuss the flux in more detail later on, but first 
proceed with the geometric construction. 
In the following we will exemplify our construction for a main example in detail,
and we list the toric and geometrical data necessary to reproduce our results. 
Further examples are relegated to appendix \ref{appa}.

The Calabi-Yau fourfolds $(X_4,\tilde X_4)$ are realized as hypersurfaces in 
a toric ambient space described by a dual pair of reflexive 
polyhedra $(\Delta_5^X,\Delta_5^{\tilde X})$. The 
reflexive polyhedron $\Delta_5^{\tilde X}$ for a fibration of the toric 
variety constructed from $\Delta_4^{\tilde Y}$ over $\P^1$ is specified as follows
\begin{equation}
	\Delta_5^{\tilde X} =
	\left(
	\begin{array}{rccc|r}
		\multicolumn{4}{c|}{\Delta_4^{\tilde Y}} & 0\\ \hline
		\text{-}1 & 0 & 2 & 3 & \text{-}1\\
		0 & 0 & 2 & 3 & \text{-}1\\
		0 & 0 & 2 & 3 & 1
	\end{array}
	\right).
	\label{eqn:delta5}
\end{equation}
By construction, one finds $\Delta_4^{\tilde Y}$ by intersecting the hyperplane $\tilde H = (p_1,p_2,p_3,p_4,0)$ with $\Delta_5^{\tilde X}$. 
Following \eqref{fibrations} this indeed identifies $\tilde X_4$ as a $\tilde Y_3$-fibration, and by performing the quotient $\Delta_5^{\tilde X}/\Delta_4^{\tilde Y}$ 
the base is readily shown  
to be the toric variety $((-1),(1))$, i.e.~a $\bbP^1$. It is crucial to note that the additional points which do not 
lie on $\tilde H$ are constrained by two important conditions. Firstly, they are chosen of a form such that the 
mirror $X_4$ is elliptically fibered. This means, that using the projection
to the third and fourth coordinate one finds the polyhedron of a torus in $\bbP^2(1,2,3)$ just as in 
the threefold case in section \ref{localThreefold}. The fact that $\tilde X_4$ is also elliptically fibered is not 
crucial in the construction. In particular, a similar construction can also be performed 
for the quintic hypersurface fibered over a $\bbP^1$, since the mirror quintic admits an 
elliptic fibration with generic elliptic fiber being a torus in $\bbP^2$. Secondly, 
the remaining entries are inserted such that one charge vector for the Calabi-Yau fourfold is 
of the form $(\hat\ell^{(1)},-,-,-)$. Adding this vector to form a higher-dimensional non-reflexive 
polyhedron was already proposed in \cite{Lerche:2001cw,Alim:2009rf}.\footnote{Note 
that the interpretation of  the construction in terms of the B-model in \cite{Lerche:2001cw, Alim:2009rf} seems different from the F-theory interpretation given here.}

In the following we will choose to realize the open string vector $\hat\ell^{(1)}$ as in \eqref{eqn:delta5}
to construct the $\P^1$-fibration.\footnote{We could have used also $\hat\ell^{(2)}$, reproducing 
the same local D5-brane limit. In this case $\hat \ell^{(1)}$ specifies the gauge flux inducing the 
D5-brane charge on the D7-brane.} 
The Calabi-Yau fourfold $\tilde X_4$ is realized in the toric space described by the polyhedron $\Delta_5^{\tilde X}$.
Its topological numbers are computed to be 
\begin{equation}
       h^{3,1}=2796\ ,\quad  h^{1,1}=4\ ,\quad h^{2,1}=0\ ,\quad
  h^{2,2} = 11244\,,\quad \chi = 16848\, .
\end{equation}
Here we first used (\ref{Hodgenumbers1}), (\ref{Hodgenumbers2}) as well as (\ref{Hodgenumbers3}) and next applied (\ref{HodgeRel}), (\ref{FFEulerNumb}).

Note that $\Delta_5^{\tilde X}$ has three triangulations, which correspond to non-singular Calabi-Yau phases which 
are connected by flop transitions.
In the following we will consider two of these phases in detail. These phases will match the 
two brane phases in figure \ref{two_branes_phases} in the local Calabi-Yau threefold geometry.

To summarize the topological data of the 
Calabi-Yau fourfold for the two phases of interest, we  
specify the generators of the Mori cone $l^{(i)}_{I}$ and $l^{(i)}_{II}$ for
$i=1,\ldots4$.
\small
\begin{equation} 
	\begin{pmatrix}[c|ccccc|cccc|cccc]
	    	& && \Delta_5^{\tilde X} &&  &           \ell_I^{(1)} & \ell_I^{(2)} &
                \ell_I^{(3)} & \ell_I^{(4)} & \ell_{II}^{(1)} &
                \ell_{II}^{(2)} & \ell_{II}^{(3)} & \ell_{II}^{(4)} \\ \hline
		v_0   & 0 & 0 & 0 & 0 & 0 	    & 0  &  -6   & 0  &0    &  0    & -6&  0  & 0    \\
		v^b_1 & 0 & 0 & 2 & 3 & 0 	    &-2  & 1   &-1  & -1    & -3    & 0&  1  &  -2    \\
		v^b_2 & 1 & 1 & 2 & 3 & 0 	    & 1  &  0   & 0  & 0    &  1    & 0&  0  &  0     \\
		v^b_3 &-1 & 0 & 2 & 3 & 0 	    & 0  &  0   & 1  & -1   &  1    & 1& -1  &  0    \\
		v^b_4 & 0 &-1 & 2 & 3 & 0 	    & 1  &  0   & 0  & 0    &  1    & 0&  0  &  0 \\
		v_1   & 0 & 0 &-1 & 0 & 0 	    & 0  &  2   & 0  & 0    &  0    & 2&  0  &  0 \\
		v_2   & 0 & 0 & 0 &-1 & 0 	    & 0  &  3   & 0  & 0    &  0    & 3&  0  &  0 \\
   \hat v_1   &-1 & 0 & 2 & 3 &-1	  & 1  &  0   &-1  & 1    &  0    & -1&  1 &  0  \\
	 \hat v_2   & 0 & 0 & 2 & 3 &-1 	&-1  &  0   & 1  & 0    &  0    & 1& -1  &  1 \\
	 \hat v_3   & 0 & 0 & 2 & 3 & 1 	& 0  &  0   & 0  & 1    &  0    & 0&  0  &  1  
	\end{pmatrix}.
	\label{eqn:vl-for-p2}
\end{equation}
\normalsize
The charge vectors are best identified in the phase II. Here $\ell^{(1)}_{II}$ and $\ell^{(2)}_{II}$
are the extensions of the threefold charge vectors $\ell^{(1)},\, \ell^{(2)}$ in \eqref{3foldellp2} to the fourfold. 
The brane vector $\hat\ell^{(1)}$ is visible in phases II as a subvector of $\ell^{(3)}_{II}$. 
The remaining vector $\ell^{(4)}_{II}$ arises since one had to complete the polyhedron 
such that it becomes reflexive implying that $\tilde X_4$ is a Calabi-Yau manifold.
It represents the curve of the $\mathds{P}^1$-basis of $\tilde{X}_4$. Phase I is related to phase II by a flop transition of the curve associated to 
$\ell^{(3)}_I$. Hence, in phase I the brane vector is identified with $-\ell^{(3)}_I$.
Furthermore, after the flop transition one has to take 
\beq \label{ell_flop}
  \ell^{(3)}_{II}=-\ell^{(3)}_{I}\ ,\qquad \ell^{(1)}_{II}=\ell^{(1)}_I + \ell^{(3)}_I\ ,\qquad 
   \ell^{(2)}_{II}=\ell^{(2)}_I + \ell^{(3)}_I\ ,\qquad  \ell^{(4)}_{II}=\ell^{(4)}_I + \ell^{(3)}_I\ .
\eeq
Note that the $\ell^{(i)}_I$ and $\ell^{(i)}_{II}$ are chosen in such a way, that they parameterize 
the Mori cone of $\tilde X_4$. The dual K\"ahler cone generators for phase I are then given by 
\begin{equation} \label{KCPhaseI}
      J_1=D_2\ ,\quad   J_2=D_1+2D_2+D_3+2D_9\ ,\quad J_3=D_3+D_9, \quad J_4=D_9\ ,
\end{equation}
where $D_i:=\{x_i=0\}$ are the nine toric divisors associated to the points $\Delta_5^{\tilde X}$
which differ from the origin. In phase II one has 
\begin{equation} \label{KCPhaseII}
	J_1=D_2,\quad J_2=D_1 + 2 D_2 + D_3 + 2 D_9 \ ,\quad  J_3=D_1 + 3 D_2 + 2 D_9,\quad J_4=D_9 \ .
\end{equation}
The $J_i$ provide a distinguished integral basis of $H^{1,1}(\tilde X_4)$ since in the 
expansion of the K\"ahler form $J$ in terms of the $J_i$ all coefficients will 
be positive when parameterizing physical volumes of cycles in $\tilde X_4$. The 
$J_i$ are also canonically used as a basis in which one determines the topological data of 
$\tilde X_4$. The complete set of topological data of $\tilde X_4$ including the intersection ring as well 
as the non-trivial Chern classes are summarized in appendix \ref{FurtherP2}.

The polyhedron $\Delta_5^{\tilde X}$ has only few K\"ahler classes which makes it possible to 
identify part of the fibration structures 
from the intersection numbers.
However, an analogous analysis is not possible for the mirror manifold $X_4$ since the dual polyhedron $\Delta_5^{X}$ has $2796$ K\"ahler classes.
Therefore, we apply the methods reviewed in section \ref{Kreuzermethods} in analyzing both $\tilde X_4$ and $X_4$.
As already mentioned above, $\Delta_5^{\tilde X}$ intersected with the 
two hyperplanes
\begin{equation}
	H_1 = (0,0,p_3,p_4,0)\ ,\qquad H_2=(p_1,p_2,p_3,p_4,0).
\end{equation}
gives two reflexive polyhedra corresponding to the generic torus fiber and the generic three-dimensional Calabi-Yau fiber $\tilde Y_3$. 
The fibration structures of the mirror Calabi-Yau $X_4$ is studied by identifying appropriate projections to $\Delta^k_{\tilde{F}}\subset\Delta_5^{\tilde X}$ .
Three relevant projections $P_i$ are 
\begin{equation}
	P_1(\underline{p})=(p_3,p_4)\ , \qquad P_2(\underline{p})=(p_1,p_2,p_3,p_4)\ , \qquad P_3(\underline{p})=(p_3,p_4,p_5)\ ,
\end{equation}
where $\underline{p}=( p_1,\dots,p_5)$ are the columns in the polyhedron $\Delta_5^{\tilde X}$.
Invoking the theorem of section \ref{Kreuzermethods}, we see from $P_1$ 
that $X_4$ is also elliptically fibered and 
since the polyhedron of $\P^2({1,2,3})$ is self-dual, the fibration is of $\P^2({1,2,3})$ type.
In addition, it is clear from $P_2$ that $X_4$ is Calabi-Yau threefold fibered.
The fiber threefold is $Y_3$, the mirror to $\tilde Y_3$. The fact, that the threefold 
fibers of $X_4$ and $\tilde X_4$ are mirror manifolds is special to this example since the subpolyhedra
obtained by $H_2$ and $P_2$ are identical. Finally, note that $X_4$ is K3 fibered as inferred 
from the projection $P_3$. This ensures the existence of a heterotic dual theory by 
fiberwise applying the duality of F-theory on K3 to heterotic strings on $T^2$. Replacing 
the K3-fiber by an elliptic fiber we find the Calabi-Yau threefold $Y_3$.

The hypersurface constraint for $X_4$ depends on the four complex structure moduli $z_i$. This dependence 
is already captured by only introducing $12$ out of the many coordinates needed to specify a non-singular 
$X_4$. This subset of points in $\Delta_5^X$ is given by
\small
\begin{equation} \label{Delta5P2}
         \Delta_5^X \supset	\left(
\begin{array}{c|rrrrr|c}
v_1   &   0 &  0 & 1 & 1 & 0 & z \\
v_2   & -12 & 6 & 1 & 1 & 0 & u_1\\
v_3   &  6  &-12 & 1 & 1 & 0 & u_2\\
v_4   &  6  &  6 & 1 & 1 & 0 & u_3\\
v_5   &  0  &  0 &-2 & 1 & 0 & x\\
v_6   &  0  &  0 & 1 &-1 & 0 & y\\
v^b_1 & -12 &  6 & 1 & 1 &-6 & x_1\\
v^b_2 & -12 &  6 & 1 & 1 & 6 & x_2\\
v^b_3 &  6  &-12 & 1 & 1 &-6 & x_3\\
v^b_4 &  6  &  6 & 1 & 1 &-6 & x_4\\
v^b_5 &  0  & -6 & 1 & 1 & 6 & x_5\\
v^b_6 &  0  &  6 & 1 & 1 & 6 & x_6
\end{array}
\right)
\end{equation}
\normalsize
where we have omitted the origin. Note that we have displayed in \eqref{Delta5P2} 
the vertices of $\Delta_5^X$ and added the inner points $v_1$ 
and $v_2$ to list all points necessary to identify the polyhedron 
$\Delta_4^Y$ with vertices \eqref{mirror3foldellp2} in the hyperplane orthogonal to $(0,0,0,0,1)$ and thus the 
Calabi-Yau fibration with fiber $Y_3$. The base of this fibration is given 
by the points labeled by a superscript $^b$.
Note that $(0,0,1,1,0)$ is also needed to observe the elliptic fibration.
The base of the elliptic fibration is obtained by performing the quotient 
$\Delta_3^{\text{base}}=\Delta_5^X/(P_1\Delta_5^{\tilde X})^*$ which 
amounts to simply dropping the third and fourth 
entry in the points of $\Delta_5^X$.
Additionally, one can also see the elliptic fibration directly on the 
defining polynomial $\tilde \mu$ of $X_4$ 
which can be written in a Weierstrass form. Indeed if we apply (\ref{Y3typeIIB}) 
for the points displayed in \eqref{Delta5P2} of $\Delta_5^X$ and all points 
$p$ of $\Delta_5^{\tilde X}$ that are not on codimension one faces we obtain a 
hypersurface of the form\footnote{The polynomial $\tilde \mu$ can 
be easily brought to the standard Weierstrass form by completing the square and the cube, i.e.\ $\tilde y=y+\frac12\tilde m_1 xz$ and 
$\tilde x=x-\frac{1}{12}\tilde m_1^2z^2$.}
\begin{equation} \label{preWeier}
	\tilde \mu=a_6 y^2+ a_5 x^3+\tilde m_1 (x_j,u_i) x y z+\tilde m_6 (x_j,u_i) z^6=0\, .
\end{equation}
Here $x_j,u_i$ are the homogeneous coordinates on the base of the elliptic fibration, while $x$, $y$, and $z$ are the homogeneous 
coordinates of the $\mathds{P}^2(1,2,3)$ fiber.
The polynomials $\tilde m_1$ and $\tilde m_6$ are given by
\begin{eqnarray}  \label{def-tildem1}
 	\tilde m_1(x_j,u_i)&=&a_{0} u_1 u_2 u_3 x_1 x_2 x_3 x_4 x_5 x_6\,,\\ 
 	\tilde m_6 (x_j,u_i)&=&
	 u_1^{18}\, (a_7 x_1^{24} x_2^{12} x_3^6 x_4^6 + a_3 x_1^{18} x_2^{18} x_5^6 x_6^6 )+a_4 u_2^{18}\, x_3^{18} x_5^{12}
	 + a_2 u_3^{18}\, x_4^{18} x_6^{12} \nn\\
	&&+ u_1^6 u_2^6 u_3^6\, ( a_1 x_1^6 x_2^6 x_3^6 x_4^6 x_5^6 x_6^6+ a_9 x_2^{12} x_5^{12} x_6^{12} + a_8  x_1^{12} x_3^{12} x_4^{12})\ , \label{def-tildem6}
\end{eqnarray}
where the $a_i$ denote coefficients encoding the complex structure deformations of $X_4$. 
However, since $h^{3,1}(X_4)=h^{1,1}(\tilde X_4)=4$ there are only four complex structure parameters rendering six of the $a_i$ redundant. It is also straightforward to compare $\tilde m_1,\tilde m_6$ for the fourfold $X_4$ with 
the corresponding threefold data in \eqref{MirrorEtale} and \eqref{defm_i}.

For the different phases we can identify the complex structure moduli in the hypersurface constraint 
by using the charge vectors $\ell^{(i)}$ in \eqref{eqn:vl-for-p2}. For phase I one finds 
\beq \label{zI}
  \zI_1=\frac{a_2a_4a_7}{a_1^2a_8}\,,\quad \zI_2=\frac{a_1 a_5^2 a_6^3}{ a_{0}^6} \,,\quad \zI_3=\frac{a_3 a_8}{a_1 a_7}\,,\quad 
  \zI_4=\frac{a_7 a_9}{a_1a_3}\ ,
\eeq
while for the phase II one finds in accord with \eqref{ell_flop} that
\beq \label{zIzII}
  \zII_1 = \zI_1 \zI_3\ , \quad 
  \zII_2 = \zI_2 \zI_3 \ , \quad 
  \zII_3 = (\zI_3)^{-1} \ ,\quad 
  \zII_4 = \zI_4 \zI_3 \ .
\eeq
In order to compare to the threefold $Y_3$ we chose the gauge $a_i=1,\, i=2,\ldots,6$ and $a_8=1$, such that
\beq \label{id_ai}
   a_0^6 = \frac{1}{(\zII_1)^{{1}/{3}} \zII_2 \zII_3 }\ ,\quad
 a_1 = \frac{1}{(\zII_1)^{1/3}}\ ,\quad
   a_7 =  \zII_3 (\zII_1)^{1/3}\ ,\quad
  a_9 = \frac{\zII_4}{(\zII_1)^{2/3}} 
\ .
\eeq
It is straightforward to find the expression for phase I by inserting \eqref{zIzII} into 
this expression for $a_0,a_1$ and $a_7,a_9$.

Having determined the defining equations for the Calabi-Yau fourfolds it is straightforward 
to evaluate the discriminant $\Delta(X_4)$ of the elliptic fibration. Using \eqref{def-j} for 
a fourfold in the Weierstrass form \eqref{Weierstrass} we find 
that 
\beq \label{disc_X4}
  \Delta(X_4) = - \tilde m_6 (432 \tilde m_6 + \tilde m_1^6)\ .
\eeq 
We conclude that there will be seven-branes on the divisors $\tilde m_6= 0$ and $432 \tilde m_6 + \tilde m_1^6=0$ 
in the base $B_3^X$. The key observation is that in addition to a moduli independent part $\tilde m_6^{0}$ the 
full $\tilde m_6$ is shifted as 
\beq \label{m6shift}
	\tilde m_6 = \tilde m_6^0+a_1 (u_1 u_2 u_3 x_1 x_2 x_3 x_4 x_5 x_6)^6+
	  a_7 u_1^{18} x_1^{24} x_2^{12} x_3^6 x_4^6
	 + a_9 u_1^6 u_2^6 u_3^6 x_2^{12} x_5^{12} x_6^{12}\ .
\eeq
The moduli dependent part is best interpreted in the phase II with $a_1,a_7$ and $a_9$ given 
in \eqref{id_ai}. In fact, when setting the fourth modulus to $\zII_4=0$, one notes
that the deformation of the seven-brane locus $\tilde m_6 = 0$ is precisely parameterized 
by $\zII_3$. By setting $x_i=1$ one fixes a point in the base of $X_4$ viewed as fibration 
with fiber $Y_3$. One is then in the position to compare the shift 
in \eqref{m6shift} with the first constraint in \eqref{divisors}
finding agreement if one identifies $\hat z_1 = \zII_3 (\zII_1)^{1/3}$. In the next section we
will show that the open string BPS numbers of the local model with D5-branes of section \ref{localThreefold}
are recovered in the $\zII_3$ direction. The 
shift of the naive open modulus $\hat z_1$ by the closed complex structure modulus $\zII_1$ 
fits nicely with a similar redefinition made for the local models in ref.~\cite{Aganagic:2001nx}. 
This leaves us with the interpretation that indeed $\zII_3$ deforms the seven-brane locus 
and corresponds to an open string modulus in the local picture. As we will 
show in the next section, a $\zII_3$-dependent superpotential is induced upon 
switching on fluxes on the seven-brane. It can be computed explicitly and matched 
with the local results for D5-branes for an appropriate choice of flux.

A second interpretation of the shifts \eqref{m6shift} by the monomials proportional to 
$\zII_3,\zII_4$ is via the heterotic dual theory on $Y_3$ and the spectral cover 
construction discussed in section \ref{het-Fdual}. To see this, we bring $\tilde \mu$
into the form \eqref{spectral_cov} by an appropriate coordinate redefinition. Setting 
$v=x_1^6 x_3^6 x_4^6 x_2^{-6}$, $\tilde u_1 = u_1 x_1 x_2$, $\tilde u_2=u_2 x_3$, 
$\tilde u_3=u_3 x_4$, and picking the local patch $x_5=x_6=1$ one rewrites \eqref{preWeier} 
as 
\beq
   \tilde \mu = p_0 + v p_+ + v^{-1} p_-\ ,
\eeq
where $p_0(y,x,z,\tilde u_1,\tilde u_2,\tilde u_3)=0$ is the threefold constraint \eqref{Weierst3fold} of $Y_3$, and 
\beq
   p_+ = (a_7 \tilde u_1^{18} + a_8 \tilde u_1^{6} \tilde u_2^{6} \tilde u_3^{6}) z^6 \ , \qquad p_- = a_9 \tilde u_1^{6} \tilde u_2^{6} \tilde u_3^{6} z^6 \ .
\eeq
Hence, in the local mirror limit in which $p_- \rightarrow 0$ \cite{Berglund:1998ej}, it is natural to interpret 
the modulus $\zII_3$ as a bundle modulus of $V_1=\text{SU}(1)$ in the heterotic dual 
theory, i.e.~as a deformation of the spectral cover as in \eqref{spectral_cov}. One might be surprised that an SU$(1)$-bundle carries any bundle moduli due to the trivial structure group. Indeed the adequate physical interpretation of this configuration is in terms of heterotic five-branes as discussed in detail in \cite{Grimm:2009sy}.

Finally, as a side remark, let us note again that \eqref{disc_X4} with \eqref{def-tildem1} and \eqref{def-tildem6} 
is not the full answer for the discriminant since we have set many of the blow-up coordinates 
to unity. However, one can use the toric methods of \cite{Bershadsky:1996nh,Candelas:1996su,Candelas:1997eh}
to determine the full minimal gauge group in the absence of flux to be 
\beq
   G_{X_4}\ =\ E_8^{25}\ \times\ F_4^{69} \ \times \ G_2^{184}\ \times \ SU(2)^{276}\ .
\eeq
Groups of such large rank are typical for elliptically fibered Calabi-Yau geometries 
with many K\"ahler moduli corresponding to blow-ups of singular fibers \cite{Candelas:1997eh}.

\section{Mirror symmetry for Calabi-Yau fourfolds} 
\label{FFMirrors}

In this section we will describe mirror symmetry on fourfolds. In its weak
formulation it states the  equivalence of the complex structure moduli space of
$X$ and the (instanton corrected) K\"ahler moduli space of its mirror
$\tilde{X}$. As was pointed out in~\cite{Witten:1991zz} this equivalence can be
formulated in physical terms by considering topological field theories called
the A- and B-model on the spaces $(\tilde{X},X)$. These theories are consistent
cohomological truncations of some particular $(2,2)$ superconformal field
theories and their physical observables are the vertical subspace of the de Rham
groups $H^{p,p}(\tilde X)$, $0\leq p\leq n$ and the horizontal subspace of
$H^p(X,\bigwedge^q TX)$, $p+q=n$, respectively.  In particular their marginal
deformations coincide with the cohomology groups $H^{1,1}(\tilde X)$ for the
A-model and $H^1(X,TX)$ for the B-model that are, in geometrical terms,
precisely the infinitesimal directions on the K\"ahler and complex structure
moduli space of $\tilde{X}$ and $X$, respectively. Therefore, the physical
statement of mirror symmetry is the equivalence of the A-model constructed from
$\tilde X$ and the B-model constructed from $X$.

Our final goal is the calculation of special holomorphic quantities $F^0(\gamma)$ for
the Calabi-Yau fourfolds. They are identified with the holomorphic
superpotentials of $\mathcal N=1$ effective actions and at the same time they are
generating functions of the genus zero Gromov-Witten invariants of the
fourfold.  In certain cases of special interest they can be  further
interpreted as generating functions of disk instantons in a dual type IIA
theory. The (covariant) double derivatives of the $F^0(\gamma)$, w.r.t.\ to the
moduli $t_a$ are three-point correlators $C^{(1,1,2)}_{ab\gamma}$. Their leading
behaviour is fixed in the A-model by the classical intersection of two
divisors with an element of $H_4(\tilde{X})$. In this sense $F^0(\gamma)$ is similar to
the familiar prepotential $F^0$ of Calabi-Yau threefolds~\eqref{pre_largeV}
whose triple derivative yields in the A-model a three-point function whose
leading behaviour is fixed by the classical intersection between three
divisors.

In order to understand the $F^0(\gamma)$ one needs  to study more fundamental
quantities, namely the two- and three-point correlation functions of the topological
A- and  B-model. They encode all other correlators of these topological
theories.

As described in section \ref{bmodel} the operators representing elements of
$H^p(X,\bigwedge^q TX)$ together with their two-point and three-point  correlators,
given by period integrals, form the B-model operator ring. In the large radius
limit the A-model is defined by another ring, namely the vertical part of the
classical cohomology ring $H^{*,*}(\tilde X)$.  Its two- and three-point correlators
are simply given by the classical intersections. Away from the large radius
point the notion of the classical intersection rings has to be extended to the
quantum cohomology ring in which the three-point correlators are corrected by
holomorphic instantons. 

All these rings with two- and three-point correlators mentioned above carry a natural
Frobenius structure, which we  describe in section \ref{frobenius}. Mirror
symmetry identifies the quantum cohomology ring of the A-model with the B-model
ring. For the precise identification one has to specify matching points in the
moduli spaces. The natural candidate for our present purposes is the large
radius point in the K\"ahler moduli space of the A-model, which is identified
with a point of maximal unipotent monodromy in the complex structure moduli space of the
B-model.  As described in section \ref{matching} the  matching makes use of the
mirror map, properties of the Picard-Fuchs system near the point of maximal
unipotent monodromy and the classical intersection ring of $\tilde{X}$.  The precise matching
is necessary for the enumerative predictions of the A-model and the
construction of a basis of the horizontal cohomology of $H^{2,2}(X)$, which
is needed to identify the flux.
It is crucial for this analysis to identify the integral basis of cohomology. One important step in this context is to determine the classical terms in the leading logarithmic period by means of analytic continuation to other points on the fourfold complex structure moduli space and a monodromy discussion. 

After presenting the general formalism we discuss the  relevant application to
the case of elliptic Calabi-Yau fourfolds. In particular we will argue how the
techniques and results of the first part help us finding the F-theory
interpretation of the prepotentials $F^0(\gamma)$. These are precisely the
quantities that we will identify with the flux superpotential \eqref{GVW-super}
and we will furthermore understand its relation to the Type IIB superpotentials
\eqref{SuperpotLimit}. We will exemplify this rather general analysis for our main
example containing local $\mathds{P}^2$ introduced in the last sections.

Since our discussion can at several points be generalized to arbitrary complex
dimensional Calabi-Yau $n$-folds \cite{Greene:1993vm} we denote a mirror pair of
Calabi-Yau $n$-folds by $(\tilde{X}_n,X_n)$ and identify with the fourfold case by
putting $n=4$.

\subsection{States and correlation function of the B-model}
\label{bmodel}

In the B-model one considers a family of $n$-folds $X_z$ fibered over the complex
structure moduli space ${\cal M}$, $X_z\rightarrow \mathcal{M}$. The
states\footnote{We use in the following the same symbol for states and
operators.} in the B-model are elements $B^{(j)}_k$ of the cohomology groups
$H^j(X_z,\bigwedge^j T)$. Their cubic forms are defined as 
\begin{equation} 
C(B^{(i)}_a,B^{(j)}_b,B^{(k)}_c)=\int_X \Omega(B^{(i)}_a\wedge B^{(j)}_b\wedge B^{(k)}_c) \wedge \Omega \ ,
\label{threepoint}    
\end{equation}   
and vanish unless $i+j+k=n$.  Here $\Omega$ is the unique holomorphic
$(n,0)$-form and $\Omega(B^{(i)}_a\wedge B^{(j)}_b\wedge B^{(j)}_c)$ is the
contraction of the $n$ upper indices of $B^{(i)}_a\wedge B^{(j)}_b\wedge
B^{(j)}_c\in H^n(X_z,\bigwedge^n TX_z)$ with $\Omega$, which produces an
anti-holomorphic $(0,n)$-form on $X_z$. Note that this is just the isomorphism
$H^i(X_z,\bigwedge^j TX_z)\cong H^{n-j,i}(X_z)$ obtained by contraction with the
holomorphic $(n,0)$-form $\Omega$. We denote the image of $B^{(i)}_k$ in
$H^{n-i,i}(X_z)$ by $b^{(i)}_k=\phantom{}^\Omega(B^{(i)}_k)$ and the inverse\footnote{The inversion is just the contraction $(b^{(i)}_k)^{\Omega}=\tfrac{1}{||\Omega||^2}\bar{\Omega}^{a_1\ldots a_{n-i}b_1\ldots b_i}(b^{(i)}_k)_{a_1\ldots a_{n-i}\bar{b}_1\ldots\bar{b}_i}$ such that $(\Omega)^{\Omega}=1$. Formally it is the multiplication with the inverse $\mcal{L}^{-1}$ of the K\"ahler line bundle $\mcal{L}=\langle\Omega\rangle$, see e.g.~\cite{Hori:2003ic}.} by
$B^{(i)}_k=(b^{(i)}_k)^\Omega$.  Now we can define the hermitian metric
\begin{equation} 
G(B^{(i)}_c,\bar B^{(i)}_d)=\int_X b^{(i)}_c \wedge \bar b^{(i)}_d  \ .    
\label{twopoint}
\end{equation} 
We consider only states  $B^{(i)}_a$ for which the image $b^{(i)}_a$ is in the
horizontal subspace $H^{n-j,i}_{H}(X_z)$ and assume that the $b^{(i)}_a$ form a
basis of this space. For $B^{(1)}_c\in H^1(X_z, TX_z)$ the image spans all of
$H^{n-1,1}(X_z)$ and (\ref{twopoint}) is the Weil-Petersson metric on ${\cal
M}$.

The integrals (\ref{threepoint}) and  (\ref{twopoint}) are calculable by 
period integrals of the holomorphic $n$-form $\Omega$. It is very hard to integrate the
periods directly. They encode however the variations
of Hodge structure of the family $X_z\rightarrow \cal{M}$, which are 
reflected by the Picard-Fuchs differential equations on ${\cal M}$. 
The periods are therefore determined as solutions of the latter up to linear 
combination. The precise identification as addressed in section \ref{matching}
is an important problem, as it determines the superpotential \eqref{GVW-super}
of our F-theory setup.

For a given base point $z=z_0$ in the complex structure moduli space ${\cal M}$
with fiber $X_{z_0}$, one fixes a graded topological basis
$\hat{\gamma}^{(p)}_{a}$ of the primary horizontal subspace
$H^{n}_H(X_{z_0},\mathbb{Z})$. Here $a=1,\ldots,h_H^{n-p,p}$ labels the basis
$\hat{\gamma}^{(p)}_a$ for fixed $p=0,\ldots,n$ of each graded piece
$H^{n-p,p}(X_z)$.  These forms can be chosen to satisfy~\eqref{def-eta} in
addition\footnote{The generalization from fourfolds to $n$-folds is the obvious
one.}. Later on in section \ref{matching} we will identify this grading by $p$
with the natural grading on the observables of the A-model which are given by the
vertical subspaces $H^{p,p}_V(\tilde X)$ of the mirror cohomology. 

Note that the $\hat\gamma^{(p)}_a$ basis serves as a local frame of the vector
bundle $\mathcal H^{n}_H(X)\rightarrow\mathcal{M}$ over the moduli space whose
fiber at the point $z\in \mathcal M$ is $H^n_H(X_z)$. However, the individual
$H^{n,n-p}(X_z)$ form no holomorphic vector bundles over ${\cal M}$ since
holomorphic and anti-holomorphic coordinates are mixed under a change of complex
structure $z$. Only the horizontal  parts of $F^k=\oplus_{p=0}^k H^{n-p,p}(X_z)$
form holomorphic vector bundles for which we introduce frames $\beta^{(k)}_a$ specified by the basis expansion  
\begin{equation} \label{FiltBasis}
     \beta^{(k)}_{a} = 
       \hat{\gamma}^{(k)}_{a} + 
       \sum_{p>k}\Pi^{(p,k)\ c}_{a} \hat{\gamma}_{c}^{(p)}\ .
\end{equation}
In special coordinates $t^a$ at the point of maximal unipotent monodromy this can be written as $\{\beta^{(0)}=\Omega,\beta^{(1)}_{a}=\partial_a\Omega,\ldots \}$, cf. \eqref{flatcoords}.
We note that for this basis choice we fixed the overall normalization of each
$\beta^{(k)}_a$ such that the coefficient of $\hat{\gamma}^{(k)}_a$ is unity.
This is needed to obtain the right inhomogeneous flat coordinates on $\mathcal{M}$ and
to make contact with enumerative predictions for the A-model, see section
\ref{matching}. For grade $k=0$ it corresponds to the familiar gauge choice 
of $\Omega$ in mirror applications to threefolds, see (\ref{flatcoords}).
We also introduce a basis of integral homology cycles $\gamma^{(p)}_{a}$ dual to $\hat \gamma_{a}^{(p)}$
as in \eqref{dualbasis}. Then, by construction of the basis \eqref{FiltBasis}, the period matrix $P$ defined by period integrals takes an upper triangular form in this basis, 
\begin{equation} 
 	 P=\int_{\gamma^{(p)}_{a}} \beta^{(k)}_{b}=
	\begin{cases}
         \Pi^{(p,k)\, a}_{b}\ ,& p>k\ ,\\ 
 	 \delta_{a b}\ ,& p=k\ ,\\
	0\ ,& p<k \ ,
	\end{cases}
\label{ffperiods} 
\end{equation}
where $(p,k)$ is the bi-grade of the non-trivial periods $\Pi^{(p,k)\, a}_{b}$.
Since $\hat\gamma^{(p)}_{a}$ are topological and thus are locally constant
on $\mathcal{M}$, the moduli dependence of \eqref{FiltBasis} is captured by the
moduli dependence of the period matrix $P\equiv P(z)$. For Calabi-Yau fourfolds we 
summarize the basis choices and the periods in Table \ref{basis_summary}.
 \begin{table}[t]
\begin{center}
	\begin{tabular}{|c|c|c|c|c|c|}
	\hline
Dimension & $1$ & $h^{3,1}(X)$ & $h^{2,2}(X)$ &$h^{3,1}(X)$& $1$ \\ \hline	
 Basis	of& $\hat{\gamma}^{(0)}_{a_0}$ & $\hat{\gamma}^{(1)}_a$ & $\hat{\gamma}^{(2)}_\beta$& $\hat{\gamma}^{(3)}_a$& $\hat{\gamma}^{(4)}_{a_0}$ \\
 $H^4_H(X,\mathbb{Z})$& $\beta^{(0)}_{a_0}$ &  $\beta^{(1)}_a$ &  $\beta^{(2)}_\beta$&$\beta^{(3)}_a$&$\beta^{(4)}_{a_0}$ \\ \hline
\end{tabular}
\caption{Topological $\hat{\gamma}$ and moduli-dependent $\beta$ basis of $H^4_H(X,\mathbb{Z})$}
\label{basis_summary}
\end{center}
 \end{table}
The $(n,0)$-form $\Omega$ can
always be expanded over the topological basis 
$\hat{\gamma}^{(p)}_{a}$ of $H_H^n(X,\mathbb{Z})$ as
\begin{equation}
 	\Omega=\Pi^{(p,0)\, a}_{1} \ \hat{\gamma}_{a}^{(p)} \equiv \Pi^{(p)\, a} \ \hat{\gamma}_{a}^{(p)}\, ,
\label{omegaexpansion} 
\end{equation}
where we introduced a simplified notation $\Pi^{(p)\, a}$ for the periods $\Pi^{(p,0)\, a}_1$ of the holomorphic $n$-form 
already given in \eqref{def-periods}.
For an arbitrarily normalized $n$-form $\Omega$, the periods
$(X^{0},X^{a})=(\Pi^{(0)\, 1},\Pi^{(1)\, a})$ with
$a=1,\ldots,h^{n,1}(X_{z_0})$ at the fixed reference point $z_0$ form for all
Calabi-Yau $n$-folds homogeneous projective coordinates of the complex moduli
space $\mcal{M}$. The choice of inhomogeneous coordinates set by
\begin{equation} \label{flatcoords}
	t^{a}=\frac{X^{a}}{X^0}=\frac{\int_{\gamma^{(1)}_{a}}\Omega}{\int_{\gamma^{(0)}_{1}}\Omega}\,, 
\end{equation} 
which agrees with the basis choice in \eqref{FiltBasis}, corresponds to a
gauge in the K\"ahler line bundle $\langle\Omega\rangle$. At the point of
maximal unipotent monodromy $X^0$ and $X^a$ are distinguished by their monodromy, as $X^0$ is holomorphic and single-valued and $X^a\sim \text{log}(z)$ has monodromy $X^a\mapsto X^a+1$. Below the $t^a$ in
(\ref{flatcoords}) are identified with the complexified K\"ahler parameters of the mirror $\tilde X$. 

The $t^a$ defined in \eqref{flatcoords} are flat coordinates for the Gauss-Manin 
connection $\nabla_a$, i.e.~the latter becomes  just the ordinary differential
$\frac{\partial}{\partial t_a}$. This can be seen from the gauge choice
reflecting in the basis \eqref{FiltBasis} combined with the Griffiths 
transversality constraint $\nabla_a(F^k)\subset F^{k+1}/F^{k}$ which together imply \cite{Greene:1993vm} $\nabla_a t^b=\delta_a^b$, i.e~$\nabla_a=\partial_a$.  
In these coordinates the three-point coupling becomes 
\begin{equation} 
C^{(1,k,n-k-1)}_{abc}=C((\beta^{(1)}_a)^{\Omega},(\beta^{(k)}_b)^{\Omega},(\beta^{(n-k-1)}_c)^{\Omega})=\int_X \beta_{c}^{(n-k-1)}\wedge \partial_{a} 
\beta_b^{(k)} \,  .
\label{3point} 
\end{equation} 
This triple coupling is a particularly important example of (\ref{threepoint}). Here we use 
(\ref{FiltBasis},\ref{ffperiods}) and the fact that $(\beta^{(1)}_a)^\Omega \wedge (\beta_b^{(k)})^\Omega$ 
can be replaced  by $(\nabla_a \beta_b^{(k)})^\Omega$ \cite{Greene:1993vm} under the integral (\ref{threepoint}) 
to obtain \eqref{3point}. 
Furthermore, one can show from the properties of the Frobenius algebra that all other triple 
couplings in (\ref{threepoint}) can be expressed in terms of (\ref{3point}), see 
section~\ref{frobenius}.   

The holomorphic topological two-point couplings of \eqref{def-eta} in this basis trivially read 
\begin{equation} 
\eta^{(k)}_{ab}=\int_X     \beta_b^{(n-k)} \wedge   \beta_a^{(k)}   
\label{toptwopoint} 
\end{equation} 
since only the lowest $\hat{\gamma}^{(p)}$ for $p=k$ in the upper-triangular
basis transformation \eqref{FiltBasis} contributes to the integral due to the
second property of \eqref{def-eta}.  In particular $\eta^{(k)}$ is moduli
independent. From the above it is easy to see the basis expansion at grade
$k+1$,
\begin{equation} 
\label{alphader}
\partial_a \beta_b^{(k)} =C^{(1,k,n-k-1)}_{abc}\eta^{cd}_{(n-k-1)}
\beta_d^{(k+1)}\ ,
\end{equation}     
where $\eta^{cd}_{(p)}$ is the inverse of $\eta_{cd}^{(p)}$.  

Let us finish this section with some comments on general properties of the
periods and their implications on the $\mathcal{N}=1$ effective action.  The
period integrals (\ref{ffperiods}) obey differential and algebraic relations,
which are different from the special geometry relations of Calabi-Yau threefold
periods. They have however exactly the same origin namely the Griffiths
transversality constraints $\int_X \Omega \wedge \partial_{i_1}\ldots
\partial_{i_k} \Omega=0$ for $k<n$. However since $a \wedge b=(-1)^n b \wedge a
$ for $a,b$ real $n$-forms one has additional algebraic relations from
$\int_X \Omega\wedge \Omega=0$ between the periods $\Pi^{(p)\,a}$ for $n$ even like $n=4$, which are
absent for $n$ odd, in particular the threefold case. $\mathcal{N}=2$
compactifications of type II on Calabi-Yau threefolds to four dimensions inherit
the structure induced by the above differential relations in the vector moduli
space. In fact such a structure is, up to minor generalizations, generic to
$\mathcal{N}=2$ supergravity theories and known as special K\"ahler geometry.
For $\mathcal{N}=1$ supergravity theories in 4d there is generically no special
structure beyond K\"ahler geometry.

\subsection{The Frobenius Algebras}
\label{frobenius}

As was mentioned above the B-model operators form a Frobenius algebra.  Since
also the A-model classical cohomology operators as well as its quantum
cohomology operators form such an algebra it is worthwhile to describe the
general structure before discussing the precise matching in the next section.

A Frobenius algebra has the following structures. It is a graded vector space
${\cal A}=\oplus_{i=1}^n {\cal A}^{(i)}$ with ${\cal A}^{(0)}=\mathbb{C}$ equipped with a
non-degenerate symmetric bilinear form  $\eta$ and a cubic form 
\beq
  C^{(i,j,k)}:{\cal A}^{(i)}\otimes {\cal A}^{(j)}\otimes {\cal A}^{(k)}\rightarrow \mathbb{C}\ ,
\eeq 
$i,j,k\ge 0$ and the following properties:
\begin{itemize}
\item[(F1)] Degree: $C^{(i,j,k)}=0$ unless $i+j+k=n$ 
\item[(F2)] Unit: $C^{(0,i,j)}_{1bc}=\eta^{(i)}_{bc}$
\item[(F3)] Nondegeneracy: $C^{(1,i,j)}$ is non-degenerate in the second slot 
\item[(F4)] Symmetry: $C^{(i,j,k)}_{abc}=C^{(\sigma(i,j,k))}_{\sigma(abc)} $ under any
  permutation of the indices.
\item[(F5)] Associativity:
  $C^{(i,j,n-i-j)}_{abp}\eta^{pq}_{(n-i-j)}C^{((i+j),k,(n-i-j-k))}_{qef}=C^{(i,k,n-i-k)}_{aeq}\eta^{qp}_{(n-i-k)} C^{(i+k,j,n-i-j-k)}_{pbf}$
where the sum over common indices is over a basis of the corresponding spaces.
\end{itemize}
The product 
\begin{equation} 
{\cal O}^{(i)}_a \cdot {\cal O}^{(j)}_b = C^{(i,j,i+j)}_{abq} \eta^{qp}_{(i+j)} {\cal O}^{(i+j)}_p \ 
\end{equation}     
defines now the Frobenius algebra for a basis of elements ${\cal O}_k^{(i)}$ of
${\cal A}^{(i)}$ which is easily seen using (F4) to be commutative. Note that the associativity implies that $n$-point correlators
can be factorized in various ways in the three-point functions.  Also the three-point
correlators are not independent and by associativity, non-degeneracy and symmetry
it can be shown\footnote{See \cite{Klemm:1996ts} for an explicit inductive
proof.} that all three-correlators can be expressed in terms of the
$C^{(1,r,n-r-1)}$ correlators defined in (\ref{3point})

It is easy to see that the $B^{(i)}$ operators of the B-model with 
the correlators defined by (\ref{threepoint}) or equivalently
(\ref{3point}) and (\ref{toptwopoint}) fulfill the axioms 
of a Frobenius algebra.      

Let us consider  the A-model operators. As already mentioned
such an operator corresponds to an element in the vertical subspace 
$H^{*,*}_{V}(\tilde X)$. It is generated by the K\"ahler forms
$J_i$, $i=1,\ldots, h^{1,1}(\tilde X)$  and naturally graded, 
\begin{equation}
A^{(p)}_\alpha =a_\alpha^{i_1,\ldots, i_p} J_{i_1}\wedge \ldots \wedge J_{i_p}\,\in H^{p,p}(\tilde X)\,.  
\label{amodeloperators}
\end{equation} 
For the classical A-model the 
correlation functions are simply  the intersections
\begin{equation} 
C_{abc}^{0\, (i,j,k)}=C(A^{(i)}_a A^{(j)}_b A^{(k)}_c)= \int_{\tilde X}  A^{
  (i)}_a\wedge  A^{(j)}_b \wedge A^{(k)}_c\ .
\label{classicalamodeltriplecouplings}  
\end{equation} 
They vanish unless $i+j+k=n$. The topological metric is similarly 
defined by 
\begin{equation} 
\eta^{(k)}_{ab} = \int_{\tilde X}  A^{(k)}_a\wedge  A^{(n-k)}_b \  
\label{amodeltopologicametric}  
\end{equation} 
and together with  (\ref{classicalamodeltriplecouplings}) this defines a
Frobenius algebra. Clearly the $A^{(p)}_k$ are not freely generated by the
$J_i$. The products $J_{i_1}\wedge \ldots \wedge J_{i_p}$  are set to zero if
their pairings (\ref{amodeltopologicametric}) with all other cohomology elements
vanish. This is easily calculated using toric techniques and reflects
geometrical properties of $\tilde X$ like for instance fibration structures.

The classical intersections are extended using mirror symmetry to the quantum
cohomological intersections\footnote{We denote the both the operators of the
classical algebra and the operators of quantum cohomology algebra by
$A^{(p)}_k$.}
\begin{equation} 
C(A^{(i)}_a A^{(j)}_b A^{(k)}_c)= C_{abc}^{0\, (i,j,k)}+
{\rm instanton\ corrections}\ , 
\label{quantumamodeltriplecouplings}  
\end{equation} 
where the instanton corrections are from holomorphic curves with meeting
conditions on the homology cycles dual to the $A^{(p)}_k$. They are such that
the correlator vanishes again unless $i+j+k=n$. Note that  the $C(A^{(i)}_a
A^{(j)}_b A^{(k)}_c)$ depend via the instantons on the complexified  K\"ahler
parameters of $\tilde X$, while $\eta^{(k)}_{ab} = \int_{\tilde X}
A^{(k)}_a\wedge  A^{(n-k)}_b$ is still purely topological.
There are no instanton corrections present
because in the moduli space for the two-pointed sphere not all zero modes are saturated due to
the one remaining conformal Killing field on the sphere.

\subsection{Matching of the A-model and B-model Frobenius algebra} 
\label{matching}

We now describe the matching of the B-model Frobenius algebra with the $A$ model
quantum Frobenius algebra. At the large radius point of the K\"ahler structure, the 
correlation functions of the classical $A$-model can be calculated using
toric intersection theory. We will match this information with the leading
logarithmic behaviour of the periods at the corresponding point in the 
complex structure moduli space, the point of large complex structure, 
which is characterized by its maximal logarithmic degeneration, which 
leads to a maximal unipotent monodromy.

Let us now discuss the Picard-Fuchs differential operators associated to 
the mirror Calabi-Yau $X$ at the large complex structure point. Since we 
are dealing throughout our paper with mirror pairs $(\tilde{X}_4,X_4)$ that 
are given torically, the derivation of the Picard-Fuchs operators simplifies 
significantly, since they are completely encoded by the toric data.
To the Mori cone generators $\ell^{(a)}$ on the A-model side one associates  
the canonical GKZ-system of differential operators on the B-model side
\begin{equation}
 	\mathcal{D}_{a}=\prod_{\ell^{(a)}_i>0}\left(\frac{\partial}{\partial a_i}\right)^{\ell^{(a)}_i}-\prod_{\ell^{(a)}_i<0}\left(\frac{\partial}{\partial a_i}\right)^{-\ell^{(a)}_i}\,,
\end{equation}
where the derivative is taken with respect to the coefficients $a_i$ of monomials in the constraint $f=0$ defining $X$.
By the methods described in \cite{Hosono:1993qy} we obtain linear Picard-Fuchs
operators $\mathcal{L}_a({\underline \theta},{\underline z})$, which are written in terms of the logarithmic derivatives
$\theta_a=z_a\frac{\partial}{\partial z_a}$ with respect 
to the canonical complex variables $z_a$ that vanish at the large complex structure point
defined by~\eqref{LargeComplexStr}. 
We extract the leading $\theta$ piece of the differential operators as the formal limit
${\cal L}_i^{\rm lim}({\underline \theta})=\lim_{z_i\rightarrow 0} {\cal
  L}_i({\underline \theta},{\underline z})$, $i=1,\ldots, r$ and consider the algebraic ring  
\begin{equation} 
{\cal R}=\mathbb{C}[{\underline \theta}]/({\cal J}=\{ {\cal L}_1^{\rm lim},\ldots,{\cal
  L}_r^{\rm lim}\})\ .
\label{ring}
\end{equation}  
This ring is graded by the degree $k$ in $\theta$ and we denote the ring at grade $k$
by $\mathcal{R}^{(k)}$ whose number of 
elements is given by $h^H_{n-k,k}(X)=h^V_{k,k}(\tilde X)$ for 
$k=0,\ldots,n$. We note that for $k=0,1,n-1,n$ there is no 
difference between the vertical (horizontal) homology and the full 
homology groups. Let us explain in more detail how this 
ring connects  the $A$- and the B-model structure  at large radius:

\begin{itemize} 
\item[(A)] 
The construction of the ring is up to normalization equivalent to the 
calculation of the intersection numbers of the $A$-model. In particular 
the $n$-point intersections appear as coefficients of the up to a 
normalization unique top ring element ${\cal R}^{(n)}=\sum_{i_1\leq\ldots \leq
 i_n} C^{0}_{i_1, \ldots, i_n } 
\theta_{i_1}\ldots \theta_{i_n}$ and similar the ${\cal R}^{(k)}$ 
are generated by ${\cal R}^{(k)}_\alpha=  \sum_{i_1\leq\ldots \leq
 i_k }a_{\alpha}^{i_1, \ldots,i_{k}}\theta_{i_1}\ldots \theta_{i_k}$, $\alpha=1,\ldots,h^V_{k,k}(\tilde X)$, 
where $a_\alpha^{i_1, \ldots,i_{k}} \kappa_{i_1, \ldots,i_{k},j_1,\ldots j_{n-k}}=C^{0}_{\alpha,j_1,\ldots j_{n-k}}$.    
\item[(B)] 
The ring $\mathcal{R}^{(k)}$ is in one-to-one correspondence to the solutions to the Picard-Fuchs equations 
at large radius. As discussed before the solutions are characterized by their monodromies around this point, 
i.e.~they are graded by their leading logarithmic structure. 
To a given ring element ${\cal R}^{(k) a }=\sum_{|{\underline \alpha}|=k}
\frac{1}{(2\pi i)^k} m^{a}_{\underline \alpha} \theta_1^{\alpha_1}\ldots
\theta^{\alpha_h}_h$ in ${\cal R}^{(k)}$ we associate a solution of the form 
\beq
\label{generalformofsolution} 
   \tilde{\mathcal{R}}^{(k) a}=X_0({\underline z})\Big[\mathbb{L}^{(k)\, a}  + 
                 {\cal O}(\log(z)^{|\alpha|-1})\Big]\ ,
\eeq 
with leading logarithmic piece of order $k$, 
\begin{equation} \label{LeadingLog}
\mathbb{L}^{(k)\,a }= \sum_{|{\underline \alpha}|=k} \frac{1}{(2\pi i)^k} 
\tilde m^{a}_{\underline \alpha} \log^{\alpha_1}(z_1)\ldots \log^{\alpha_h}(z_h)\,, 
\end{equation}
by assigning $\tilde m^{a}_{\underline\alpha} (\prod_i \alpha_i !) = m^{a}_{\underline \alpha}$. 
In particular we map the unique element $1$ of $\mathcal{R}^{(0)}$ to the unique holomorphic solution $X_0({\underline z})=1+{\cal O}(z)$.
The above map follows from the fact that all ${\cal L}_s^{\rm lim}$ in the ideal $\mathcal{J}$ must annihilate the leading
logarithmic terms for $\Pi^{(k) a}$ to be a solution which yields the same conditions as for $\mathcal{R}^{(k)}$ to be normal to $\mathcal{J}$ in \eqref{ring}.

\end{itemize} 
The two facts (A) and (B) imply mirror symmetry at the level of the classical 
couplings and can be proven for toric hypersurfaces by matching the 
toric intersection calculation with the toric derivation of the 
Picard-Fuchs operators as it was argued in the threefold case~\cite{Hosono:1994ax}.  
The identification   
\begin{equation}  
\theta_i \leftrightarrow J_i
\label{matchingmap}  
\end{equation} 
provides a map between ${\cal R}^{(k)}_a$ and the classical A-model operators
$A^{(k)}_a$ defined in (\ref{amodeloperators}). This provides also the matching of the A- and 
B-model Frobenius structures at the large radius limit by identifying the periods of $\Omega$ and the 
solutions of the Picard-Fuchs system in the following way. To a given element 
${\cal R}^{(k)}_a$ we can associate an A-model operator  $A^{(k)}_a$ by the 
replacement (\ref{matchingmap}) and wedging of the $J_i$. Similarly we can construct the dual B-model 
operators $\beta^{(k)}_a$ in $F^{k}$ by applying the elements of the ring $\mcal{R}^{(k)}$ as differentials in 
the $z_i$ to the holomorphic form $\Omega$.  We obtain the map
\begin{equation}  \label{Atobeta_map}
A^{(k)}_a\mapsto\beta^{(k)}_a|_{z=0}={\mcal R}^{(k)}_a \Omega|_{z=0} \ .
\end{equation}
This determines the Frobenius algebra of the B-model completely.
However, to relate the two- and three-point functions to the period integrals of the $\beta^{(k)}_a$ along the lines of section \ref{bmodel} we have to specify the topological basis $\gamma^{(k)}_a$ in terms of the A-model operator $A^{(k)}_a$ as well.
First we select a basis of solutions $\Pi^{(k)\,a}(z)$ of the Picard-Fuchs system with leading logarithm $\mathbb{L}^{(k)\,a}$ that is dual 
to the $A^{(k)}_a$ at large radius, i.e.
\beq
   {\mcal R}^{(k)}_a\,  \mathbb{L}^{(k)\, b }=\delta^{b}_a \ ,
\eeq 
in the limit $z_i\rightarrow 0$ \cite{Mayr:1996sh}. Then, the $\gamma^{(k)}_a$ are fixed
by setting the periods $\Pi^{(k,0)\ a}_1\equiv \Pi^{(k)\,a}(z)$ in the expansion \eqref{omegaexpansion} 
of $\Omega=\Pi^{(k)\,a}(z)\hat{\gamma}^{(k)}_a$ that provides a map between the $\mathbb{L}^{(k)\, a}$ and 
$\hat \gamma^{(k)}_a$. With these definitions the requirements (\ref{ffperiods}) on the upper triangular basis
$\beta^{(k)}_a$ are trivially fulfilled since $\beta^{(q)}_b={\mcal R}^{(q)}_b \Omega= {\mcal R}^{(q)}_b (\Pi^{(q)\,a}\hat{\gamma}^{(q)}_a+\ldots)\rightarrow\hat{\gamma}^{q}_b+\ldots$ 
where the dots indicate forms $\hat{\gamma}^{(k)}_a$ at grade $k>q$ with higher logarithms.

Let us exploit this matching by e.g.~considering the B-model coupling $C^{(1,1,n-2)}_{abc}$. 
We obtain in (\ref{3point}) for fourfolds that
\begin{equation}  
C^{(1,1,2)}_{ a b \gamma}=\partial_{t^a} \Pi_{b}^{(2,1) \, \delta }
\eta^{(2)}_{\delta \gamma}=\partial_{t^a} \partial_{t^b} \Pi^{(2)\, \delta}
\eta^{(2)}_{\delta \gamma}=:\partial_{t^a} \partial_{t^b} F^{g=0}(\gamma)  \ ,
\label{simple3point} 
\end{equation}   
where $a,b = 1,\ldots,h^{3,1}$ and $\gamma=1,\ldots, h^H_{2,2}$. 
Here we  used the upper triangular form (\ref{FiltBasis}) of
$\beta^{(k)}_a$ and the intersection properties (\ref{def-eta}) of the $\hat
\gamma_i$ for the first equality. Then we replaced 
$\partial_{t_a}\beta^{(0)}=\beta^{(1)}_a$ for general dimension $n$ as follows from 
(\ref{alphader}) and property (F2) in flat coordinates.
If we now let $z_i\rightarrow 0$ and use the flat coordinates (\ref{flatcoords}), 
which are given by (\ref{generalformofsolution}) as $t^i\sim
\log(z_i)+\text{hol.}\rightarrow\log(z_i)$, we see that in this limit the classical intersection $C^{0\,
  (1,1,2)}_{ a b \gamma}$ of \eqref{classicalamodeltriplecouplings} are reproduced. 
Once the matching is established in  this large radius limit we can define the full quantum cohomological
Frobenius structure by (\ref{simple3point}) in the coordinates
(\ref{flatcoords}). The latter can be viewed as the classical 
topological intersections deformed by instanton corrections.

For the case at hand the intersections are obtained from the second derivative 
of the holomorphic quantities $F^0(\gamma)$ that were introduced in \eqref{simple3point} 
for each basis element $\beta^{(2)}_\gamma$, $\gamma=1,\ldots h_H^{2,2}(X_4)$. 
These are the analogues of the holomorphic prepotential $F^0$ familiar from the threefold case
and they are obtained in the general discussion of section 
\ref{EnumGeo} from the generating functionals\footnote{We note that the terms $b^{0}_{a\gamma }$, $a^{0}_{\gamma}$ are irrelevant 
for the quantum cohomology, but important for the large radius limit of the 
superpotential (\ref{GVW-super}).} (\ref{g=0multicovering}) for $k=1$. 
The relation \eqref{simple3point} tells us that we obtain them simply from the Picard-Fuchs equation
as double-logarithmic solutions, that we will identify below.
However as  mentioned above the identification 
using the ring structure fixes the solutions of the Picard-Fuchs system so far only up to normalizations.
The normalization of the unique holomorphic solution is determined by the fact that 
the leading term in $X_0(z)=1+{\cal O}(z)$  has to be one. Also the dual period can be 
uniquely normalized by the classical $n$-point intersections.
The single logarithmic solutions are normalized to reproduce the effect of a shift 
of the NS-background field $B$ on $t^i= \int_{\cal C}(B+i \omega)$, where ${\cal C}$ 
is a generator of $H_2(\tilde X,\mathbb{Z})$ and $\omega$ is the K\"ahler form. The shift is then 
$t^i\rightarrow t^i+1$. This corresponds to the monodromy around $z_i=0$ and implies 
according to (\ref{flatcoords}) that 
$\tilde{m}_{\underline{\alpha}}^a=1$ for $|\alpha|=1$ in (\ref{generalformofsolution}) .

All further $t$-dependent quantities are restricted further by the monodromy of the period vector $\Pi$ 
of the holomorphic $(4,0)$-form, $\Pi=(\Pi^{(0)},\Pi^{(1) *},\Pi^{(2) *},\Pi^{(3) *}, \Pi^{(4)})^T$ around  
$z_i=0$.  Let $\Sigma$ be the matrix representing the intersection 
$K=\int_X\Omega\wedge \bar \Omega=\Pi\Sigma \Pi^\dagger$. Using (\ref{omegaexpansion}, \ref{def-eta}) 
it is easy to see that the  anti-diagonal of $\Sigma$ is given by the blocks 
$(1,(\eta^{(1)})^T, \eta^{(2)},\eta^{(1)},1)$. The monodromies act by $\Pi\rightarrow M_i\Pi$, 
where $M_i$ is a  $(h^H_4\times h^H_4)$ matrix.  The monodromy invariance of 
$K$ and $\hat \gamma_a^{(p)}\in H^4_H(X_4,\mathbb{Z})$ implies 
\begin{equation} \label{monodromyconstraint}
M_i^T \Sigma M_i=\Sigma
\end{equation}
with $M_i$ an integer matrix.  Using the monodromy at other points in the moduli space and analytic 
continuation, one can fix all a priori undetermined constants in the solutions to the Picard-Fuchs system. 

However, this is tedious and  useful information  about some of the irrational constants 
that appear e.g.~in the leading logarithmic solution follow from the Frobenius method~\cite{Hosono:1994ax,Libgober}.  
By this method the leading logarithmic solution can be obtained by applying the operator 
$D^{(4)}=\frac{1}{4! (2 \pi i)^4} \mathcal{K}_{i_1 i_2 i_3 i_4}\partial_{\rho_{i_1}}
\partial_{\rho_{i_2}}\partial_{\rho_{i_3}} \partial_{\rho_{i_4}}$ on the 
fundamental solution
\begin{equation} 
\omega_0({\underline{z}},{\underline{\rho}}) = \sum_{{\underline{n}} } c({\underline{n}},{\underline{\rho}}) z^{{\underline{n}}+{\underline{\rho}}}
\end{equation} 
with 
\begin{equation} 
 c({\underline{n}},{\underline{\rho}})=
\frac{\Gamma(-\sum_\alpha l_0^{(\alpha)} ( n_\alpha+\rho_\alpha)+1)}{\prod_i \Gamma(\sum_{\alpha} l_i^{(\alpha)}( n_\alpha+\rho_\alpha)+1)}\  
\end{equation}
and setting ${\underline{\rho}}=0$.    
The general leading logarithmic solution, i.e.~with all possible admixtures 
of lower logarithmic solutions, for $X^0=\omega_0({\underline{z}})|_{{\underline \rho}=0}$ reads
\begin{equation}  \label{Pi4}
\Pi^{(4)}=X^0(\tfrac{1}{4!}\mathcal{K}_{ijkl} t^{i}t^{j}t^{k}t^{l} +\tfrac{1}{3!} a_{ijk} t^{i}t^{j}t^{k} +\tfrac{1}{2!} a_{ij} t^{i}t^{j}+a_{i} t^{i}+a_0)\,,
\end{equation}
 where as in the threefold case $\mathcal{K}_{ijkl}$ is the 
classical top intersection.
It was observed in~\cite{Hosono:1994ax} for the threefold case that the Frobenius method reproduces 
some of the topological constants in (\ref{pre_largeV}). In particular 
$\int_{\tilde Y_3} c_2\wedge J_i=\frac{3}{\pi^2} \mathcal{K}_{i j k}\partial_{\rho_{j}} \partial_{\rho_{k}} c({\underline{0}},{\underline{\rho}})_ {{\underline \rho}=0}$ and 
$\int_{\tilde Y_3} c_3=\frac{1}{3!\zeta(3)} \mathcal{K}_{i j k}\partial_{\rho_{i}}\partial_{\rho_{j}} \partial_{\rho_{k}} c({\underline{0}},
{\underline{\rho}})_ {{\underline \rho}=0}$, where we write $c_i$ for the $i$-th Chern class of the Calabi-Yau manifold. If we generalize these to fourfolds, we get 
\begin{equation}
 	\int_{\tilde X_4} \frac{3}{4} c_2^2+c_4
=\frac{1}{4!\zeta(4)} \mathcal{K}_{i j k l}\partial_{\rho_{i}}\partial_{\rho_{j}} \partial_{\rho_{k}} \partial_{\rho_{l}} c({\underline{0}},
{\underline{\rho}})_ {{\underline \rho}=0}\,.
\end{equation}
These constants are expected to appear as coefficients of the lower order 
logarithmic solutions in (\ref{Pi4}). Similar as in the threefold 
case one can also use the induced K-theory charge 
formula~\cite{Minasian,Cheung} in combination with central charge formula with $A$ being the bundle on the brane wrapping $\tilde X_4$
\begin{equation} 
\vec Q\cdot \vec \Pi=-\int_{\tilde X_4} e^{-J} {\rm ch}(A)\sqrt{{\rm td}(\tilde X_4)}=Z(A)\ 
\end{equation} 
and mirror symmetry to obtain information about the subleading logarithmic terms 
in the periods.

Let us apply a more direct argument and use properties of the simplest
Calabi-Yau fourfold, the sextic in $\mathds{P}^5$. The mirror 
has the Picard-Fuchs equation (see e.g. \cite{Klemm:2007in})
\begin{equation}
\theta^5 - 6 z \prod_{k=1}^5(6 \theta+k)\ . 
\end{equation}    
We can easily construct solutions at $z=0$ using the Frobenius 
method, but let us first give a different basis of logarithmic 
solutions namely
\begin{equation}
\hat \Pi_k=\frac{1}{(2\pi i)^k}\sum_{l=0}^k\left(k\atop l\right) \text{log}(z)^l s_{k-l}(z)\ ,
\end{equation}
where
\bea
   X^0 &=& s_0=1+720z+748440 z^2+\ldots \ ,\quad
   s_1 = 6246 z+ 7199442z^2+\ldots\ , \nn \\  
   s_2 &=& 20160z+327001536 z^2+\ldots,\quad
   s_3 =-60480 z-111585600 z^2+\ldots \nn \\ 
   s_4 &=&-2734663680z^2
-57797926824960 z^3+\ldots\ .
\eea 
The point is that under the mirror 
map one obtains $\hat \Pi_k=t^k+{\cal O}(q)$, so that these solutions correspond
to the leading volume term of branes of real dimension $2k$. 
The ``conifold'' locus of the sextic is at $\Delta=1-6^6 z=0$. Near that point the 
Picard-Fuchs equation has the indicials $\left(0,1,2,3,\frac{3}{2}\right)$. It 
is easy to construct solutions and we choose a basis in which 
the solution  to indicial $k\in \mathbb{Z}$ has the next power $z^4$.
The only unique solution is the one with the branch cut 
\begin{equation}
\nu=\Delta^\frac{3}{2}+\frac{17}{18} \Delta^\frac{5}{2}
+\frac{551}{648} \Delta^\frac{7}{2}+\ldots\ .
\end{equation}

The situation at the universal conifold is crucial for mirror 
symmetry in various dimensions $n$. At this point the non-trivial 
monodromy is between a cycle of topology $\mathbb{T}^n$ that 
corresponds to the solution $X^0$, i.e. the zero-dimensional brane 
in the $A$-model, which is uniquely defined at $z=0$, and a cycle of topology 
$\mathbb{S}^n$ that corresponds to the solution $\Pi^{(n)}$, i.e.\ the top
dimensional brane in the $A$-model. The topological intersection 
between these cycles is $1$ and their classes in the homology are 
the fiber and the base of the SYZ-fibration respectively \cite{Strominger:1996it}. In odd dimensional Calabi-Yau manifolds
the conifold monodromy acts on the vector $\Pi_{\text{red}}=(\Pi^{(n)},X_0)^T$ 
as $M_{2\times 2}=\left(\begin{array}{cc} 1&0\\ 1&1 \end{array}\right)$. This corresponds to the Lefschetz 
formula  with vanishing  $\Pi^{(n)}$, i.e. the 
quantum volume of $\tilde X$ vanishes. 

In four dimensions we have a monodromy of order two and the only 
way to  write an integral idempotent monodromy compatible with the 
intersection (\ref{monodromyconstraint}) is  
$M_{2\times 2}=\left(\begin{array}{cc} 0&1\\ 1&0 \end{array}\right)$. It is 
noticeable that the zero- and  the highest dimensional brane get exchanged by the 
conifold monodromy in even dimensions.  

This implies that $X^0=\eta-c \nu$ and\footnote{The sign is chosen so that the $t^4$ 
term in $\Pi^{(4)}$ comes out with positive sign.}  $\Pi^{(4)}=\eta+c \nu$. Here $\eta$ 
is a combination of solutions at $\Delta=0$  without branch cut. We 
can determine the latter by analytic continuation of $X^0$ to the conifold. 
While the precise combination is easily obtained, the only constant that 
matters below is $c$, which turns out to be $c=\frac{1}{\sqrt{3}\pi^2}$. 
Now we can determine the combination which corresponds to the correct 
integral choice of the geometric period $\Pi^{(4)}$ as
\begin{equation}
 \Pi^{(4)}=2 c \nu +X^0\ 
\label{canonical}
\end{equation}
from the uniquely defined periods $(X^0,\nu)$ at $z=0$ and 
$\Delta=0$. The analytic continuation of $\nu$ to $z=0$  fits nicely with our 
expectation from above and fixes most of the numerical coefficients in (\ref{Pi4}) 
universally 
\beq
  a_0=\frac{\zeta(4)}{2^4 (2\pi i)^6} \int_{\tilde X_4} 5 c_2^2\ , \qquad 
\eeq
and
\beq
   a_{i} =- \frac{\zeta(3)}{(2\pi i)^3} \int_{\tilde X_4} c_3 \wedge J_i \, ,\quad a_{ij}=\frac{\zeta(2)}{2(2\pi i)^2} \int_{\tilde X_4} c_2 \wedge J_i \wedge J_j\, , 
   \quad a_{ijk} = \tilde c \int_{\tilde X_4} \imath_*(c_1(J_i)) \wedge J_j \wedge J_k\ ,
\eeq
where as before $c_i = c_i (T_{\tilde X_4})$ and $c_1(J_i)$ is the first Chern class of the 
divisor associated to $J_i$ which is mapped to a four-form via the Gysin homomorphism $\imath_*$ of the embedding map 
of this divisor into $\tilde X_4$. This is the generalization of \eqref{classic_terms_threefold} to the case of 
Calabi-Yau fourfolds. To be precise, the coefficients $a_{ijk}$ are not fixed by the sextic example, because 
it turns out to be zero in this case, and the canonical choice of $\Pi^{(4)}$ (\ref{canonical}). 
This does not mean that it is absent in general. Rather it implies that it is physically irrelevant for the sextic 
because the divisibility of the correctly normalised solution, which is cubic in the logarithms, allows an 
integral symplectic choice of the periods in which this term can be set to zero. This might not be 
in general the case and  other hypersurface in weighted projective space indicate that $\tilde c=1$.
It is similarly possible to use the orbifold monodromy to fix the exact integral choice 
of the other periods. The principal form of the terms should again follow from the 
Frobenius method.

We conclude with some remarks about the enumerative geometry of the prepotentials $F^0(\gamma)$
of \eqref{simple3point}. As in the threefold case there is an enumerative geometry or counting
interpretation of mirror symmetry in higher dimensions for the A-model
\cite{Klemm:2007in}. The results and formulas necessary for our analysis are
summarized in section \ref{EnumGeo} to which we refer at several points.  As
will be discussed there, the flux $\gamma$ is necessary to reduce to a counting
of curves with the prepotential $F(\gamma)$ as a generating function,
cf.~(\ref{g=0multicovering}).  The prepotential furthermore has a
$\Li_2$-structure and it is now possible to calculate the genus zero BPS
invariants $n^0_\beta(\gamma)$ of section \ref{EnumGeo} for a suitable basis of
$H^{2,2}_{V}(\tilde X_4)$ and $\beta$ in $H_2(\tilde X,\mathbb{Z})$.

\subsection{Application to elliptic fourfolds}

For the matching of the flux and brane superpotentials \eqref{oc_super} from the perspective of F-theory, we use the following strategy. We identify the periods of the threefold fiber $Y_3$ of $X_4$ among the fourfold periods. This implies a matching of all instanton numbers as well as the classical terms on the mirror $\tilde Y_3$. Furthermore we explicitly identify fourfold periods that reproduce the physics of branes on the local geometry of $\tilde Y_3$ discussed in section \ref{localThreefold}, namely all disk instantons calculated in \cite{Aganagic:2001nx}.
This is equivalent to the statement that we calculate the superpotential (\ref{wgromovwitten}) and the D7-brane superpotential \eqref{wgromovwitten} for a specific brane flux from the fourfold perspective of F-theory where the closed BPS states of the fourfold are encoded in $F^0(\gamma)$. We explicitly show that there is an element $\hat{\gamma}\in H^{2,2}(M)$ such that the enumerative geometry on the threefold mirror pair $(\tilde Y_3,Y_3)$ with and without Harvey-Lawson branes is reproduced.
The results presented below are of further importance for the discussion of the F-theory/heterotic string duality in section \ref{het-Fdual}, where the space $Y_3$ is promoted to the background geometry of the heterotic string.

Here we will discuss the geometry $X_4$ introduced in section \ref{fourfolds} and refer to appendix \ref{appa} for further examples.
The Calabi-Yau geometry at hand has four complex moduli. We find 
that the moduli dependence of the fourfold periods is determined by a complete set of six Picard-Fuchs operators which are linear differential operators ${\cal L}_\alpha$, $\alpha=1,\ldots, 6$ of order $(3,2,2,2,3,2)$, 
that can be obtained from the $\mathbb{C}^*$ symmetries of period integrals 
associated to the charge vectors
$\ell_I^{(1)},\ell_I^{(2)},\ell_I^{(3)},\ell_I^{(4)},\ell_I^{(1)}+\ell_I^{(3)},\ell_I^{(3)}+\ell_I^{(4)}$, 
by the methods described in \cite{Hosono:1993qy}. 
We use logarithmic derivatives $\theta_a=z_a \frac{d}{d z_a}$
in the canonical complex variables~\cite{Hosono:1993qy} and 
write down only the leading piece of the differential equations
${\cal L}_\alpha^{\rm lim}=\lim_{z_a\rightarrow 0} {\cal
  L}_\alpha(\theta_a,z_a)$, $a=1,\ldots, 4$. E.g.~for the case at hand we have
\begin{equation} 
\begin{array}{rlrlrl}  
{\cal L}_1^{\rm lim}&=\theta_1^2(\theta_3 -\theta_1 - \theta_4), \ \  &
{\cal L}_2^{\rm lim}&=  \theta_2(\theta_2- 2 \theta_1- \theta_3 - \theta_4),&
{\cal L}_3^{\rm lim}&= (\theta_1 -\theta_3)(\theta_3-\theta_4), \\
{\cal L}_4^{\rm lim}&=\theta_4 (\theta_1 - \theta_3 + \theta_4 ),\ \ & 
{\cal L}_5^{\rm lim}&=   \theta_1^2 (\theta_4 - \theta_3),&
{\cal L}_6^{\rm lim}&=  \theta_4 (\theta_1 - \theta_3)\ .
\end{array}
\end{equation} 
For the complete Picard-Fuchs equations as well as the cohomology basis we extract from them we refer to appendix \ref{FurtherP2}.  

Applying \eqref{ring} it is easy to see that the there are $(1,4,6,4,1)$ 
generators of the ring ${\cal R}$ of degree $0,\ldots,4$, which are  
\begin{equation} 
\begin{array}{l|c}  
\cR^{(0)} & 1 \\[2mm]
\cline{1-1}
\rule[.5cm]{0cm}{.2cm}\cR^{(1)}_a & \theta_1, \ \  \theta_2, \ \ \theta_3, \ \ \theta_4, \\[2 mm] 
\cline{1-1}
\rule[.5cm]{0cm}{.2cm} \cR^{(2)}_\alpha & \theta_1^2, \ \   (\theta_1+\theta_3) \theta_4, \ \ (\theta_1+\theta_3)\theta_3,\ \ (\theta_1+2 \theta_2) \theta_2, \ \
(\theta_2+\theta_4 ) \theta_2,\ \ (\theta_2+\theta_3) \theta_2\\[2 mm]
\cline{1-1}
\rule[.5cm]{0cm}{.2cm} \cR^{(3)}_a &  \left(\theta_3+\theta_4\right) \left(\theta_1^2+\theta_1 \theta_3+\theta_3^2\right),\ \ \theta_2 \left(\theta_3^2+3 \theta_2 \theta_3+5 \theta_2^2+\theta_1 \left(\theta_2+\theta_3\right)\right),\\ 
& \theta_2 \left(\theta_1 \left(\theta_2+\theta_4\right)+\theta_4 \left(\theta_3+3 \theta_2\right)+\theta_2 \left(\theta_3+6 \theta_2\right)\right),\ \ 
\theta_2 \left(\theta_1^2+2 \theta_1 \theta_2+4 \theta_2^2\right)\\[2 mm]
\cline{1-1}
\rule[.5cm]{0cm}{.2cm} \cR^{(4)} & \theta_4(\theta _1^2 \theta _2+3 \theta _1 \theta _2^2+9 \theta _2^3+\theta _1 \theta _2 \theta _3+3 \theta _2^2 \theta _3+\theta _2 \theta _3^2)\\
&+ \theta _2 \left(46 \theta _2^3+15 \theta _2^2 \theta _3+4 \theta _2 \theta _3^2+\theta _3^3+\theta _1^2 \left(2 \theta _2+\theta _3\right)+\theta _1 \left(11 \theta _2^2+4 \theta _2 \theta _3+\theta _3^2\right)\right)\ .
\end{array}
\label{ringffp2}
\end{equation} 
These can be associated to solutions of the Picard-Fuchs equations and to a
choice of basis elements of the Chow ring as explained in section \ref{matching}. 
At grade $k=2$ the leading solutions $\bbL^{(k)\, \alpha}$ of the Picard-Fuchs system \eqref{PFOFFP2T1} 
which are normalized to obey $\mathcal{R}^{(k)}_\alpha \mathbb{L}^{(k)\, \beta}=\delta^{\beta}_\alpha$ 
are then given by
\begin{eqnarray}
 &\bbL^{(2)\, 1}= l_1^2\,,\ \ \bbL^{(2)\, 2}=\frac{1}{2} l_4 \left(l_1+l_3\right)\,,\ \ \bbL^{(2)\, 3}=\frac{1}{2} l_3 \left(l_1+l_3\right)\,,&\nn\\
&\bbL^{(2)\, 4}=\frac{1}{7}  l_2\left(3 l_1-2 \left(l_3+l_4-l_2\right)\right) \,,\ \ \bbL^{(2)\, 5}=\frac{1}{7} l_2 \left(-2 l_1+l_2+6 l_4-l_3\right)\,,&\nn\\
&\bbL^{(2)\, 6}=\frac{1}{7}  l_2 \left(-2 l_1+l_2+6 l_3-l_4\right)\,,& \label{Sbasis}
\end{eqnarray}
where we used the abbreviation $\text{log}(z_k)\equiv l_k$ and omitted the prefactor $X_0$. In comparison to the 
complete solutions $\Pi^{(2)\, \alpha}$ of the Picard-Fuchs equations we omitted terms of order $\mathcal{O}(l)$
as in \eqref{generalformofsolution} and \eqref{LeadingLog}. 
Since we are calculating the holomorphic potentials $F(\gamma)$ of \eqref{simple3point} and the 
corresponding BPS-invariants we have to change the basis of solutions such that to any 
operator $\mathcal{R}^{(2)}_\alpha$ in \eqref{ringffp2} we associate a solution with leading logarithm 
determined by the classical intersection $C^{0\ (1,1,2)}_{ab\alpha}$ of \eqref{classicalamodeltriplecouplings},
\begin{equation} \label{intersecsLL}
 	\mathbb{L}^{(2)}_\alpha=\tfrac12X_0 C^0_{\alpha ab}l_al_b\,.
\end{equation}
From the above classical intersection data in $\mathcal{R}^{(4)}$ we obtain the leading terms $\bbL^{(2)}_\alpha$
which are related to the leading periods $\bbL^{(2)\ \alpha}$ of the four-form $\Omega$ in \eqref{Sbasis} 
by $\bbL^{(2)}_\alpha=\bbL^{(2)\, \beta}\eta^{(2)}_{\alpha\beta}$.

As in the discussion after eq.~\eqref{Atobeta_map} the choice of periods $\Pi^{(2)\, \alpha}$ with leading terms $\bbL^{(2)\, \alpha}$ corresponds 
to a particular choice of a basis $\hat \gamma^{(2)}_\alpha $ of $H^{2,2}_V(\tilde X)$. In fact, by construction 
one finds 
\beq
  \hat \gamma^{(2)}_\alpha =  \mathcal{R}^{(k)}_\alpha \Omega|_{z=0}\ .
\eeq
However, this choice of basis for $H^{2,2}_V(\tilde X)$ is not necessarily a basis of integral cohomology. 
An integral basis can, however, be determined by an appropriate basis change. 
We first note that the K\"ahler generator $J_4$ can be identified as the class of the 
Calabi-Yau threefold fiber $\tilde Y_3$ (see appendix \ref{FurtherP2} for more details on this identification).
Moreover, one finds the identification of the fourfold K\"ahler generators $J_i$ with the threefold generators 
$J_k(\tilde Y_3)$ as
\beq  \label{exchangeJs}
   J_1+J_3\ \leftrightarrow\ J_1(\tilde Y_3)\ , \qquad J_2\ \leftrightarrow\ J_2 (\tilde Y_3)\ ,
\eeq 
by comparing the coefficient of $J_4$ in the intersection form $\cC_0(\tilde X_4)$, given in \eqref{IntX_TextEx},
with the threefold intersections $\mathcal{C}_{0}(\tilde Y_3)$ in \eqref{intersectionsY}.
A subset of the basis elements of the fourfold integral basis are now naturally induced from the threefold 
integral basis. In order to do this one identifies the threefold periods $\partial_i  F_{\tilde Y_3}$, 
with derivatives in the directions $J_1(\tilde Y_3)$ and $J_2(\tilde Y_3)$, with an appropriate linear combination 
of the fourfold periods $\Pi^{(2)\, \alpha}$ \cite{Mayr:1996sh}. In other words one determines a new basis $\hat \gamma_i^{(2)}$ such 
that
\begin{equation} \label{matchPrepot}
 	\partial_i F_{\tilde Y_3} = F^0(\hat{\gamma}_i^{(2)})|_{z_4=0} \equiv \Pi^{(2)}_i|_{z_4=0}\ . 
\end{equation}
In this matching both the classical part of the periods as well as the threefold 
BPS invariants $n_{d_1,d_2}$ and fourfold BPS invariants $n_{d_1,d_2,d_1,0}(\gamma)$ have to 
match in the large $\bbP^1$-base limit. 

The match \eqref{matchPrepot} is most easily performed by first comparing the classical parts 
of the periods. In fact, using the classical intersections of $\tilde Y_3$ in \eqref{intersectionsY}
one deduces that the leading parts of the threefold periods are
\beq \label{threefoldperiods}
   \bbL_1(Y_3) =\tfrac{1}{2} X_0\tilde l_2 \big(2 \tilde l_1+3 \tilde l_2\big)\,\quad 
   \bbL_2(Y_3) =\tfrac{1}{2} X_0\big(\tilde l_1+3 \tilde l_2\big)^2\, ,
\eeq
where $\tilde l_i = \log \tilde z_i$ correspond to the two threefold directions $J_k(\tilde Y_3)$ in \eqref{exchangeJs}.
Using \eqref{exchangeJs} and \eqref{matchPrepot} one then finds the appropriately normalized leading fourfold periods 
\beq \label{LeadPer}
   \bbL^{(2)}_2=\tfrac{1}{2} X_0l_2 \left(2 l_1+3 l_2+2 l_3\right)\,\quad \bbL^{(2)}_5=\tfrac{1}{2} X_0\left(l_1+3 l_2+l_3\right)^2\,.
\eeq
A direct computation also shows that the threefold BPS invariants $d_i n_{d_1,d_2}$ and fourfold BPS invariants $n_{d_1,d_2,d_1,0}(\hat{\gamma}_i)$ match in the large $\bbP^1$-base limit, such that \eqref{matchPrepot} is established on the classical as well as quantum level.
This match fixes corresponding integral basis elements of $H^{2,2}_V(\tilde{X}_4)$ as follows. First we determine those two ring elements $\tilde{\cR}^{(2)}_\alpha$, $\alpha=2,5$, such that we obtain \eqref{LeadPer} from them using \eqref{intersecsLL} . 
We complete them to a new basis of ring elements $\tilde{\cR}^{(2)}_\alpha$ by choosing
\begin{eqnarray} \label{newring}
 &\tilde{\cR}^{(2)}_1= \theta_1^2\,,\ \ \tilde{\cR}^{(2)}_2=\frac{1}{2}  \theta_4 \left( \theta_1+ \theta_3\right)\,,\ \ \tilde{\cR}^{(2)}_3=\frac{1}{2}  \theta_3 \left( \theta_1+ \theta_3\right)\,,&\nn\\
&\tilde{\cR}^{(2)}_4=\frac{1}{7}   \theta_2\left(3  \theta_1-2 \left( \theta_3+ \theta_4- \theta_2\right)\right) \,,\ \ \tilde{\cR}^{(2)}_5=\frac{1}{7}  \theta_2 \left(-2  \theta_1+ \theta_2+6  \theta_4- \theta_3\right)\,,&\nn\\
&\tilde{\cR}^{(2)}_6=\frac{1}{7}   \theta_2 \left(-2  \theta_1+ \theta_2+6  \theta_3- \theta_4\right)\, .&
\end{eqnarray}

These operators fix the two integral basis elements 
\beq
  \hat \gamma^{(2)}_2 = \tilde{\cR}^{(2)}_2 \Omega |_{z=0}\ ,\qquad  \hat \gamma^{(2)}_5 = \tilde{\cR}^{(2)}_5 \Omega |_{z=0}\ .
\eeq
which reproduce the corresponding part of the flux superpotential \eqref{flux_ori} on $Y_3$ for $\hat{N}_i=0$ when turning on four-form flux on $X_4$ in these directions,
\begin{eqnarray} \label{FP2flux}
 	W_{\rm flux}\equiv M^1F^{0}(\hat{\gamma}_2^{(2)})+M^2F^{0}(\hat{\gamma}_5^{(2)})=\int_{X_4}\Omega\wedge G_4
	=M^1\Pi^{(2)}_2+ M^2\Pi^{(2)}_5\,
\end{eqnarray}
for the $G_4$-flux choice
\begin{equation}
 	G_4=M^1\hat{\gamma}^{(2)}_2+M^2\hat{\gamma}^{(2)}_5\,.
\end{equation}
For the choices $M^{i}=1$ we extract the invariants $d_i n_{d_1,d_2}$ from this superpotential, i.e.~from the prepotentials $F^{0}(\hat{\gamma}_2^{(2)})$ and $F^{0}(\hat{\gamma}_5^{(2)})$. 
We note that the above grade $k=2$ basis elements \eqref{newring} become under $\theta_i \leftrightarrow l_i$ exactly the 
leading solutions of the Picard-Fuchs system \eqref{Sbasis}. 
Using the same identification we find $\bbL^{(2)\, 2}=X_0(l_1+l_3 ) l_4$ and $\bbL^{(2)\, 5}=X_0(l_2+l_4 ) l_2$ as the leading behaviour of corresponding  
periods $\int_{\gamma^{2}_\alpha}\Omega=\Pi^{(2)\,\alpha}$ of the holomorphic four-form. 
This agrees with the naive expectation from the large base limit that a partial factorization of the periods occurs as 
$t_4\cdot t^{\tilde Y_3}_i$ for $t^{\tilde Y_3}_i$, $i=1,2$, the two classes in $\tilde Y_3$ \cite{Mayr:1996sh}.

It is one crucial point of our whole analysis that we can extend this matching of threefold invariants even for disk invariants 
counting curves with boundaries on Lagrangian cycles $L$ in $\tilde Y_3$. Having explained the F-theory origin of this fact before we will here explicitly find the flux choice in $H^{(2,2)}_H(X_4)$ for which the flux superpotential \eqref{GVW-super} on the fourfold reproduces the brane superpotential \eqref{wgromovwitten}. 
By construction our fourfold $\tilde X_4$ inherits information of the fiber $\tilde Y_3$ and in particular the local limit geometry $\mathcal{O}(-3)\rightarrow \mathds{P}_2$
for which the disk invariants have been computed in \cite{Aganagic:2001nx}. As noted earlier the brane data is translated to the F-theory picture of the fourfold $\tilde X_4$ by 
the Mori cone generator $\ell^{(3)}$ and its dual divisor $J_3$. Therefore, we expect to reproduce all classical terms as well as extract the disk instantons of \cite{Aganagic:2001nx} from the Gromov-Witten 
invariants $n_{d,0,d+k,0}(\hat{\gamma}_3)$ of a period that we construct via \eqref{intersecsLL} from operators of \eqref{Sbasis} of the form $\cR^{(2)}_{\gamma}=\theta_3(\theta_1+\theta_3)+\ldots$. 
However, the geometry is more complicated and the ring element $\cR^{(2)}_{\gamma}$ with this property is not unique. 
It takes the form
\begin{equation} 
\cR^{(2)}_{\gamma}=-\cR^{(2)}_1+\tfrac{1}{3}\cR^{(2)}_2+ \cR^{(2)}_3=-\theta_1^2+\tfrac{1}{2}\theta_3(\theta_1+\theta_3)+\tfrac{1}{6}\theta_4(\theta_1 +\theta_3)
\end{equation} 
that is the most convenient choice by setting the arbitrary coefficients of $\cR^{(2)}_\alpha$, $\alpha=4,5,6$, to zero. 
We note that only the coefficient in front of $\cR^{(2)}_3$ was fixed to unity by the requirement of reproducing the disk instanton invariants. The 
two further coefficients $(-1,\tfrac13)$ were fixed by the requirement of reproducing the Gromov-Witten invariants $n_{d}$ of local $\mathds{P}^2$ \cite{Haghighat:2008gw} 
by the fourfold invariants $n_{d,0,d,0}\equiv n_d$, i.e.~for $m=0$, as explained below. 
The relation between $\cR^{(2)}_{\gamma} $ and the corresponding solution is via $\hat \gamma={\cal R}_{\gamma} \Omega|_{z=0}$ and 
$\Pi^{(2)\gamma}=\int_{\gamma} \Omega$, i.e.~$\cR^{(2)}_\gamma\Pi^{(2)\gamma}=1$, so that    
\begin{equation}
\mathbb{L}^{(2)\, \gamma}=- X_0 l_1^2\ ,\quad
\mathbb{L}^{(2)}_{\gamma}=\tfrac{1}{6} X_0l_2 \left(8 l_1+9 l_2+2 l_3\right)\,.
\end{equation}
This implies that we have explicitly calculated the D7-brane superpotential \eqref{wgromovwitten} from the fourfold superpotential \eqref{GVW-super} by turning on the flux $G_4=\hat{\gamma}$,
\begin{equation} \label{FP2branepot}
 	W_{\rm D7}\equiv F^{0}(\hat{\gamma})=\int\Omega\wedge\hat{\gamma}=\Pi^{(2)}_\gamma\,.
\end{equation}

If we list the numbers $n_{d_1, 0,d_3, 0}(\hat{\gamma})$ extracted from $F^{0}(\hat{\gamma})$ we get the following table. 
\begin{table}[htbp]
\centering
$
 \begin{array}{|c|rrrrrrr|}
\hline
\rule[-0.2cm]{0cm}{0.6cm}  \text{d}_1&d_3=0&d_3=1&d_3=2&d_3=3&d_3=4&d_3=5&d_3=6\\
\hline
0&  0& 1& 0& 0& 0& 0& 0\\ 
1&  1& n_1& -1& -1& -1& -1& -1\\ 
2&  -1& -2& 2 n_2& 5& 7& 9& 12\\ 
3&   1& 4& 12& 3 n_3& -40& -61& -93\\ 
4&   -2& -10& -32& -104& 4 n_4& 399& 648\\ 
5&   5& 28& 102& 326& 1085& 5 n_5& -4524\\ 
6& -13& -84& -344& -1160& -3708& -12660& 6 n_6\\
\hline
\end{array}
$
\caption{BPS invariants $n_{d_1, 0, d_3, 0}(\hat{\gamma})$ for the disks. With the 
identification $d_3-d_1=m$ (winding) and $d_1=d$ ($\mathds{P}^2$ degree) this agrees with Tab. 5 
in~\cite{Aganagic:2001nx}.}
\end{table}
The BPS invariants of the holomorphic disks depend only  on the relative
homology class of the latter. In the table $d_3-d_1=m$ labels the winding number 
of the disks and $d_1=d$ the degree with respect to canonical class 
of $\mathds{P}^2$, i.e. if the open string disk superpotential is 
in terms of the closed string parameter $q=e^{2\pi it}$ and the open string
string parameter $Q=e^{2\pi i\hat t}$ for the outer brane  defined as 
\begin{equation} 
W=a_{tt} t^2 + a_{t \hat t} t \hat t  + a_{\hat t \hat t}\hat t^2 + a_t t + a_{\hat t} \hat t+ a_0  + \sum_{d=1}^\infty \sum_{m=-d}^\infty n_{d,m} {\rm Li}_2(q^d Q^m),
\end{equation} 
then $n_{d_1,0,d_3,0}=n_{d_1,d_3-d_1}$. Note that the numbers $n_{d_1,0}$ are not
calculated in the framework of \cite{Aganagic:2001nx}. However it is natural and calculable  
in the topological vertex formalism that they should be identified 
with  $d n_d$, where $n_d$ is the closed string  genus zero BPS invariant, 
defined via the prepotential as $F=\sum_{d=1}^\infty n_d {\rm  Li}_3(q^d)$. 
The factor of $d$ comes from the fact that we identify $W= \frac{d}{dt} F$.
This interpretation as $n_{d,0,d,0}=d n_d$ could be consistently imposed and yields  
two further conditions as mentioned above.

To obtain the open BPS invariants of phase III of \cite{Aganagic:2001nx}, we use the phase II of \eqref{eqn:vl-for-p2}.
In this phase the fiber class is not realized as a generator of the K\"ahler cone. However we readily recover the classes of $\tilde Y_3$ as
\begin{equation} \label{3foldMatchT2}
 	J_1\ \leftrightarrow\ J_1(\tilde Y_3)\,\qquad J_2+J_3\ \leftrightarrow\ J_2(\tilde Y_3)\,
\end{equation}
by comparison of the Mori cone \eqref{eqn:vl-for-p2} with the Mori cone \eqref{3foldellp2} of $\tilde Y_3$.
Then we fix a basis $\mathcal{R}^{(2)}_\alpha$ of the ring at grade two as
\begin{eqnarray}
	\theta _1^2,\,\,\,\,2 \theta _2 \left(\theta _1+3 \theta _3\right),\,\,\,\,\theta _3 \left(\theta _1+3 \theta _3\right),\,\,\,\,\theta _1 \theta _4,\,\,\,\,\theta _2^2,\,\,\,\,\left(\theta _2+\theta _3\right) \left(2 \theta _3+\theta _4\right)\,,
\end{eqnarray}
from which we obtain a basis of dual solutions $\mathbb L^{(k)\,\alpha}$ to the Picard-Fuchs system \eqref{PFOFFP2T2}
\begin{eqnarray} \label{LeadPerFFP2T2}
\nn	&\mathbb L^{(2)\, 1}=l_1^2,\,\,\,\,\mathbb L^{(2)\, 2}=\frac{1}{140} \left(l_1 \left(16 l_2+9 l_3\right)+3 \left(l_2 \left(6 l_3-5 l_4\right)-l_3 \left(l_3+5 l_4\right)\right)\right),&\\
	&\mathbb L^{(2)\, 3}=\frac{1}{70} \left(l_1 \left(9 l_2+16 l_3\right)-3 \left(l_3 \left(-6 l_3+5 l_4\right)+l_2 \left(l_3+5 l_4\right)\right)\right),\,\,\,\,\mathbb L^{(2)\, 4}=l_1 l_4,&\\
\nn	&\mathbb L^{(2)\, 5}=l_2^2,\,\,\,\,\mathbb L^{(2)\, 6}=\frac{1}{14} \left(l_2+l_3\right) \left(-3 l_1+l_3+5 l_4\right).&
\end{eqnarray}
Next we construct two solutions with leading logarithms matching the two threefold periods of \eqref{threefoldperiods} for which we are able to match the threefold invariants $d_in_{d_1,d_2}$ in the large base limit as well. The leading logarithms of these fourfold periods read
\begin{eqnarray}
\nn	\mathbb L_4^{(2)}&=&\tfrac{1}{2} X_0\left(l_1+3 \left(l_2+l_3\right)\right){}^2,\\
	\mathbb L_6^{(2)}&=& \tfrac{1}{2} X_0\left(l_2+l_3\right) \left(2 l_1+3 \left(l_2+l_3\right)\right).
\end{eqnarray}
which is in perfect agreement with \eqref{threefoldperiods} under the identification \eqref{3foldMatchT2}.
We fix the corresponding operators $\tilde{\cR}^{(2)}_4$, $\tilde{\cR}^{(2)}_6$ by matching the above two leading logarithms by the classical intersections $C^0_{\alpha ab}$ via \eqref{intersecsLL}. We complete them to a basis of $\tilde{\cR}^{(2)}$ as follows
\begin{eqnarray}
\nn&\tilde{\cR}_1^{(2)}=\theta _1^2,\,\,\,\,\tilde{\cR}_2^{(2)}=\frac{1}{140} \left(\theta _1 \left(16 \theta _2+9 \theta _3\right)+3 \left(\theta _2 \left(6 \theta _3-5 \theta _4\right)-\theta _3 \left(\theta _3+5 \theta _4\right)\right)\right),&\\
\nn	&\tilde{\cR}_3^{(2)}=\frac{1}{70} \left(\theta _1 \left(9 \theta _2+16 \theta _3\right)-3 \left(\theta _3 \left(-6 \theta _3+5 \theta _4\right)+\theta _2 \left(\theta _3+5 \theta _4\right)\right)\right),\,\,\,\,\tilde{\cR}_4^{(2)}=\theta _1 \theta _4,&\\
\nn	&\tilde{\cR}_5^{(2)}=\theta _2^2,\,\,\,\,\tilde{\cR}_6^{(2)}=\frac{1}{14} \left(\theta _2+\theta _3\right) \left(-3 \theta _1+\theta _3+5 \theta _4\right),&
\end{eqnarray}
where again this basis relates to the leading periods \eqref{LeadPerFFP2T2} by $\theta_i\leftrightarrow l_i$.
The corresponding integral basis elements of $H^{2,2}_H(X_4)$ read 
\begin{equation}
 	\hat{\gamma}^{(2)}_4=\tilde{\cR}^{(2)}_4\Omega|_{z=0}\ ,\qquad  \hat{\gamma}^{(2)}_6=\tilde{\cR}^{(2)}_6\Omega|_{z=0}\,.
\end{equation}

Furthermore, we determine the ring element $\mathcal{R}^{(2)}_\gamma$ that matches the open superpotential by turning on four-form flux in the direction $\hat{\gamma}=\mathcal{R}^{(2)}_\gamma\Omega|_{z=0}$. Again we fix
\begin{equation}
	\mathcal R^{(2)}_\gamma=a_1\mathcal R^{(2)}_2-\tfrac{1}{10}(1+6a_2)\mathcal R^{(2)}_3+\mathcal R^{(2)}_{4}+a_3\mathcal R^{(2)}_5+a_2\mathcal R^{(2)}_6
\end{equation}
by extracting the disk invariants from the solution associated to it via \eqref{intersecsLL} which reads
\begin{eqnarray}
 	\nn\mathbb{L}^{(2)\,\gamma}&=&c(a_1) X_0l_2 \left(l_1+3 l_3\right)\ ,\\ \nn\mathbb{L}^{(2)}_\gamma&=&\tfrac{1}{6} \left(l_2+l_3\right) \left(2 l_1+3 \left(l_2+l_3\right)\right)-\tfrac{1}{10} \left(l_1+3 \left(l_2+l_3\right)\right) \left(3 l_1+29 l_2+29 l_3+10 l_4\right)\,.
\end{eqnarray}
Here we explicitly displayed the dependence on the three free parameters $a_i$ for $\mathbb{L}^{(2)\,\gamma}$ by $c(a_1)=\tfrac{7}{9+140 a_1}$ and evaluated $\mathbb{L}^{(2)}_\gamma$ for the convenient choice $a_i=0$. 

\begin{table}[htbp]
	\centering
	$
\begin{array}{|c|rrrrrrr|}
	\hline
	\rule[-0.2cm]{0cm}{0.6cm} \text{d}&k=0&k=1&k=2&k=3&k=4&k=5&k=6\\
	\hline
 0&0 & n_1 & 2n_2 & 3n_3 & 4n_4 & 5n_5 & 6n_6\\
 1&-1 & 2 & -5 & 32 & -286 & 3038 & -35870 \\
 2&0 & 1 & -4 & 21 & -180 & 1885 & -21952 \\
 3&0 & 1 & -3 & 18 & -153 & 1560 & -17910 \\
 4&0 & 1 & -4 & 20 & -160 & 1595 & -17976 \\
 5&0 & 1 & -5 & 26 & -196 & 1875 & -20644 \\
 6&0 & 1 & -7 & 36 & -260 & 2403 & -25812\\
 \hline
\end{array}
$
\caption{BPS invariants $n_{k, 0, i, 0}(\gamma)$ for the disks of the second triangulation. }
	\label{tab:phase3table}
\end{table}

\section{Basics of enumerative geometry} \label{EnumGeo}

In this section we want to describe from the A-model perspective the relevant
enumerative quantities, which are calculated in this paper in the B-model using
mirror symmetry. Important circumstantial evidence for the open-string/fourfold
duality approach~\cite{Lerche:2001cw, Alim:2009rf} advocated in this paper is
the identical integral structure of the generating functions.

However as we will review in the following, the absence of higher genus
invariants on smooth fourfolds as opposed to the open string setting might be a
hint that this duality discussed relies merely on an embedding of the
open/closed moduli space into the closed moduli space as discussed, e.g.\ in
\cite{Grimm:2008dq}, rather than on a full duality of physical theories. A possibility to
avoid this conclusion would be that one has in general to consider singular
fourfolds, typical for an F-theory compactification with degenerate elliptic
fiber along the zero-locus of the discriminant.

\subsection{Closed GW invariants} 

First we review the theory of closed Gromov-Witten invariants, i.e.\ the theory
of holomorphic maps
\begin{equation}
	\phi:\Sigma_g \rightarrow \tilde X
\end{equation}
from an oriented closed curve $\Sigma_g$ into a  Calabi-Yau manifold $\tilde X$.
We do not consider marked points.
It can be defined mathematically rigorously in general and explicitly calculated
using localization techniques if $\tilde X$ is represented e.g.\ by a
hypersurface in a toric variety. Here $g$ is the genus of the domain curve and
we denote by $\beta\in H_2(\tilde X,\mathbb{Z})$ the homology class of the image
curve. One measures the multi-degrees of the latter $\beta=\sum_{i=1}^{h^{1,1}}
d_i \beta_i$ w.r.t.\ to an ample polarization $L$ of $\tilde X$, i.e.\ ${\rm
deg} (\beta)=\int_{\beta} c_1(L)=\sum_{i=1}^{h^{1,1}} d_i t_i$ with
$d_i\in\mathbb{N}_+$.  In string theory and in the context of the mirror
symmetry the volume of the curve $\beta_i$ is complexified by an integral over
the antisymmetric two-form field $B$.  Thus, one defines the complexified closed
K\"ahler moduli $t_i=\int_{\beta_i} (B+ic_1(L))$.   

For smooth $\tilde X$ the virtual (complex) dimension of the moduli space of 
these maps are computed by an index theorem and reads  
\begin{equation}
{\rm vir} \ {\rm dim} \ {\cal M}_g(\tilde X,\beta)= \int_\beta c_1(\tilde X)+ ({\rm dim}
\tilde X-3)(1-g)\ .
\label{virtdim} 
\end{equation}            
In particular for Calabi-Yau fourfolds one obtains ${\rm vir} \ {\rm dim} \
{\cal M}_g(\tilde X_4,\beta)=1-g$. Thus in order to define genus zero
Gromov-Witten invariants one requires an incidence relation of the curve with
$k=({\rm dim}(\tilde X)-3)$ surfaces to reduce the dimension to zero
in order to arrive at a well-defined counting problem. For fourfolds one thus
needs one incidence surface and we denote the dual cycle of the surface by
$\gamma\in H^{2,2}(\tilde X_4)$.
Note that for $\dim \tilde X\ge 4$ and $g\ge 2$ the
dimension of the moduli space is negative and
no holomorphic maps exist. The Calabi-Yau threefolds are critical in the sense
that the dimension of the moduli space for all genera is zero. Thus, in general
invariants associated to the maps are non-zero for all values of $g$.

We define a generating function for each genus $g$ Gromov-Witten invariant as follows:
\begin{equation}
F^g(\gamma_1)=\sum_{\beta\in H_2(\tilde X,\mathbb{Z})}
r^g_\beta(\gamma_1,\dots,\gamma_k)  q^\beta \ .
\end{equation}
They
are labelled by $g$, $\beta$ and for $\dim \tilde X\ge 4$ also by cycles
$\gamma_i$ dual to the incidence surfaces. 
Here $q^\beta$ is a shorthand notation for $q^\beta=\prod_{i=1}^{h^{1,1}} e^{2\pi it_i
d_i}$. We note  that this is not just a formal power series\footnote{This is 
important for the interpretation of such terms in the effective action. In fact,
analyticity allows to define such terms beyond the large radius limit in terms
of period integrals on the mirror geometry.}, but rather has  finite region of
convergence for large volumes of the curves $\beta_i$, i.e.\ for ${\rm
Im}(t_i)\gg 0$. This puts a bound on the growth of the Gromov-Witten invariants
$r^g_\beta(\gamma_1)$. 
The contributions of the
maps is divided by their automorphism groups and the associated Gromov-Witten
invariants $r^g_\beta(\gamma_1,\ldots, \gamma_k)$ are in general rational.

Although the discussion of \eqref{virtdim} indicates that the Gromov-Witten
theory on higher dimensional Calabi-Yau manifolds  
is less rich than in the threefold case, one has a remarkable integrality
structure associated to the 
invariants. In particular at genus zero one can define integer  
invariants $n^g_\beta(\gamma_1,\ldots \gamma_k)\in \mathbb{Z}$ 
for arbitrary ${\rm  dim}(\tilde X)=k+3$  dimensional manifolds as
\begin{equation} 
F^0(\gamma_1,\ldots, \gamma_k)=\tfrac{1}{2}C^{0\, (1,1,n-2)}_{ab\gamma_1\cdots\gamma_k} 
t^a t^b + b^{0}_{a \gamma_1\cdots\gamma_k }t^a + a^{0}_{\gamma_1\cdots\gamma_k} +  \sum_{\beta>0} n^g_\beta(\gamma_1,\ldots,\gamma_k) {\rm Li}_{3-k} (q^\beta)\ ,
\label{g=0multicovering}
\end{equation}  
where ${\rm Li}_{p}(q)=\sum_{d=1}^\infty \tfrac{q^d}{d^{p}}$ and $C^{0\,
  (1,1,n-2)}_{ab\gamma_1\cdots\gamma_k}$ are the classical triple intersections.   
For threefolds an analogous formula was found in~\cite{Candelas:1990rm} and
the multicovering was explained in~\cite{Aspinwall:1991ce}. Note that
$b^{0}_{a \gamma_1\cdots\gamma_k }$, $a^{0}_{\gamma_1\cdots\gamma_k}$ are
irrelevant for the quantum cohomology, as the latter is defined by the 
second derivative of $F^0(\gamma_1,\ldots, \gamma_k)$.

Genus one Gromov-Witten invariants exist on Calabi-Yau manifold of all 
dimensions with the need of incidence conditions as discussed above. For fourfolds 
the authors of \cite{Klemm:2007in} define the following integrality 
condition    
\begin{equation}
\begin{array}{rl}  
F^1 =
&\ds{\sum_{\beta>0} n^1_\beta \tfrac{\sigma(d)}{d} q^{d\beta}} \\[3 mm ]
&\ds{ +\tfrac{1}{24} \sum_{\beta>0} n^0_\beta(c_2(\tilde X)) \log(1-q^\beta)}\\[3 mm] 
& \ds{-\tfrac{1}{24} \sum_{\beta_1,\beta_2} m_{\beta_1,\beta_2} \log(1-q^{{\beta_1+\beta_2}})}\ . 
\end{array}  
\end{equation} 
Here the $m_{\beta_1,\beta_2}$ are so called meeting invariants, which
are likewise integer as the $n^g_\beta(\cdot)$
and the function $\sigma$ is defined by $\sigma(d)=\sum_{i|d}i$.

Note for threefolds one also has the
BPS state counting formula~\cite{Gopakumar:1998jq}
\begin{equation}
F(\lambda,q)=\sum_{g=0}^\infty \lambda^{2g-2}F^g = \sum_{\beta>0,g\ge
  0,k>0} n^g_\beta\tfrac{1}{k} \left(2 \sin \tfrac{k
  \lambda}{2}\right)^{2g-2} q^{k\beta}\ .
\label{closedBPS}
\end{equation}   

\subsection{Open GW invariants} 

Let us come now to the open Gromov-Witten invariants on Calabi-Yau 
threefolds $\tilde Y$. They arise in the open topological A-model on $\tilde Y$. 
We consider a Calabi-Yau manifold $\tilde Y$ together with a special Lagrangian
submanifold $L$ and consider a map from an oriented open Riemann surface,
i.e.~Riemann surface with boundary
\begin{equation} 
\psi:\Sigma_{g,h}\rightarrow (\tilde Y,L) 
\end{equation}      
into the Calabi-Yau manifold $\tilde Y$. Here the Riemann surface is mapped with
a given winding number into $L$ such that the $h$ boundary circles $B_i$ of 
$\Sigma_{g,h}$ are mapped on non-trivial elements $\vec \alpha =(\alpha_1,\ldots ,\alpha_h)\in
H^1(L,\mathbb{Z})^{\oplus h}$. As in the closed case we do not 
consider marked points. For threefolds the moduli space  
${\cal  M}_{g,h}(\tilde Y,L,\beta,\vec \alpha, \mu)$ with the Maslov index $\mu$
 has virtual dimension zero \cite{MLiu1}. If
$H^1(L,\mathbb{Z})$ is non-trivial, the special Lagrangian has a
geometric deformation moduli. The open string moduli $\hat{t}_i$ are 
 complexifications of the geometric moduli by the Wilson-Loop integrals 
of the flat $U(1)$ gauge connection on the brane.      

The open BPS state counting formula analogous to (\ref{closedBPS}) 
was given in \cite{Ooguri:1999bv} and reads
\begin{equation} 
F(t,u)=\sum_{g=0,h=1}^\infty \lambda^{2g-2+h} F^g_{n_1,\ldots,n_h} (t) 
{\rm tr} U^{n_1}  \cdots  U^{n_h}=i \sum_{n=1}^{\infty}
\tfrac{n^g_{\beta,{\cal R}}}{2 n \sin\left(\tfrac{n\lambda}{2}\right)}
q^{ng}_\lambda q^{n\beta} {\rm Tr}_{\cal R} U^n\ .  
\end{equation}    
In particular the disk amplitude, which gives rise to the 
superpotential, is given by
\begin{equation} 
W=F^0_{h=1}=\sum_{\beta, m} {\rm Li}_2(q^\beta Q^m)
\label{wopgromovwitten}
\end{equation}     
with $Q=e^{2\pi i\hat{t}}$. Comparison with (\ref{g=0multicovering}) suggest 
that the counting problem of specific disks amplitudes can
be mapped to the counting of rational curves in fourfolds since the integrality
structure is the same and given by the $\Li_2$-structure.

\section{Conclusions}

In this work we have studied the holomorphic flux superpotential in F-theory 
compactifications on Calabi-Yau fourfolds. In F-theory the complex structure 
moduli of the elliptically fibered fourfold contain both the
closed string and seven-brane moduli of the associated IIB theory. Thus, a non-trivial 
$G_4$-flux induces a superpotential for both closed and open moduli of the Type IIB background.
We made use of this unified description of open and closed moduli to explicitly 
present the splitting of the fourfold complex structure moduli into threefold complex 
structure and brane moduli for a given example. This identification is further 
confirmed by heterotic/F-theory duality where we consistently matched F-theory 
and heterotic moduli. In particular, we used this to physically argue that the 
periods of the threefold $Y_3$ have to be contained in the periods of $X_4$,
since $Y_3$ arises as the compactification space of the heterotic string by duality. 
Furthermore, we recovered the flux superpotential of the Type IIB string and 
the open-closed superpotential on the seven-brane with gauge flux inducing a five-brane 
charge from the fourfold perspective and comment on its heterotic interpretation. 

The presence of a non-trivial superpotential for the complex structure moduli 
of the fourfold $X_4$ is of crucial importance in the study of F-theory vacua.
In particular, as initiated in refs.~\cite{Beasley:2008dc,Donagi:2008ca}
, F-theory provides 
a promising framework to model supersymmetric GUT models with a remarkably  
realistic phenomenology on non-compact Calabi-Yau fourfolds. 
In such non-compact scenarios one can tune the complex structure moduli 
by hand to obtain realistic settings. However, eventually one has to 
compactify these models as recently done in \cite{Blumenhagen:2009yv,Marsano:2009ym}. 
This yields a large set of dynamical complex structure 
moduli and only a detained study of the F-theory flux superpotential 
will show whether phenomenologically preferred settings can indeed be 
stabilized by fluxes. The compact setups of ref.~\cite{Blumenhagen:2009yv}
are realized on explicit Calabi-Yau fourfolds in a toric ambient space, 
and it would be interesting to analyze the superpotential for a relevant 
subset of the complex structure moduli by the guideline presented in this work.

Let us summarize the methods applied in this work. 
Since the whole analysis of this work is based on the extensive use of mirror symmetry, 
we presented an account of basic facts and methods of closed and open mirror symmetry 
in section \ref{SuperpotsMirrors}. After discussing the general structure of flux 
and D7-brane superpotentials in Calabi-Yau orientifolds we introduced 
the necessary technical machinery to construct Calabi-Yau hypersurfaces in compact toric 
Fano varieties. For open mirror symmetry we recalled the non-compact setting with Harvey-Lawson 
type branes dual to D5-branes which was crucial for our later re-interpretation of the open 
Gromov-Witten invariants as
invariants originating from a seven-brane in the B-model geometry. 
In our F-theory analysis of section \ref{ffConstruction} the whole Riemann surface of 
the local mirror geometry was contained in the discriminant locus of the elliptic 
fibration that encodes seven-branes in F-theory. The open 
superpotential was induced on this seven-brane since a world-volume gauge flux induced 
five-brane charge.

Before delving into the details of these calculations we briefly 
introduced in section \ref{fluxinF} the general concepts of F-theory and the toric 
construction of elliptic fourfolds with particular fibration structures encoding the 
non-perturbative physics of seven-branes as well as heterotic dual geometries.
Here we put special emphasis on the flux superpotential and the supersymmetric F-theory 
fluxes. We completed our picture by briefly mentioning the concepts of the spectral cover 
construction for heterotic/F-theory duality, that we later on applied to identify moduli on both sides.
After this preparation we started to embed the local A-model geometries with Harvey-Lawson type
branes into compact threefolds in section \ref{ffConstruction}. For the sake of clarity we chose as an
explicit example local $\P^2$. We described the geometry of the
elliptically fibered threefold $\tilde Y_3$ and its mirror
$Y_3$ which is also elliptically fibered. We put strong emphasis on the
Weierstrass model of $Y_3$ and checked that in the limit of large
elliptic fiber, we obtained the local B-model geometry, a conic over an
elliptic curve.
Next, we gave a construction to
associate a compact fourfold $\tilde X_4$, i.e.\ a five-dimensional
reflexive polyhedron to the compact Calabi-Yau threefold $\tilde Y_3$.
The fourfold $\tilde X_4$ contained $\tilde Y_3$ as generic fiber with
base $\P^1$. There is a rich fibration structure on $\tilde X_4$ and we
showed that the mirror geometry $X_4$ which was relevant for our study is
also elliptically fibered such that F-theory is well-defined. We studied
the geometry of $X_4$ in great detail and showed that the discriminant
locus of $X_4$ contains a component which corresponds to the seven-brane in F-theory.
Furthermore, both $X_4$ and its mirror are fibered by an elliptic $K3$ which 
allows for a heterotic dual compactified on the elliptic threefold $Y_3$. 
For both the F-theory and the heterotic side we could determine the splitting 
of the complex structure moduli of $X_4$ into complex structure moduli and brane 
respectively spectral cover moduli.

To actually compute the superpotential on fourfolds and to identify the threefold periods, we described mirror symmetry on Calabi-Yau fourfolds in more detail and also more
formal aspects related to the topological A- and B-model defined on
fourfolds.  We took also a closer look on the properties of the periods
of the fourfolds since the flux superpotential could be expressed as a
linear sum of them. Here we put special emphasis on the determination of the 
integral base of $H^{(2,2)}_V(\tilde X_4)$. This made necessary to determine 
the classical terms in the maximally logarithmic period \eqref{Pi4} as well as 
to perform the analytic continuation and monodromy analysis for the periods of sextic hypersurfaces. 
We used the identification of prepotentials \eqref{matchPrepot} implying a 
matching of classical terms as well as instanton numbers of the corresponding 
periods to match the periods of $Y_3$.
This way we determined a $G_4$-flux to match the flux superpotential in the form of \eqref{FP2flux}. A similar analysis allowed for a $G_4$-flux matching the brane superpotential in \eqref{FP2branepot}.
Thus we have
explicitly checked, on the level of computing the superpotential and identification 
of periods, that the complex structure moduli space or the flux superpotential of
F-theory contains the closed/open moduli space of the Type IIB theory
with seven-branes. Equivalent statements could of course be inferred for the moduli 
space of the heterotic dual theory. 

We concluded with section \ref{EnumGeo} where we provided a basic background in enumerative geometry and introduced some of the notions we used throughout our work. Further examples of fourfolds with more moduli were discussed in two appendices.

For possible future directions it would be interesting to analyze the structure of periods on 
fourfolds at other points of the moduli space including a general analysis of monodromies. 
Physical interpretations of their different structure compared to the threefold case might 
shed some light on additional massless states tightly related to singularities in the moduli space. 
Furthermore it might repay further studies to deduce the classical terms of the maximally 
logarithmic period from first principles. 
From a physical point of view a more thorough analysis on the other $G_4$-flux choices not 
considered in our work would be useful. In particular we note that the F-theory superpotential 
intrinsically contains non-perturbative corrections due to a non-trivial dilaton 
dependence of the fourfold periods that treat it on an equal footing to ordinary complex structure moduli.

\subsection*{Acknowledgements}

We gratefully acknowledge discussions with Ralph Blumenhagen, Babak Haghighat, Hans Jockers, Wolfgang Lerche, Adrian Mertens, Marco Rauch, Johannes Walcher and Timo Weigand.
TG~would like to thank the KITP Santa Barbara, the MPI Munich and the 
Simons Center for Geometry and Physics, Stony Brook, for hospitality and support.
This work was supported in parts by the European Union 6th 
framework program MRTN-CT-2004-503069 ``Quest for unification'', 
MRTN-CT-2004-005104 ``ForcesUniverse'', MRTN-CT-2006-035863 ``UniverseNet'',
SFB-Transregio 33 ``The Dark Universe'' by the DFG. 
The work of T.-W.H.\ and D.K.\ is supported by the German Excellence Initiative via the graduate school ``Bonn Cologne Graduate School". The work of D.K.\ is supported by a scholarship of the ``Deutsche Telekom Stiftung".

\appendix

\section{Further topological data of the main example} \label{FurtherP2}

Here we supply the topological data of the fourfold $\tilde{X}_4$ that was omitted in the main text for convenience. Besides the intersection rings we will also present the full Picard-Fuchs system at the large radius/large complex structure point. These determine as explained in section \ref{FFMirrors} the primary vertical subspace $H^{p,p}_{V}(\tilde X_4)$ of the A-model. 

As was mentioned before there are four triangulations whereas only three yield non-singular varieties. Again we restrict our exposition to the two triangulations mentioned in section \ref{fourfolds}.
For the following we label the points in the polyhedron 
$\Delta_5^{\tilde X}$ given in \eqref{eqn:vl-for-p2} consecutively by $\nu_i$, $i=0,\ldots,9$
and associated coordinates $x_i$ to each $\nu_i$. Then the toric divisors are given by 
$D_i:=\{x_i=0\}$. 

\textbf{Phase I:} In phase I of the toric variety defined by the polyhedron $\Delta_5^{\tilde X}$ in \eqref{eqn:vl-for-p2} one has the following Stanley-Reisner ideal 
\begin{equation}
	SR=\{D_3 D_8, D_7 D_9, D_8 D_9, D_1 D_5 D_6,
        D_2 D_3 D_4, D_2 D_4 D_7\}\ .
\end{equation}
From this we compute by standard methods of toric geometry the intersection numbers 
\begin{align} \label{IntX_TextEx}
	\nn\mcal C_0&= 
                 J_4(J_1^2 J_2 + J_1 J_3 J_2 + J_3^2 J_2 + 3 J_1 J_2^2 + 3 J_3 J_2^2+9 J_2^3) + J_1^2 J_3 J_2  
                 + J_1 J_3^2 J_2 +  J_3^3 J_2 \\
 \nn&\quad +2 J_1^2 J_2^2  + 4 J_1 J_3 J_2^2  + 4 J_3^2 J_2^2 + 11 J_1 J_2^3 + 15 J_3 J_2^3 + 46 J_2^4,\\
 \mcal C_2&= 24J_1^2+36 J_1 J_4 + 48 J_1 J_3 + 36 J_4 J_3 + 48J_3^2+ 12 8 J_1 J_2 +102 J_2 J_4 \\
    &\phantom{=}+172 J_2 J_3 + 530 J_2^2 \ , \nn \\
 \nn\mcal C_3&= -660 J_1-540J_4-900 J_3-2776 J_2\ .
\end{align}
Here we denoted generators of the K\"ahler cone of \eqref{KCPhaseI} dual to the Mori cone by $J_i$ as before.
The notation for the $\mathcal{C}_k$ is as follows. Denoting 
the dual two-forms to $J_i$  by $\omega_i$ the coefficients 
of the top intersection ring $\mcal C_0$ are the quartic
intersection numbers $J_i\cap J_j\cap J_k\cap J_l=\int_{\tilde X_4}
\omega_i\wedge\omega_j\wedge\omega_k\wedge \omega_l$, while the coefficients
of $\mcal C_2$ and $\mcal C_3$ are $[c_2({\tilde X_4})]\cap J_i \cap J_j= \int_{\tilde X_4} c_2\wedge
\omega_i\wedge \omega_j$ and  $[c_3({\tilde X_4})]\cap J_i= \int_{\tilde X_4} c_3\wedge
\omega_i$ respectively. 

As reviewed in section \ref{matching} the Picard-Fuchs operators of the mirror fourfold $X_4$ at the large complex structure point are calculated by the methods described in \cite{Hosono:1993qy}. In the appropriate coordinates $z_i$ defined by \eqref{LargeComplexStr} and evaluated in \eqref{zI} we obtain the full Picard-Fuchs system on $\tilde X_4$ given by
\begin{eqnarray} \label{PFOFFP2T1}
	\nn\mcal L_1^I&=& -\theta_1^2 (\theta_1+\theta_4-\theta_3)\\
	\nn&\phantom{=}&-(-1+\theta_1-\theta_3) (-2+2 \theta_1+\theta_4+\theta_3-\theta_2) (-1+2 \theta_1+\theta_4+\theta_3-\theta_2) z_1\,,\\
	\mcal L_2^I&=& \theta_2 (-2 \theta_1-\theta_4-\theta_3+\theta_2)-12 (-5+6 \theta_2) (-1+6 \theta_2) z_2\,,\\
	\nn\mcal L_3^I&=& (\theta_1-\theta_3) (-\theta_4+\theta_3)-(1+\theta_1+\theta_4-\theta_3) (-1+2 \theta_1+\theta_4+\theta_3-\theta_2) z_3\,,\\
	\nn\mcal L_4^I&=& \theta_4 (\theta_1+\theta_4-\theta_3)-(-1+\theta_4-\theta_3) (-1+2 \theta_1+\theta_4+\theta_3-\theta_2) z_4\,.	
\end{eqnarray}
Now we calculate the ring $\mathcal{R}$ given by the orthogonal complement of the ideal of Picard-Fuchs operators defined in \eqref{ring}. Using the isomorphism $\theta_i\mapsto J_i$ discussed in section \ref{matching} we obtain the topological basis of $H^{p,p}_{V}(\tilde X_4)$ by identification with the graded ring $\mcal{R}^{(p)}$. Since $J_i$ form the trivial basis of $H^{1,1}(\tilde X_4)$ and $H^{3,3}(\tilde X_4)$ is fixed by duality to $H^{1,1}(\tilde X_4)$, the non-trivial part is the cohomology group $H^{2,2}_{V}(X)$. We calculate the ring $\mathcal{R}^{(2)}$ by choosing the basis
\begin{eqnarray}
\nn	& \cR_1^{(2)}=\theta_1^2,\quad \cR_2^{(2)}=\theta_4(\theta_1 +\theta_3),\quad \cR_3^{(2)}=\theta_3( \theta_1+\theta_3),\quad \cR_4^{(2)}=\theta_2(\theta_1+2 \theta_2),\quad \\
	& \cR_5^{(2)}=\theta_2 (\theta_4+\theta_2),\quad \cR_6^{(2)}=\theta_2(\theta_3+\theta_2)\,.&
\end{eqnarray}
Then we can use the intersection ring $\mathcal{C}_0$ to determine the topological metric $\eta^{(2)}$ of \eqref{amodeltopologicametric} given by
\begin{equation}
	\eta^{(2)}_I=
\left(
\begin{array}{cccccc}
 0 & 0 & 0 & 4 & 3 & 3 \\
 0 & 0 & 0 & 14 & 6 & 8 \\
 0 & 0 & 0 & 18 & 10 & 10 \\
 4 & 14 & 18 & 230 & 124 & 137 \\
 3 & 6 & 10 & 124 & 64 & 73 \\
 3 & 8 & 10 & 137 & 73 & 80
\end{array}
\right)\,.
\end{equation}
The entries are just the values of the integrals
\begin{equation}
 	\cR^{(2)}_\alpha \cR^{(2)}_\beta=\int_{\tilde{X}_4}(\cR^{(2)}_\alpha \cR^{(2)}_\beta)|_{\theta_i\mapsto J_i}\,,
\end{equation}
where we think of it in terms of the Poincar\'e duals and the quartic intersections are given as the coefficients of monomials in $\mathcal{C}_0$.
The basis $\cR^{(3)}_i$ at grade $p=3$ is determined by requiring $\eta^{(3)}_{ab}=\delta_{a,h^{1,1}-b+1}$ where $h^{1,1}=4$ for the case at hand.
Then the basis reads 
\begin{eqnarray}
& \cR_1^{(3)}=\theta_1(-\theta_1 \theta_4-\theta_2 \theta_4+\theta_2 \theta_3),\quad \cR_2^{(3)}=\theta_1(-\theta_1 \theta_4+\theta_1 \theta_2+\theta_2 \theta_4- \theta_2 \theta_3),&\nn \\ 
& \cR_3^{(3)}=\theta_1^2 \theta_4,\quad \cR_4^{(3)}=\theta_1(-2 \theta_1 \theta_4-\theta_1 \theta_2+\theta_2 \theta_3)\,.&
\end{eqnarray}
Finally, we choose a basis of $\cR^{(4)}$ by $\cR^{(4)}=\tfrac{1}{103}\mathcal C_0|_{J_i\mapsto \theta_i}$ such that $\eta^{(4)}_{a_0,b_0}=1$ for $\cR^{(0)}=1$.

\textbf{Phase II:} Turning to the phase II of \eqref{eqn:vl-for-p2} the Stanley-Reisner ideal and the intersection numbers read
\begin{eqnarray}
	\nn SR&=& \{D_1 D_7,D_7 D_9,D_8 D_9, D_1 D_5 D_6, D_2 D_3 D_4, D_2  D_4 D_7, D_3 D_5 D_6 D_8\},\\
  \nn\mathcal C_0&=& J_1^2 J_4 J_3+2 J_1^2 J_3^2+3 J_1 J_4 J_3^2+12 J_1 J_3^3+9 J_4 J_3^3+54 J_3^4+J_1^2 J_2 J_4+2 J_1^2 J_3 J_2\\
\nn&\phantom=&+3 J_1 J_2 J_3 J_4+12 J_1 J_3^2 J_2 +9 J_2 J_3^2 J_4+54 J_3^3 J_2+2 J_1^2 J_2^2+3 J_1 J_4 J_2^2+12 J_1 J_3 J_2^2\\
 &\phantom=&+9 J_4 J_3 J_2^2+54 J_3^2 J_2^2+11 J_1 J_2^3+9 J_4 J_2^3+51 J_3 J_2^3+46 J_2^4\,, \nn \\
\nn\mathcal C_2 &=&24 J_1^2+36 J_1 J_4+138 J_1 J_3+102 J_4 J_3+618 J_3^2+128 J_1 J_2 +102 J_2 J_4\\
 &&+588 J_3 J_4+530 J_4^2\,,\\
\nn\mathcal C_3&=& 660J_1- 540 J_4-3078 J_3-2776 J_2\,,
\end{eqnarray}
where the K\"ahler cone generators were given in \eqref{KCPhaseII}.

The complete Picard-Fuchs system consists of four operators given by
\begin{eqnarray} \label{PFOFFP2T2}
 	\nn\mcal L_1^{II}&=&-\theta _1^2 \left(\theta _1+\theta _2-\theta _3\right)\\
	&\nn\phantom{=}&- \left(-3+3 \theta _1-\theta _3+2 \theta _4\right) \left(-2+3 \theta _1-\theta _3+2 \theta _4\right) \left(-1+3 \theta _1-\theta _3+2 \theta _4\right)z_1\,,\\
	\nn\mcal L_2^{II}&=&-\theta _2 \left(\theta _1+\theta _2-\theta _3\right) \left(\theta _2-\theta _3+\theta _4\right)-12 \left(-5+6 \theta _2\right) \left(-1+6 \theta _2\right) \left(-1+\theta _2-\theta _3\right)z_2\,, \\
	\mcal L_3^{II}&=&-\left(\theta _2-\theta _3\right) \left(-3 \theta _1+\theta _3-2 \theta _4\right)-\left(1+\theta _1+\theta _2-\theta _3\right) \left(1+\theta _2-\theta _3+\theta _4\right)z_3\,,\\
	\nn\mcal L_4^{II}&=&
\theta _4 \left(\theta _2-\theta _3+\theta _4\right)-\left(-2+3 \theta _1-\theta _3+2 \theta _4\right) \left(-1+3 \theta _1-\theta _3+2 \theta _4\right)z_4 \,.
\end{eqnarray}
This enables us to calculate $H^{p,p}_{V}(\tilde X_4)$ as before. The basis at grade $p=2$ reads
\begin{eqnarray}
\nn	& \cR_1^{(2)}=\theta_1^2,\quad \cR_2^{(2)}=\theta_2(2 \theta_1+6 \theta_3),\quad \cR_3^{(2)}=\theta_3(\theta_1+3 \theta_3),\quad \cR_4^{(2)}=\theta_1 \theta_4,\quad \\
	& \cR_5^{(2)}=\theta_2^2,\quad \cR_6^{(2)}=\theta_3(2 \theta_2 +2 \theta_3+\theta_4)+\theta_2 \theta_4\,,&
\end{eqnarray}
for which the topological metric $\eta^{(2)}$ is given by
\begin{equation}
 	\eta^{(2)}=\left(
\begin{array}{cccccc}
 0 & 12 & 6 & 0 & 2 & 10 \\
 12 & 2240 & 1120 & 20 & 328 & 1512 \\
 6 & 1120 & 560 & 10 & 174 & 756 \\
 0 & 20 & 10 & 0 & 3 & 12 \\
 2 & 328 & 174 & 3 & 46 & 228 \\
 10 & 1512 & 756 & 12 & 228 & 1008
\end{array}
\right)\,.
\end{equation}
Again the basis of $H^{3,3}(\tilde X_4)$ is fixed by  $\eta^{(3)}_{ab}=\delta_{a,h^{1,1}-b+1}$ to be
\begin{eqnarray}
& \cR_1^{(3)}=-\tfrac{1}{91} \left(182 \theta_1^2+25 \theta_2^2+\theta_1 (-225 \theta_2+85 \theta_3)\right) (\theta_1+\theta_2+\theta_3+\theta_4)\,,&\nn \\ 
& \cR_2^{(3)}=\tfrac{1}{91} \left(91 \theta_1^2+10 \theta_2^2+\theta_1 (\theta_2-57 \theta_3)\right) (\theta_1+\theta_2+\theta_3+\theta_4)\,, &\\\
&\cR_3^{(3)}=-\theta_1 (\theta_2-\theta_3) (\theta_1+\theta_2+\theta_3+\theta_4)\,,\nn&\\
& \nn \cR_4^{(3)}=-\tfrac{1}{91} \left(273 \theta_1^2+23 \theta_2^2+\theta_1 (-207 \theta_2+60 \theta_3)\right) (\theta_1+\theta_2+\theta_3+\theta_4)\,.&
\end{eqnarray}
We conclude with the basis of $H^{4,4}(\tilde X_4)$ fixed by $\cR^{(0)}=1$ as $\cR^{(4)}=\tfrac{1}{359}\mcal{C}_0|_{J_i\mapsto \theta_i}$.

\section{Further examples of fourfolds}
\label{appa}

Here we consider a broader class of Calabi-Yau fourfolds $(\tilde{X}_4,X_4)$ that are constructed as described in section \ref{ffConstruction} by fibering Calabi-Yau threefolds $\tilde{Y}_3$ over $\mathds{P}^1$. The threefolds we consider here are itself elliptically fibered over the two-dimensional base of the Hirzebruch surfaces $F_n$ for $n=0,1$,
\begin{equation}
 \begin{array}{cccc} F_n  & \rightarrow & \tilde{Y}_3 & \\
         && \downarrow &\\ 
         && \mathds{P}^1 & 
   \end{array}\,
\end{equation} 
Therefore, we will distinguish the constructed mirror pairs $(\tilde{X}_4,X_4)$ by the two-dimensional base $F_n$ we used to construct the threefold $\tilde{Y}_3$. 

In the following we will present the toric data of the threefolds $\tilde{Y}_3$ and fourfolds $\tilde{X}_4$ including some of their topological quantities. Then we will determine the complete system of Picard-Fuchs differential operators at the large complex structure point of the mirror Calabi-Yau fourfold and calculate the holomorphic prepotential $F^0$. From this we extract the invariants $n^g_\beta$ which are integer in all considered cases. Furthermore we show that there exists a subsector for these invariants that reproduces the closed and open Gromov-Witten invariants of the local Calabi-Yau threefolds obtained by a suitably decompactifying the elliptic fiber of the original compact threefold. 
This matching allows us to determine the four-form flux $G_4$ for the F-theory compactification on these fourfolds such that the superpotential \eqref{GVW-super} admits the split \eqref{SuperpotLimit} into flux and brane superpotential. 

\subsection{Fourfold with $F_0$} \label{FF0}

We start with an elliptically fibered Calabi-Yau threefold $\tilde{Y}_3$ with base given by the toric Fano basis of the zeroth Hirzebruch surface $F_0=\mathds{P}^1\times \mathds{P}^1$. Its polyhedron and charge vectors read
\begin{equation}\label{3foldellf0}
	\begin{pmatrix}[c|cccc|ccc]
	    	&   &  \Delta_4^{\tilde Y} &   &   	&  \ell^{(1)} & \ell^{(2)}& \ell^{(3)}\\ \hline
		v_0 & 0 & 0 & 0 & 0 	& -6   & 0 & 0  \\
		v^b_1 & 0 & 0 & 2 & 3 	&  1   &-2 &-2  \\
		v^b_2 & 1 & 0 & 2 & 3 	&  0   & 1 & 0  \\
		v^b_3 &-1 & 0 & 2 & 3 	&  0   & 1 & 0  \\
		v^b_4 & 0 & 1 & 2 &  3 	&  0   & 0 & 1  \\
		v^b_5 & 0 &-1 & 2 & 3 	&  0   & 0 & 1  \\
	        v^1   & 0 & 0 &-1 & 0 	&  2   & 0 & 0  \\
		v_2   & 0 & 0 & 0 &-1 	&  3   & 0 & 0
	\end{pmatrix}\,,
\end{equation}
where points in the base are again labelled by a superscript $^b$. There is one triangulation for which the Stanley-Reisner ideal in terms of the toric divisors $D_i=\{x_i=0\}$ takes the form
\begin{equation} 
 	SR=\{D_2D_3,D_4 D_5,D_1 D_6D_7\}.
\end{equation}
This threefold $\tilde{Y}_3$ has Euler number $\chi=-480$, $h_{1,1}=3$ and $h_{2,1}=243$, where the three K\"ahler classes correspond to the elliptic fiber and the two $\mathds{P}^1$'s of the base $F_0$. The intersection ring for this Calabi-phase in terms of the K\"ahler cone generators
\begin{equation}
 	J_1=D_1+2D_2+2D_4,\quad J_2=D_2,\quad J_3=D_4
\end{equation}
reads $\mathcal{C}_0=8J_1^3+2J_1^2J_3+2J_1^2J_2+J_1J_2J_3$ and $\mathcal{C}_2=92J_1+24J_2+24J_3$.

In the local limit $\mathcal{O}(K)\rightarrow F_0$ Harvey-Lawson type branes described by the brane charge vectors $\hat\ell^{(1)}=(-1,0,1,0,0)$ and $\hat\ell^{(1)}=(-1,0,0,1,0)$ were studied in \cite{Aganagic:2001nx}. To construct the Calabi-Yau fourfold $\tilde{X}_4$ we use the construction of section \ref{ffConstruction} with the brane vector $\hat\ell^{(1)}$ and expand $\Delta_4^{\tilde Y}$ to the polyhedron $\Delta_5^{\tilde X}$ and determine the Mori cone generators $\ell^{(i)}$ with $i=1,\ldots5$ for the four different triangulations of the corresponding Calabi-Yau phases. Here we display one of the four triangulations on which we focus our following analysis:

\begin{equation}
	\begin{pmatrix}[c|ccccc|ccccc]
	    	&   &   &\Delta_5^{\tilde X}  &   &   & \ell^{(1)} & \ell^{(2)} & \ell^{(3)} & \ell^{(4)} & \ell^{(5)} \\ \hline
		v_0 & 0 & 0 & 0 & 0 & 0          &-6 & 0   & 0 & 0 & 0  \\
		v_1 & 0 & 0 & 2 & 3 & 0          & 1 &-1   &-2 &-1 &-1  \\
		v_2 & 1 & 0 & 2 & 3 & 0          & 0 & 1   & 0 & 0 & 0  \\
		v_3 &-1 & 0 & 2 & 3 & 0          & 0 & 0   & 0 & 1 &-1  \\
		v_4 & 0 & 1 & 2 & 3 & 0          & 0 & 0   & 1 & 0 & 0  \\
		v_5 & 0 &-1 & 2 & 3 & 0          & 0 & 0   & 1 & 0 & 0  \\
		v_6 & 0 & 0 &-1 & 0 & 0          & 2 & 0   & 0 & 0 & 0  \\
		v_7 & 0 & 0 & 0 &-1 & 0          & 3 & 0   & 0 & 0 & 0  \\
		v_8 &-1 & 0 & 2 & 3 &-1          & 0 & 1   & 0 &-1 & 1  \\
		v_9 & 0 & 0 & 2 & 3 &-1          & 0 &-1   & 0 & 1 & 0  \\
		v_{10} & 0 & 0 & 2 & 3&1         & 0 & 0   & 0 &0 & 1  
	\end{pmatrix}\,.
	\label{eqn:vl-for-f0}
\end{equation}
In this triangulation the Stanley-Reisner ideal takes the form
\begin{equation}
	SR=\{\Div_2 \Div_3,~ \Div_2 \Div_8,~ \Div_3 \Div_9,~ \Div_4 \Div_5,~ \Div_8 \Div_{10},~ \Div_9 \Div_{10},~ \Div_1 \Div_6 \Div_7\}.
\end{equation}
The generators of the K\"ahler cone of the fourfold $\tilde{X}_4$ in the given triangulation are 
\begin{equation}
	\Jiv_1 = \Div_1+2\Div_{10}+\Div_2+\Div_3+2\Div_4,\quad \Jiv_2 = \Div_{10},\quad \Jiv_3 = \Div_4,\quad \Jiv_4 = \Div_{10}+\Div_3,\quad \Jiv_5 = \Div_2\,,
\end{equation}
for which the intersections are determined to be
\begin{align} \label{intsFFF0}
	\nonumber \mcal C_0&=42 J_1^4+8 J_1^3 J_2+7 J_1^3 J_3+2 J_1^2 J_2 J_3+12 J_1^3 J_4+2 J_1^2
J_2 J_4+3 J_1^2 J_3 J_4+J_1 J_2 J_3 J_4\nn\\
	&\quad +2 J_1^2 J_4^2+J_1 J_3 J_4^2+8
J_1^3 J_5+2 J_1^2 J_2 J_5+2 J_1^2 J_3 J_5+J_1 J_2 J_3 J_5+2 J_1^2 J_4
J_5+J_1 J_3 J_4 J_5,\\
 	\nonumber \mcal C_2&=92\Jiv_1\Jiv_2+486\Jiv_1^2+24\Jiv_2\Jiv_3+82\Jiv_1\Jiv_3+24\Jiv_3\Jiv_5+92\Jiv_1\Jiv_5+24\Jiv_2\Jiv_5\\
	&\quad +24\Jiv_2\Jiv_4+138\Jiv_1\Jiv_4+36\Jiv_3\Jiv_4+24\Jiv_4\Jiv_5+24\Jiv_4^2,\\
	\nonumber \mcal C_3&=-2534\Jiv_1-480\Jiv_2-420\Jiv_3-720\Jiv_4-480\Jiv_5.
\end{align}
We calculate the core topological quantities to be
\begin{equation}
 	\chi=15408\ ,\quad h_{3,1}=2555\ , \quad h_{2,1}=0\ ,\quad h_{1,1}=5\ .
\end{equation}
 
Furthermore, we note that these intersections reveal the fibration structure of $\tilde{X}_4$. We recognize the Euler number of the threefold $\tilde{Y}_3$ as the coefficient of $J_2$ and $J_5$ in $\mathcal{C}_3$ and the fact that both $J_2$ and $J_5$ appear at most linear in $\mathcal{C}_0$, $\mathcal{C}_2$. This is consistent with the fact that the fiber $F$ of a fibration has intersection number $0$ with itself which implies $c_3(F)=c_3(\tilde{X}_4)$ using the adjunction formula as well as $c_1(F)+c_1(N_{\tilde{X}_4} F)=c_1(N_{\tilde{X}_4}F)=0$ for $\tilde{X}_4$ Calabi-Yau. Thus we observe a fibration of $\tilde{Y}_3$ represented by the classes $J_2$ and $J_5$ over the base curves corresponding to $\ell^{(2)}$, $\ell^{(5)}$, respectively.   

The Picard-Fuchs operators are determined as before and read
\begin{eqnarray}
	\nn\mcal L_1 &=& \theta _1 \left(\theta _1-\theta _2-2 \theta _3-\theta _4-\theta _5\right)-12  \left(-5+6 \theta _1\right) \left(-1+6 \theta _1\right)z_1,\\
	\nn\mcal L_2 &=& \theta _2 \left(\theta _2-\theta _4+\theta _5\right)-\left(-1+\theta _2-\theta _4\right) \left(-1-\theta _1+\theta _2+2 \theta _3+\theta _4+\theta _5\right)z_2 ,\\
	\mcal L_3 &=& \theta _3^2-\left(1+\theta _1-\theta _2-2 \theta _3-\theta _4-\theta _5\right) \left(2+\theta _1-\theta _2-2 \theta _3-\theta _4-\theta _5\right)z_3 ,\\
	\nn\mcal L_4 &=& \left(\theta _2-\theta _4\right) \left(\theta _4-\theta _5\right)- \left(1+\theta _2-\theta _4+\theta _5\right) \left(-1-\theta _1+\theta _2+2 \theta _3+\theta _4+\theta _5\right)z_4,\\
	\nn\mcal L_5 &=& \theta _5 \left(\theta _2-\theta _4+\theta _5\right)- \left(1+\theta _1-\theta _2-2 \theta _3-\theta _4-\theta _5\right) \left(1+\theta _4-\theta _5\right)z_5\ .
\end{eqnarray}
Then we can proceed with fixing the basis of $H^{(p,p)}_{V}(\tilde X_4)$ at each grade $p$ by determining the ring $\mathcal{R}$ of \eqref{ring}.
We choose a basis at grade $p=2$ as
\begin{eqnarray}
\nn &\cR_1^{(2)}=\theta_1 \left(\theta_1+\theta_5\right),\quad \cR_2^{(2)}=\theta_1 \left(\theta_1+\theta_2\right),\quad \cR_3^{(2)}=\theta_1 \left(2 \theta_1+\theta_3\right),\quad \cR_4^{(2)}=\theta_1 \left(\theta_1+\theta_4\right), &\\
& \cR_5^{(2)}=\theta_2 \theta_3,\quad \cR_6^{(2)}= \left(\theta_2+\theta_4\right) \left(\theta_4+\theta_5\right), \quad \cR_7^{(2)}=\theta_3 \theta_4, \quad \cR_8^{(2)}=\theta_3 \theta_5\,.&
\end{eqnarray}
The basis of solution dual to this basis choice is given by
\begin{eqnarray} \label{FFF0dual}
 	\nn&\mathbb{L}^{(2)}_1=\tfrac{1}{8} l_1 \left(l_1-l_2-2 l_3-l_4+7 l_5\right)\ ,\quad \mathbb{L}^{(2)}_2=\tfrac{1}{8} l_1 \left(l_1+7 l_2-2 l_3-l_4-l_5\right)\ , &\\ \nn
&\mathbb{L}^{(2)}_3=\tfrac{1}{4} l_1 \left(l_1-l_2+2 l_3-l_4-l_5\right)\ ,\quad \mathbb{L}^{(2)}_4=\tfrac{1}{8} l_1 \left(l_1-l_2-2 l_3+7 l_4-l_5\right)\ ,\quad \mathbb{L}^{(2)}_5=l_2 l_3\ , &\\ &\mathbb{L}^{(2)}_6=\tfrac{1}{4} \left(l_2+l_4\right) \left(l_4+l_5\right)\ ,\quad
\mathbb{L}^{(2)}_7=l_3 l_4\ ,\quad \mathbb{L}^{(2)}_8=l_3 l_5\ .&
\end{eqnarray}
The topological two-point coupling between the $\cR^{(2)}_\alpha$ in the chosen basis reads
\begin{equation}
\eta^{(2)}=\left(
\begin{array}{cccccccc}
 58 & 60 & 109 & 64 & 3 & 8 & 4 & 2 \\
 60 & 58 & 109 & 64 & 2 & 8 & 4 & 3 \\
 109 & 109 & 196 & 118 & 4 & 20 & 6 & 4 \\
 64 & 64 & 118 & 68 & 3 & 8 & 4 & 3 \\
 3 & 2 & 4 & 3 & 0 & 0 & 0 & 0 \\
 8 & 8 & 20 & 8 & 0 & 0 & 0 & 0 \\
 4 & 4 & 6 & 4 & 0 & 0 & 0 & 0 \\
 2 & 3 & 4 & 3 & 0 & 0 & 0 & 0
\end{array}
\right).
\end{equation}
The basis of $\mathcal{R}^{(3)}$ determining $H^{3,3}(\tilde{X}_4)$ that is fixed by Poincar\'e duality to the K\"ahler cone generators satisfying $\eta^{(3)}_{ab}=\delta_{a,h^{1,1}-b+1}$ is chosen to be
\begin{eqnarray}
\nn	&\cR_1^{(3)}=\tfrac{1}{4}(9 \theta_1 \theta_5- 2\theta_1 \theta_3-\theta_3^2)\theta_3+\theta_2 \theta_3^2 -\theta_1 \theta_2 \theta_5,&\\
\nn&\quad \cR_2^{(3)}=\tfrac{1}{8}( \theta_1 \theta_3+2\theta_3^2-10 \theta_1  \theta_5)\theta_3-\theta_2\theta_3^2 -\theta_1 \theta_2 \theta_5,\quad \cR_3^{(3)}=\theta_1 (\tfrac{1}{2} \theta_3^2- \theta_3 \theta_5-2 \theta_2 \theta_5),&\\
& \cR_4^{(3)}=\theta_1 \theta_2 \theta_5,\quad \cR_5^{(3)}=\tfrac{1}{8} \theta_3(2\theta_3^2-3\theta_1 \theta_3-10 \theta_1  \theta_5-4 \theta_2\theta_3 )-\theta_1 \theta_2 \theta_5\ .&
\end{eqnarray}
We choose the basis of $H^{4,4}(\tilde{X}_4)$ such that the volume is normalized as $\eta^{(4)}_{a_0,b_0}=1$ for $\cR^{(0)}=1$, i.e.~ $\cR^{(4)}=\tfrac{1}{96}\mcal C_0|_{J\mapsto \theta}$.

In order to fix the integral basis of $H^{(2,2)}_V(\tilde X_4)$ we again match the threefold periods from the fourfold periods via \eqref{matchPrepot}. The first step is to identify the K\"ahler classes of $\tilde Y_3$. As discussed above $J_5$ represents the class of the Calabi-Yau fiber $\tilde Y_3$. The intersections of $\tilde Y_3$ are obtained from \eqref{intsFFF0} upon the identification 
\begin{equation} \label{FFF0match}
 	J_1\ \leftrightarrow \ J_1(\tilde Y_3)\quad J_2+J_4\ , \leftrightarrow \ J_2(\tilde Y_3)\ ,\quad J_3\ \leftrightarrow \ J_3(\tilde Y_3)\,.
\end{equation}
With this in mind we calculate the leading logarithms $\mathbb{L}_\alpha( Y_3)$ on the threefold given by
\begin{equation}
 	\mathbb{L}_1( Y_3)=\tfrac12X_0(2 \tilde{l}_1+\tilde{l}_2) (2 \tilde{l}_1+\tilde{l}_3)\ ,\quad \mathbb{L}_2( Y_3)=\tfrac12X_0\tilde{l}_1 (\tilde{l}_1+\tilde{l}_3)\ , \quad \mathbb{L}_3( Y_3)=\tfrac12X_0\tilde{l}_1 (\tilde{l}_1+\tilde{l}_2)\  .
\end{equation}
This together with the requirement of matching the instanton numbers\footnote{We note here that by just matching the threefold instantons the solution on the fourfold could not be fixed. The two free parameters could only be determined by matching the classical terms, too.} $n_{d_1,d_2,d_3}$ of $\tilde Y_3$ via $n_{d_1,d_2,d_3,d_2,0}$ on $\tilde X_4$ fixes unique solutions of the Picard-Fuchs system
\begin{equation}
 	\mathbb{L}^{(2)}_1=\tfrac12X_0(2 l_1+l_3) (2 l_1+l_2+l_4)\ ,\quad \mathbb{L}^{(2)}_6= \tfrac12X_0l_1 (l_1+l_3)\ ,\quad \mathbb{L}^{(2)}_8=\tfrac12X_0l_1 (l_1+l_2+l_4)\, ,
\end{equation}
that upon \eqref{FFF0match} coincide with the threefold solutions.
This fixes three ring elements $\tilde{\cR}^{(2)}_{\alpha}$, $\alpha=1,6,8$, by the map induced from \eqref{intersecsLL} that we complete to a new basis 
\begin{eqnarray}
 	\nn&\tilde{\cR}^{(2)}_1=\tfrac{1}{8} \theta_1 \left(\theta_1-\theta_2-2 \theta_3-\theta_4+7 \theta_5\right)\ ,\quad \tilde{\cR}^{(2)}_2=\tfrac{1}{8} \theta_1 \left(\theta_1+7 \theta_2-2 \theta_3-\theta_4-\theta_5\right)\ , &\\ \nn
&\tilde{\cR}^{(2)}_3=\tfrac{1}{4} \theta_1 \left(\theta_1-\theta_2+2 \theta_3-\theta_4-\theta_5\right)\ ,\quad \tilde{\cR}^{(2)}_4=\tfrac{1}{8} \theta_1 \left(\theta_1-\theta_2-2 \theta_3+7 \theta_4-\theta_5\right)\ ,  &\\ &\tilde{\cR}^{(2)}_5=\theta_2 \theta_3\ ,\quad\tilde{\cR}^{(2)}_6=\tfrac{1}{4} \left(\theta_2+\theta_4\right) \left(\theta_4+\theta_5\right)\ ,\quad
\tilde{\cR}^{(2)}_7=\theta_3 \theta_4\ ,\quad \tilde{\cR}^{(2)}_8=\theta_3 \theta_5\ .&
\end{eqnarray}
Then the integral basis elements are given by
\begin{equation}
 	\hat{\gamma}^{(2)}_1=\tilde{\cR}^{(2)}_1\Omega|_{z=0}\ ,\quad \hat{\gamma}^{(2)}_6=\tilde{\cR}^{(2)}_6\Omega|_{z=0}\ ,\quad \hat{\gamma}^{(2)}_8=\tilde{\cR}^{(2)}_8\Omega|_{z=0}\, ,
\end{equation}
where again the new grade $p=2$ basis is obtained by replacing $l_i\leftrightarrow \theta_i$ in the dual solutions of \eqref{FFF0dual}. We conclude by presenting the leading logarithms of the periods $\Pi^{(2)\, \alpha}$ when integrating $\Omega$ over the duals $\gamma^{(2)\, \alpha}$ for $\alpha=1,6,8$. They are then as well given by 
$\mathbb{L}^{(2)\, 1}=X_0l_1 \left(l_1+l_5\right)$, $\mathbb{L}^{(2)\, 6}=X_0\left(l_2+l_4\right)\left(l_4+l_5\right)$ and $\mathbb{L}^{(2)\, 8}=X_0l_3l_5$.

Finally we determine a $\hat \gamma$ flux in $H^{2,2}_{H}(X_4)$ such that we match the disk invariants of \cite{Aganagic:2001nx} for both classes of the local geometry $\mathcal{O}(K)\rightarrow F_0$ with the brane class. 
Furthermore we reproduce the closed invariants of \cite{Haghighat:2008gw} for the two $\mathds{P}^1$-classes for zero brane winding $m=0$. 
First we identify in the polyhedron \eqref{eqn:vl-for-f0} the vector $\ell^{(4)}$ as corresponding to the brane vector. Then we expect to recover the disk invariants from the fourfold invariants $n_{0,d_1,d_2,d_1+m,0}$.
Then the flux $\hat{\gamma}$ deduced this way still contains a freedom of three parameters and takes the form
\begin{eqnarray}
 	\hat{\gamma}=(-\cR^{(2)}_5+\tfrac14\cR^{(2)}_6+\cR^{(2)}_7+\tfrac12\cR^{(2)}_8)\Omega|_{z=0}
\end{eqnarray}
 where we choose the free parameters $a_i$ in front of $\cR^{(2)}_1$, $\cR^{(2)}_2$, $\cR^{(2)}_3$ and $\cR^{(2)}_4$ to be zero. Note that $a_7=1$ is fixed by the requirement of matching the disk invariants. For this parameter choice the leading logarithmic structures of the corresponding period $\int_\gamma\Omega$ and of the solution matching the invariants are respectively given by
\begin{equation}
 	\mathbb{L}^{(2)\, \gamma}=X_0\left(l_2+l_4\right) \left(l_4+l_5\right)\ ,\quad \mathbb{L}^{(2)}_{\gamma}=\tfrac{1}{2} X_0l_1 \left(4 l_1+3 l_2+2 l_3+l_4\right)\ .
\end{equation}

\subsection{Fourfold with $F_1$} \label{FF1}

Here we consider an elliptically fibered Calabi-Yau threefold $\tilde{Y}_3$ with base twofold given by $F_1=\mathds{P}(\mathcal{O}\oplus \mathcal{O}(1))$ which is the blow-up of $\mathds{P}^2$ at one point. The polyhedron and charge vectors read
\begin{equation}\label{3foldellf1}
	\begin{pmatrix}[c|cccc|ccc]
	    	&   &  \Delta_4^{\tilde Y} &   &   	&  \ell^{(1)} & \ell^{(2)}& \ell^{(3)}\\ \hline
		v_0 & 0 & 0 & 0 & 0 	&  0   &-6 & 0  \\
		v^b_1 & 0 & 0 & 2 & 3 	& -1   & 0 &-2  \\
		v^b_2 & 1 & 1 & 2 & 3 	&  1   & 0 & 0  \\
		v^b_3 &-1 & 0 & 2 & 3 	&  1   & 0 & 0  \\
		v^b_4 & 0 & 1 & 2 &  3 	& -1   & 0 & 1  \\
		v^b_5 & 0 &-1 & 2 & 3 	&  0   & 0 & 1  \\
	        v^1   & 0 & 0 &-1 & 0 	&  0   & 2 & 0  \\
		v_2   & 0 & 0 & 0 &-1 	&  0   & 3 & 0
	\end{pmatrix}.
\end{equation}
where the labels by a superscript $^b$ again denote points in the base. There are two Calabi-Yau phases and for the triangulation given above the Stanley-Reisner ideal reads
\begin{equation} 
 	SR=\{D_2 D_3,D_4 D_5,D_1 D_6 D_7\}.
\end{equation}
This threefold has Euler number $\chi=480$, $h_{1,1}=3$ and $h_{2,1}=243$, where the three K\"ahler classes correspond to the elliptic fiber and the two $\mathds{P}^1$'s of the base $F_1$. The intersection ring for this Calabi-phase in terms of the K\"ahler cone generators
\begin{equation}
 	J_1=D_2,\quad J_2=D_1+3D_2+2D_4,\quad J_3=D_2+D_4
\end{equation}
reads $\mathcal{C}_0=2J_1J_2^2+8J_2^3+J_1J_2J_3+3J_2^2J_3+J_2J_3^2$ and $\mathcal{C}_2=24J_1+92J_2+36J_3$.

For the second Calabi-Yau phase we have the following data:
\begin{eqnarray}
\nn &
	\left(
\begin{array}{l|rrrrrrrrrrr}
 \ell^{(1)} & -6 & 0 & 1 & 1 & -1 & 0 & 2 & 3 \\
 \ell^{(2)} & 0 & -3 & 1 & 1 & 0 & 1 & 0 & 0 \\
 \ell^{(3)} & 0 & 1 & -1 & -1 & 1 & 0 & 0 & 0
\end{array}
\right), &\\
&SR=\{D_1\cdot D_4,D_4\cdot D_5,D_1\cdot D_6\cdot D_7,D_2\cdot D_3\cdot D_5,D_2\cdot  D_3\cdot D_6\cdot D_7\},&\\
\nn & J_1=D_1+3D_2+2D+4,\quad J_2=D_2+D_4,\quad J_3=D_1+3D_2+3D_4, &\\
\nn & \mathcal C_0=8J_1^3+3J_1^2J_2+J_1J_2^2 + 9J_1^2J_3+3J_1J_2J_3+J_2^2J_3+9J_1J_3^2+3J_2J_3^2+9J_3^3, &\\
\nn & \mathcal C_2=92J_1+36J_2+102J_3. &
\end{eqnarray}

Harvey-Lawson type branes were considered in \cite{Aganagic:2001nx} for the brane charge vectors $\hat\ell^{(1)}=(-1,1,0,0,0)$ and $\hat\ell^{(1)}=(-1,0,0,1,0)$ for the non-compact model $\mathcal{O}(K)\rightarrow F_1$. The Calabi-Yau fourfold $\tilde{X}_4$ is constructed from the brane vector $\hat\ell^{(1)}$ for which there are eleven triangulations.  Again we restrict our attention to one triangulation with the following data
\begin{eqnarray} \label{ellFFF1}
 	&	\left(
\begin{array}{l|rrrrrrrrrrr}
 \ell^{(1)} & 0 & -1 & 0 & -1 & 0 & 0 & 0 & 0 & 1 & 0 & 1 \\
 \ell^{(2)} & 0 & -1 & 0 & 1 & 0 & 0 & 0 & 0 & -1 & 1 & 0 \\
 \ell^{(3)} & 0 & 0 & 1 & 0 & -1 & 0 & 0 & 0 & 1 & -1 & 0 \\
 \ell^{(4)} & 0 & -2 & 0 & 0 & 1 & 1 & 0 & 0 & 0 & 0 & 0 \\
 \ell^{(5)} & -6 & 1 & 0 & 0 & 0 & 0 & 2 & 3 & 0 & 0 & 0
\end{array}
\right), &\nn\\
&SR=\{D_2\cdot D_3,~ D_2\cdot D_8,~ D_3\cdot D_9,~ D_4\cdot D_5,~ D_8\cdot D_{10},~ D_9\cdot D_{10},~ D_1\cdot D_6\cdot D_7\}\ ,&\nn\\
&\Jiv_1 = \Div_2,~ \Jiv_2 = \Div_1+2\Div_{10}+\Div_2+\Div_3+2\Div_4,~ \Jiv_3 = \Div_4,~ \Jiv_4 = \Div_{10},~ \Jiv_5 = \Div_{10}+\Div_3&\nn
\end{eqnarray}
with intersections
\begin{align} \label{intsFFF1}
 	\mcal C_0&=J_1 J_2 J_4 J_5 + J_2^2 J_4 J_5 + J_1 J_3 J_4 J_5 + J_2 J_3 J_4 J_5 + J_1 J_4^2 J_5 + J_2 J_4^2 J_5 \\
\nn&\quad +2 J_1 J_2 J_5^2 + 2 J_2^2 J_5^2 + 2 J_1 J_3 J_5^2 +2 J_2 J_3 J_5^2 + 3 J_1 J_4 J_5^2 + 4 J_2 J_4 J_5^2  \\
\nn &\quad +2 J_3 J_4 J_5^2 + 2 J_4^2 J_5^2 + 8 J_1 J_5^3  +12 J_2 J_5^3 + 8 J_3 J_5^3 + 11 J_4 J_5^3 + 42 J_5^4,	\\
\nn \mcal C_2 &= 24J_1J_2+24J_2^2+24J_1J_3 +24J_2J_3+36J_1J_4+48J_2J_4+24J_3J_4+24J_4^2\\
\nn &\quad +92J_1J_5+138J_2J_5+92J_3J_5+128J_4J_5+486J_5^2, \\
\nn \mcal C_3 &= -480J_1-270J_2-480J_3-660J_4-2534J_5\ .
\end{align}
Furthermore, we determine 
$\chi=15408$, $h^{3,1}=2555$, $h^{2,1}=0$ and $h^{1,1}=5$.

Again the Euler number of  the threefold $\tilde{Y}_3$ appears in $\mcal C_3$ in front of $J_1$ and $J_3$ confirming the fibration structure. By comparing the coefficient polynomial of $J_1$, $J_3$ with the threefold intersection rings presented in appendix \ref{FF0}, \ref{FF1} we infer that $J_1$ is precisely $\tilde Y_3=\cE\rightarrow F_1$, whereas $J_3$ is $\tilde Y_3'=\cE\rightarrow F_0$. Since we discussed $F_0$ in detail before we will just concentrate on the fibration structure involving $F_1$. 

The Picard-Fuchs operators of $X_4$ read as
\begin{eqnarray}
	\nn\mcal L_1&=& \theta_1 (\theta_1-\theta_2+\theta_3)-(-1+\theta_1-\theta_2) (-1+\theta_1+\theta_2+2 \theta_4-\theta_5) z_1,\\
	\nn\mcal L_2&=& (\theta_1-\theta_2) (\theta_2-\theta_3)-(1+\theta_1-\theta_2+\theta_3) (-1+\theta_1+\theta_2+2 \theta_4-\theta_5) z_2,\\
	\mcal L_3&=& -\theta_3 (\theta_1-\theta_2+\theta_3)-(1+\theta_2-\theta_3) (-1+\theta_3-\theta_4) z_3,\\
	\nn\mcal L_4&=& \theta_4 (-\theta_3+\theta_4)-(-2+\theta_1+\theta_2+2 \theta_4-\theta_5) (-1+\theta_1+\theta_2+2 \theta_4-\theta_5) z_4,\\
	\nn\mcal L_5&=& \theta_5 (-\theta_1-\theta_2-2 \theta_4+\theta_5)-12 (-5+6 \theta_5) (-1+6 \theta_5) z_5\,,
\end{eqnarray}
from which we determine the basis of $\cR^{(2)}$ as
\begin{eqnarray}
	 \nn&\cR_1^{(2)}=\left(\theta_1+\theta_2\right)\left(\theta_2+\theta_3\right),\quad \cR_2^{(2)}=\theta_1 \theta_4,\quad \cR_3^{(2)}=\theta _5 \left(\theta _1+\theta _5\right),\quad \cR_4^{(2)}=\theta_2 \theta_4,&\\
	&\quad \cR_5^{(2)}=\theta _5 \left(\theta _2+\theta _5\right),\quad \cR_6^{(2)}=\theta _4 \left(\theta _3+\theta _4\right),\quad \cR_7^{(2)}=\theta_3 \theta_5,\quad \cR_8^{(2)}=\theta _5 \left(\theta _4+2 \theta _5\right)\ \ \ \ \ \ &
\end{eqnarray}
with the two-point coupling
\begin{equation}
	\eta^{(2)}=	\left(
\begin{array}{cccccccc}
 0 & 0 & 8 & 0 & 8 & 0 & 0 & 20 \\
 0 & 0 & 3 & 0 & 4 & 0 & 1 & 7 \\
 8 & 3 & 58 & 5 & 64 & 6 & 10 & 114 \\
 0 & 0 & 5 & 0 & 5 & 0 & 1 & 9 \\
 8 & 4 & 64 & 5 & 68 & 6 & 10 & 123 \\
 0 & 0 & 6 & 0 & 6 & 0 & 0 & 8 \\
 0 & 1 & 10 & 1 & 10 & 0 & 0 & 18 \\
 20 & 7 & 114 & 9 & 123 & 8 & 18 & 214
\end{array}
\right)\ .
\end{equation}
The dual basis of solutions reads
\begin{eqnarray}
\nn&\mathbb{L}^{(2)\,1}=\tfrac{1}{4} (l_1+l_2) (l_2+l_3)\ ,\quad\mathbb{L}^{(2)\,2}=l_1 l_4\ ,\quad\mathbb{L}^{(2)\,3}=\tfrac{1}{7} l_5 (6 l_1-l_2-2 l_4+l_5)\ ,&\\\nn& \mathbb{L}^{(4)\,1}=l_2 l_4\ ,\quad \mathbb{L}^{(2)\,5}=\tfrac{1}{7} l_5 (-l_1+6 l_2-2 l_4+l_5)\ ,\quad\mathbb{L}^{(2)\,6}=\tfrac{1}{2} l_4 (l_3+l_4)\ ,\quad\mathbb{L}^{(2)\,7}=l_3 l_5\ ,&\\ &\mathbb{L}^{(2)\,8}=\tfrac{1}{7} l_5 (-2 l_1-2 l_2+3 l_4+2 l_5)\ .&
\end{eqnarray}
We determine $H^{(3,3)}(\tilde{X}_4)$ by duality to the canonical basis of $H^{(1,1)}(\tilde{X}_4)$ by the basis choice of $\cR^{(3)}$ given as
\begin{eqnarray}
 	&\cR_1^{(3)}=\theta_1 \theta_2 \theta_4,\quad \cR_2^{(3)}=-2 \theta_1 \theta_2 \theta_4 + \theta_1 \theta_2 \theta_5,\quad \cR_3^{(3)}=-\theta_1 \theta_2 \theta_5 + 
 \theta_2 \theta_4 \theta_5 - \theta_3 \theta_4 \theta_5,&\nn\\&\cR_4^{(3)}=-\theta_1 \theta_2 \theta_4 + \theta_1 \theta_4 \theta_5 - 
 \theta_2 \theta_4 \theta_5 + \theta_3 \theta_4 \theta_5,\quad \cR_5^{(3)}=-\theta_1 \theta_2 \theta_4 - \theta_1 \theta_4 \theta_5 + 
 \theta_2 \theta_4 \theta_5 \ .&\nn
\end{eqnarray}
Our choice for a basis of $H^{(4,4)}(\tilde{X}_4)$ is given by $\cR^{(4)}=\tfrac{1}{106}\mathcal C_0|_{J_i\mapsto \theta_i}$.

Again we fix the integral basis of $H^{(2,2)}(\tilde X_4)$ by the requirement of recovering the threefold periods from the fourfold ones. We readily identify the K\"ahler classes of the threefold $\tilde Y_3$ among the fourfold classes as
\begin{equation} \label{matchFF1}
 	J_2+J_3\ \leftrightarrow\ J_1(\tilde Y_3)\,,\quad J_5\ \leftrightarrow\ J_2(\tilde Y_3)\,,\quad J_4\ \leftrightarrow\ J_3(\tilde Y_3)\,,
\end{equation}
which matches the threefold intersections by identifying $J_1\equiv \tilde Y_3$ in the fourfold intersections \eqref{intsFFF1}.
Then we calculate the classical terms of the threefold periods to be
\begin{equation}
 	\mathbb{L}_1(\tilde Y_3)=\tilde{l}_2 (\tilde{l}_2+\tilde{l}_3 )\ , \quad \mathbb{L}_2(\tilde Y_3)=\tfrac{1}{2} (2\tilde{l}_2+\tilde{l}_3) (2 \tilde{l}_1+4 \tilde{l}_2+\tilde{l}_3)\ ,\quad \mathbb{L}_3(\tilde Y_3)=\tfrac{1}{2} l_2 (2 l_1+3 l_2+2 l_3)\,.
\end{equation}
On the fourfold $X_4$ we determine the periods that match this leading logarithmic structure. They are given by 
\begin{eqnarray}
 	\nn&\mathbb{L}^{(2)}_1=X_0l_5(l_4+l_5)\ ,\quad \mathbb{L}^{(2)}_2=\tfrac{1}{2}X_0 (l_4+2 l_5) (2 (l_2+ l_3)+l_4+4 l_5)\ ,&\\
&\quad \mathbb{L}^{(2)}_3=\tfrac{1}{2} X_0l_5 (2 (l_2+ l_3)+2 l_4+3 l_5)\,& 
\end{eqnarray}
and immediately coincide with the threefold result using \eqref{matchFF1}. It can be shown explicitly that the instanton series contained in the corresponding full solution matches the series on the threefold as well. The threefold invariants $n_{d_1,d_2,d_3}$ are obtained as $n_{0,d_1,d_1,d_3,d_2}$ from the fourfold invariants. To these solutions we associate using \eqref{intersecsLL} ring elements $\cR^{2}_\alpha$, $\alpha=1,3,2$, that we complete to a new basis as
\begin{eqnarray}
 	\nn&\tilde{\cR}^{(2)}_1=\tfrac{1}{4} (\theta _1+\theta _2) (\theta _2+\theta _3)\ ,\quad\tilde{\cR}^{(2)}_2=\theta _1 \theta _4\ ,\quad\tilde{\cR}^{(2)}_3=\tfrac{1}{7} \theta _5 (6 \theta _1-\theta _2-2 \theta _4+\theta _5)\ ,&\\ \nn
	&\tilde{\cR}^{(2)}_4=\theta _2 \theta _4\ ,\quad\tilde{\cR}^{(2)}_5=\tfrac{1}{7} \theta _5 (-\theta _1+6 \theta _2-2 \theta _4+\theta _5)\ ,\quad\tilde{\cR}^{(2)}_6=\tfrac{1}{2} \theta _4 (\theta _3+\theta _4)\ ,&\\
&\tilde{\cR}^{(2)}_7=\theta _3 \theta _5\ ,\quad\tilde{\cR}^{(2)}_8=\tfrac{1}{7} \theta _5 (-2 \theta _1-2 \theta _2+3 \theta _4+2 \theta _5)\ ,&
\end{eqnarray}
where we again note that the basis of dual solutions and the new ring basis coincide by $l_i\leftrightarrow \theta_i$.
Then the integral basis elements read
\begin{equation}
 	\hat{\gamma}^{(2)}_1=\tilde{\cR}^{(2)}_1\Omega|_{z=0}\ ,\quad \hat{\gamma}^{(2)}_2=\tilde{\cR}^{(2)}_2\Omega|_{z=0}\ ,\quad \hat{\gamma}^{(2)}_3=\tilde{\cR}^{(2)}_3\Omega|_{z=0}\ ,
\end{equation}
such that we obtain the full solution with the above leading parts $\mathbb{L}^{(2)}_\alpha$ as $\Pi^{(2)}_\alpha=\int\Omega\wedge\hat{\gamma}_\alpha$.
The leading behaviour of the periods $\Pi^{(2)\,\alpha}$ is then given as $\mathbb{L}^{(2)\, 1}=X_0(l_1+l_2)(l_2+l_3)$, $\mathbb{L}^{(2)\, 2}=X_0l_1 l_4$, $\mathbb{L}^{(2)\, 3}=X_0l _5 (l_1+l_5)$, respectively

We conclude by determining the flux element $\hat \gamma$ in $H^{(2,2)}_H(X_4)$
that reproduces the disk invariants in the phase II of \cite{Aganagic:2001nx}, where the local geometry $\mathcal{O}(K)\rightarrow F_1$ is considered. First we identify $\ell^{(2)}$ of the toric data in \eqref{ellFFF1} as the vector encoding the brane physics. Therefore, we expect the fourfold invariants $n_{0,m+d_1,d_1,d_2,0}$ to coincide with the disk invariants what can be checked in a direct calculation.
The ring element yielding this result reads $\hat{\gamma}= \cR_4^{(2)}$
where the free coefficients in front of the other ring elements were chosen to vanish.
The leading logarithmic parts of the period $\int_\gamma \Omega$ and of the solution $\Pi^{(2)}_{\gamma}=\int\Omega\wedge\hat{\gamma}\equiv W_\text{D7}$ respectively read
\begin{equation}
 	\mathbb{L}^{(2)}_{\gamma}=X_0l_5 (l_1+l_2+l_3+l_4+2 l_5)\ ,\quad \mathbb{L}^{(2)\,\gamma}=X_0l_2l_4\, .
\end{equation}


\begin{thebibliography}{99}

\bibitem{Lust:2004ks}
  D.~L\"ust,
  ``Intersecting brane worlds: A path to the standard model?,''
  Class.\ Quant.\ Grav.\  {\bf 21}, S1399 (2004)
  [arXiv:hep-th/0401156];\\[.1cm]
  R.~Blumenhagen, M.~Cvetic, P.~Langacker and G.~Shiu,
  ``Toward realistic intersecting D-brane models,''
  Ann.\ Rev.\ Nucl.\ Part.\ Sci.\  {\bf 55} (2005) 71
  [arXiv:hep-th/0502005].

\bibitem{fluxrev}
  M.~R.~Douglas and S.~Kachru,
  ``Flux compactification,''
  Rev.\ Mod.\ Phys.\  {\bf 79} (2007) 733
  [arXiv:hep-th/0610102].

 \bibitem{blumenhagen-flux}
  R.~Blumenhagen, B.~Kors, D.~Lust and S.~Stieberger,
  ``Four-dimensional String Compactifications with D-Branes, Orientifolds   and
  Fluxes,''
  Phys.\ Rept.\  {\bf 445}, 1 (2007)
  [arXiv:hep-th/0610327].

\bibitem{Denef:2008wq}
  F.~Denef,
  ``Les Houches Lectures on Constructing String Vacua,''
  arXiv:0803.1194 [hep-th].


\bibitem{Witten:1997ep}
  E.~Witten,
  ``Branes and the dynamics of {QCD},''
  Nucl.\ Phys.\  B {\bf 507} (1997) 658
  [arXiv:hep-th/9706109].


\bibitem{Witten:1992fb}
  E.~Witten,
  ``Chern-Simons Gauge Theory As A String Theory,''
  Prog.\ Math.\  {\bf 133} (1995) 637
  [arXiv:hep-th/9207094].

  
\bibitem{AV}
  M.~Aganagic and C.~Vafa,
  ``Mirror symmetry, D-branes and counting holomorphic discs,''
  arXiv:hep-th/0012041.

\bibitem{Aganagic:2001nx}
  M.~Aganagic, A.~Klemm and C.~Vafa,
  ``Disk instantons, mirror symmetry and the duality web,''
  Z.\ Naturforsch.\  A {\bf 57}, 1 (2002)
  [arXiv:hep-th/0105045].



\bibitem{LMW}
  W.~Lerche, P.~Mayr and N.~Warner,
  ``N = 1 special geometry, mixed Hodge variations and toric geometry,''
  arXiv:hep-th/0208039;\\[.1cm]
  W.~Lerche, P.~Mayr and N.~Warner,
  ``Holomorphic N = 1 special geometry of open-closed type II strings,''
  arXiv:hep-th/0207259.

\bibitem{Walcher}
  J.~Walcher,
  ``Calculations for Mirror Symmetry with D-branes,''
  arXiv:0904.4905 [hep-th];\\[.1cm]
  D.~Krefl and J.~Walcher,
  ``Real Mirror Symmetry for One-parameter Hypersurfaces,''
  JHEP {\bf 0809}, 031 (2008)
  [arXiv:0805.0792 [hep-th]];\\[.1cm]
  D.~R.~Morrison and J.~Walcher,
  ``D-branes and Normal Functions,''
  arXiv:0709.4028 [hep-th];\\[.1cm]
  J.~Walcher,
  ``Opening mirror symmetry on the quintic,''
  Commun.\ Math.\ Phys.\  {\bf 276}, 671 (2007)
  [arXiv:hep-th/0605162].

\bibitem{KSch}
  J.~Knapp and E.~Scheidegger,
  ``Matrix Factorizations, Massey Products and F-Terms for Two-Parameter
  Calabi-Yau Hypersurfaces,''
  arXiv:0812.2429 [hep-th];\\[.1cm]
  J.~Knapp and E.~Scheidegger,
  ``Towards Open String Mirror Symmetry for One-Parameter Calabi-Yau
  Hypersurfaces,''
  arXiv:0805.1013 [hep-th].

\bibitem{Jockers}
  H.~Jockers and M.~Soroush,
  ``Relative periods and open-string integer invariants for a compact
  Calabi-Yau hypersurface,''
  arXiv:0904.4674 [hep-th];\\[.1cm]
  H.~Jockers and M.~Soroush,
  ``Effective superpotentials for compact D5-brane Calabi-Yau geometries,''
  arXiv:0808.0761 [hep-th].


\bibitem{Lerche:2001cw}
  W.~Lerche and P.~Mayr,
  ``On N = 1 mirror symmetry for open type II strings,''
  arXiv:hep-th/0111113;\\[.1cm]
  P.~Mayr,
  ``N = 1 mirror symmetry and open/closed string duality,''
  Adv.\ Theor.\ Math.\ Phys.\  {\bf 5} (2002) 213
  [arXiv:hep-th/0108229].


\bibitem{Alim:2009rf}
  M.~Alim, M.~Hecht, P.~Mayr and A.~Mertens,
  ``Mirror Symmetry for Toric Branes on Compact Hypersurfaces,''
  arXiv:0901.2937 [hep-th].


\bibitem{Grimm:2008dq}
  T.~W.~Grimm, T.~W.~Ha, A.~Klemm and D.~Klevers,
  ``The D5-brane effective action and superpotential in N=1
  compactifications,''
  Nucl.\ Phys.\  B {\bf 816} (2009) 139
  [arXiv:0811.2996 [hep-th]].


\bibitem{Baumgartl:2007an}
  M.~Baumgartl, I.~Brunner and M.~R.~Gaberdiel,
  ``D-brane superpotentials and RG flows on the quintic,''
  JHEP {\bf 0707} (2007) 061
  [arXiv:0704.2666 [hep-th]].


\bibitem{Gukov:1999ya}
  S.~Gukov, C.~Vafa and E.~Witten,
  ``CFT's from Calabi-Yau four-folds,''
  Nucl.\ Phys.\  B {\bf 584} (2000) 69
  [Erratum-ibid.\  B {\bf 608} (2001) 477]
  [arXiv:hep-th/9906070].

\bibitem{Lust:2005bd}
  D.~Lust, P.~Mayr, S.~Reffert and S.~Stieberger,
  ``F-theory flux, destabilization of orientifolds and soft terms on
  D7-branes,''
  Nucl.\ Phys.\  B {\bf 732} (2006) 243
  [arXiv:hep-th/0501139].


\bibitem{Giddings:2001yu}
  S.~B.~Giddings, S.~Kachru and J.~Polchinski,
  ``Hierarchies from fluxes in string compactifications,''
  Phys.\ Rev.\  D {\bf 66}, 106006 (2002)
  [arXiv:hep-th/0105097].

\bibitem{Thomas:2001ve}
  R.~P.~Thomas,
  ``Moment maps, monodromy and mirror manifolds,''
  arXiv:math/0104196.


\bibitem{Hosono:1993qy}
  S.~Hosono, A.~Klemm, S.~Theisen and S.~T.~Yau,
  ``Mirror Symmetry, Mirror Map And Applications To Calabi-Yau Hypersurfaces,''
  Commun.\ Math.\ Phys.\  {\bf 167}, 301 (1995)
  [arXiv:hep-th/9308122].


\bibitem{Candelas:1990rm}
  P.~Candelas, X.~C.~De La Ossa, P.~S.~Green and L.~Parkes,
  ``A pair of Calabi-Yau manifolds as an exactly soluble superconformal
  theory,''
  Nucl.\ Phys.\  B {\bf 359} (1991) 21.


\bibitem{Hori}
  K.~Hori and C.~Vafa,
  ``Mirror symmetry,''
  arXiv:hep-th/0002222.

\bibitem{Batyrev}
  V.~V.~Batyrev,
  ``Dual polyhedra and mirror symmetry for Calabi-Yau hypersurfaces in toric
  varieties,''
  J.\ Alg.\ Geom.\  {\bf 3}, 493 (1994).


\bibitem{Klemm:1996ts}
  A.~Klemm, B.~Lian, S.~S.~Roan and S.~T.~Yau,
  ``Calabi-Yau fourfolds for M- and F-theory compactifications,''
  Nucl.\ Phys.\  B {\bf 518}, 515 (1998)
  [arXiv:hep-th/9701023].

\bibitem{Witten:1993yc}
  E.~Witten,
  ``Phases of N = 2 theories in two dimensions,''
  Nucl.\ Phys.\  B {\bf 403} (1993) 159
  [arXiv:hep-th/9301042].


\bibitem{Leung:1997tw}
  N.~C.~Leung and C.~Vafa,
  ``Branes and toric geometry,''
  Adv.\ Theor.\ Math.\ Phys.\  {\bf 2} (1998) 91
  [arXiv:hep-th/9711013].


\bibitem{Strominger:1996it}
  A.~Strominger, S.~T.~Yau and E.~Zaslow,
  ``Mirror symmetry is T-duality,''
  Nucl.\ Phys.\  B {\bf 479} (1996) 243
  [arXiv:hep-th/9606040].


\bibitem{Vafa:1996xn}
  C.~Vafa,
  ``Evidence for F-Theory,''
  Nucl.\ Phys.\  B {\bf 469}, 403 (1996)
  [arXiv:hep-th/9602022].

\bibitem{Bershadsky:1996nh}
  M.~Bershadsky, K.~A.~Intriligator, S.~Kachru, D.~R.~Morrison, V.~Sadov and C.~Vafa,
  ``Geometric singularities and enhanced gauge symmetries,''
  Nucl.\ Phys.\  B {\bf 481}, 215 (1996)
  [arXiv:hep-th/9605200].


\bibitem{Sen:1997gv}
  A.~Sen,
  ``Orientifold limit of F-theory vacua,''
  Phys.\ Rev.\  D {\bf 55} (1997) 7345
  [arXiv:hep-th/9702165].

 
\bibitem{Candelas:1996su}
  P.~Candelas and A.~Font,
  ``Duality between the webs of heterotic and type II vacua,''
  Nucl.\ Phys.\  B {\bf 511}, 295 (1998)
  [arXiv:hep-th/9603170].

\bibitem{Candelas:1997eh}
  P.~Candelas, E.~Perevalov and G.~Rajesh,
  ``Toric geometry and enhanced gauge symmetry of F-theory/heterotic  vacua,''
  Nucl.\ Phys.\  B {\bf 507}, 445 (1997)
  [arXiv:hep-th/9704097].



\bibitem{Witten:1996md}
  E.~Witten,
  ``On flux quantization in M-theory and the effective action,''
  J.\ Geom.\ Phys.\  {\bf 22} (1997) 1
  [arXiv:hep-th/9609122].


\bibitem{Haack:2001jz}
  M.~Haack and J.~Louis,
  ``M-theory compactified on Calabi-Yau fourfolds with background flux,''
  Phys.\ Lett.\  B {\bf 507}, 296 (2001)
  [arXiv:hep-th/0103068].

\bibitem{Grimm:2009sy}
  T.~W.~Grimm, T.~W.~Ha, A.~Klemm and D.~Klevers,
  ``Five-Brane Superpotentials and Heterotic/F-theory Duality,''
  arXiv:0912.3250 [hep-th].

\bibitem{Greene:1993vm}
  B.~R.~Greene, D.~R.~Morrison and M.~R.~Plesser,
  ``Mirror manifolds in higher dimension,''
  Commun.\ Math.\ Phys.\  {\bf 173} (1995) 559
  [arXiv:hep-th/9402119].

 
\bibitem{Mayr:1996sh}
  P.~Mayr,
  ``Mirror symmetry, N = 1 superpotentials and tensionless strings on
  Calabi-Yau four-folds,''
  Nucl.\ Phys.\  B {\bf 494}, 489 (1997)
  [arXiv:hep-th/9610162].


\bibitem{Kreuzer}
  A.~C.~Avram, M.~Kreuzer, M.~Mandelberg and H.~Skarke,
  ``Searching for K3 fibrations,''
  Nucl.\ Phys.\  B {\bf 494}, 567 (1997)
  [arXiv:hep-th/9610154].


\bibitem{Sen:1996vd}
  A.~Sen,
  ``F-theory and Orientifolds,''
  Nucl.\ Phys.\  B {\bf 475}, 562 (1996)
  [arXiv:hep-th/9605150].


\bibitem{Andreas:1998zf}
  B.~Andreas,
  ``N = 1 heterotic/F-theory duality,''
  Fortsch.\ Phys.\  {\bf 47} (1999) 587
  [arXiv:hep-th/9808159].



\bibitem{MV}
  D.~R.~Morrison and C.~Vafa,
  ``Compactifications of F-Theory on Calabi--Yau Threefolds -- I,''
  Nucl.\ Phys.\  B {\bf 473} (1996) 74
  [arXiv:hep-th/9602114];\\[.1cm]
  D.~R.~Morrison and C.~Vafa,
  ``Compactifications of F-Theory on Calabi--Yau Threefolds -- II,''
  Nucl.\ Phys.\  B {\bf 476} (1996) 437
  [arXiv:hep-th/9603161].


\bibitem{Bershadsky:1997zs}
  M.~Bershadsky, A.~Johansen, T.~Pantev and V.~Sadov,
  ``On four-dimensional compactifications of F-theory,''
  Nucl.\ Phys.\  B {\bf 505} (1997) 165
  [arXiv:hep-th/9701165].

\bibitem{Berglund:1998ej}
  P.~Berglund and P.~Mayr,
  ``Heterotic string/F-theory duality from mirror symmetry,''
  Adv.\ Theor.\ Math.\ Phys.\  {\bf 2}, 1307 (1999)
  [arXiv:hep-th/9811217];\\
  P.~Berglund and P.~Mayr,
  ``Stability of vector bundles from F-theory,''
  JHEP {\bf 9912}, 009 (1999)
  [arXiv:hep-th/9904114].




\bibitem{Friedman:1997yq}
  R.~Friedman, J.~Morgan and E.~Witten,
  ``Vector bundles and F theory,''
  Commun.\ Math.\ Phys.\  {\bf 187} (1997) 679
  [arXiv:hep-th/9701162].


\bibitem{Katz:1997eq}
  S.~Katz, P.~Mayr and C.~Vafa,
  ``Mirror symmetry and exact solution of 4D N = 2 gauge theories. I,''
  Adv.\ Theor.\ Math.\ Phys.\  {\bf 1}, 53 (1998)
  [arXiv:hep-th/9706110].

\bibitem{Curio:1998bva}
  G.~Curio and R.~Y.~Donagi,
  ``Moduli in N = 1 heterotic/F-theory duality,''
  Nucl.\ Phys.\  B {\bf 518} (1998) 603
  [arXiv:hep-th/9801057].




\bibitem{Candelas:1994hw}
  P.~Candelas, A.~Font, S.~H.~Katz and D.~R.~Morrison,
  ``Mirror symmetry for two parameter models. 2,''
  Nucl.\ Phys.\  B {\bf 429}, 626 (1994)
  [arXiv:hep-th/9403187].



\bibitem{Witten:1991zz}
  E.~Witten,
  ``Mirror manifolds and topological field theory,''
  arXiv:hep-th/9112056.


\bibitem{Hori:2003ic}
  K.~Hori {\it et al.},
  ``Mirror symmetry,''
{\it  Providence, USA: AMS (2003) 929 p}

\bibitem{Hosono:1994ax}
  S.~Hosono, A.~Klemm, S.~Theisen and S.~T.~Yau,
  ``Mirror symmetry, mirror map and applications to complete intersection
  Calabi-Yau spaces,''
  Nucl.\ Phys.\  B {\bf 433} (1995) 501
  [arXiv:hep-th/9406055].


\bibitem{Libgober} A. Libgober, ``Chern Classes and the periods of mirrors'',  arXiv:math/9803119. 



\bibitem{Minasian}
  R.~Minasian and G.~W.~Moore,
  ``K-theory and Ramond-Ramond charge,''
  JHEP {\bf 9711}, 002 (1997)
  [arXiv:hep-th/9710230].


\bibitem{Cheung}
  Y.~K.~Cheung and Z.~Yin,
  ``Anomalies, branes, and currents,''
  Nucl.\ Phys.\  B {\bf 517}, 69 (1998)
  [arXiv:hep-th/9710206].


\bibitem{Klemm:2007in}
  A.~Klemm and R.~Pandharipande,
  ``Enumerative geometry of Calabi-Yau 4-folds,''
  Commun.\ Math.\ Phys.\  {\bf 281}, 621 (2008)
  [arXiv:math/0702189].

\bibitem{Haghighat:2008gw}
  B.~Haghighat, A.~Klemm and M.~Rauch,
  ``Integrability of the holomorphic anomaly equations,''
  JHEP {\bf 0810} (2008) 097
  [arXiv:0809.1674 [hep-th]].

\bibitem{Aspinwall:1991ce}
  P.~S.~Aspinwall and D.~R.~Morrison,
  ``Topological field theory and rational curves,''
  Commun.\ Math.\ Phys.\  {\bf 151}, 245 (1993)
  [arXiv:hep-th/9110048].
\bibitem{Gopakumar:1998jq}
  R.~Gopakumar and C.~Vafa,
  ``M-theory and topological strings. II,''
  arXiv:hep-th/9812127.
\bibitem{MLiu1} 
C.-C. M. Liu, {\sl Moduli of J-holomorphic curves with Lagrangian
Boundary Conditions and open Gromov-Witten Invariants for an 
$S^1$-Equivariant Pair}, arXiv:math/0210257.         
\bibitem{Ooguri:1999bv}
  H.~Ooguri and C.~Vafa,
  ``Knot invariants and topological strings,''
  Nucl.\ Phys.\  B {\bf 577}, 419 (2000)
  [arXiv:hep-th/9912123].

\bibitem{Beasley:2008dc}
  C.~Beasley, J.~J.~Heckman and C.~Vafa,
  ``GUTs and Exceptional Branes in F-theory - I,''
  JHEP {\bf 0901} (2009) 058
  [arXiv:0802.3391 [hep-th]];\\[.1cm]
  C.~Beasley, J.~J.~Heckman and C.~Vafa,
  ``GUTs and Exceptional Branes in F-theory - II: Experimental Predictions,''
  JHEP {\bf 0901} (2009) 059
  [arXiv:0806.0102 [hep-th]].

\bibitem{Donagi:2008ca}
  R.~Donagi and M.~Wijnholt,
  ``Model Building with F-Theory,''
  arXiv:0802.2969 [hep-th].

\bibitem{Blumenhagen:2009yv}
  R.~Blumenhagen, T.~W.~Grimm, B.~Jurke and T.~Weigand,
  ``Global F-theory GUTs,''
  arXiv:0908.1784 [hep-th].

\bibitem{Marsano:2009ym}
  J.~Marsano, N.~Saulina and S.~Schafer-Nameki,
  ``F-theory Compactifications for Supersymmetric GUTs,''
  JHEP {\bf 0908}, 030 (2009)
  [arXiv:0904.3932 [hep-th]].



\end{thebibliography}
\end{document}